\documentclass[prd,preprint,floatfix,nofootinbib,superscriptaddress,showpacs]{revtex4} 

\pdfoutput=1
\usepackage{epsfig}
\usepackage[caption=false]{subfig}
\usepackage{amsmath}
\usepackage{amsfonts}
\usepackage{amssymb}
\usepackage{float}
\usepackage{color}
\usepackage{graphicx}
\usepackage{graphics}
\usepackage{hyperref}
\usepackage{adjustbox,array}
\hypersetup{}
\usepackage[utf8x]{inputenc} 
\usepackage{epstopdf}
\usepackage{multirow}
\usepackage{slashed}
\usepackage{booktabs}
\definecolor{rosso}{cmyk}{0,1,1,0.4}
\definecolor{rossos}{cmyk}{0,1,1,0.55}
\definecolor{rossoc}{cmyk}{0,1,1,0.2}
\definecolor{blu}{cmyk}{1,1,0,0.3}
\definecolor{blus}{cmyk}{1,1,0,0.6}
\definecolor{bluc}{cmyk}{1,1,0,0.1}
\definecolor{verde}{cmyk}{0.92,0,0.59,0.25}
\definecolor{verdec}{cmyk}{0.92,0,0.59,0.15}
\definecolor{verdes}{cmyk}{0.92,0,0.59,0.4}

\AtBeginDocument{
\heavyrulewidth=.08em
\lightrulewidth=.05em
\cmidrulewidth=.03em
\belowrulesep=.65ex
\belowbottomsep=0pt
\aboverulesep=.4ex
\abovetopsep=0pt
\cmidrulesep=\doublerulesep
\cmidrulekern=.5em
\defaultaddspace=.5em
}

\hypersetup{
 colorlinks=true,
 linkcolor=verdec,
 urlcolor=verdec,
 citecolor=verdec
}

\newcommand{ \eq}[1]{Eq.~(\ref{#1})}

\newcommand{\gsim}{\gtrsim}
\newcommand{\lsim}{\lesssim}

\newcommand{\lf}{\left(}
\newcommand{\ri}{\right)}

\newcommand{\nn}{\nonumber}

\newcommand{\sqt}{\sqrt{2}}

\newcommand{\rr}{{\gamma\gamma}}

\newcommand{\mco}{\mathcal{O}}

\newcommand{\br}{\mathcal{B}}
\newcommand{\hc}{{\rm H.c.}}

\newcommand{\sm}{{\rm SM}}

\newcommand{\pb}{{\;{\rm pb}}}
\newcommand{\fb}{{\;{\rm fb}}}

\newcommand{\gev}{{\;{\rm GeV}}}
\newcommand{\tev}{{\;{\rm TeV}}}

\newcommand{\beq}{\begin{equation}}
\newcommand{\eeq}{\end{equation}}
\newcommand{\bea}{\begin{eqnarray}}
\newcommand{\eea}{\end{eqnarray}}
\newcommand{\barr}{\begin{array}}
\newcommand{\earr}{\end{array}}
\newcommand{\bc}{\begin{center}}
\newcommand{\ec}{\end{center}}
\newcommand{\bit}{\begin{itemize}}
\newcommand{\eit}{\end{itemize}}
\newcommand{\ben}{\begin{enumerate}}
\newcommand{\een}{\end{enumerate}}

\newcommand{\al}{\alpha}
\newcommand{\bt}{\beta}

\newcommand{\dt}{\delta}
\newcommand{\Dt}{\Delta}

\newcommand{\sg}{\sigma}
\newcommand{\es}{\epsilon}

\newcommand{\gm}{\gamma}
\newcommand{\Gm}{\Gamma}
\newcommand{\lm}{\lambda}

\newcommand{\damu}{\Delta a_\mu}
\newcommand{\damuo}{\Delta a_\mu^{\rm obs}}
\newcommand{\dae}{\Delta a_e}

\newcommand{\hsm}{{h_{\rm SM}}}
\newcommand{\ch}{H^\pm}

\newcommand{\mh}{M_{h}}
\newcommand{\mch}{M_{H^\pm}}
\newcommand{\mhh}{M_{H}}
\newcommand{\ma}{M_{A}}

\newcommand{\tb}{t_\beta}
\newcommand{\tbi}{\frac{1}{t_\beta}}
\newcommand{\cb}{c_\beta}
\renewcommand{\sb}{s_\beta}

\newcommand{\cba}{c_{\beta-\alpha}}
\newcommand{\sba}{s_{\beta-\alpha}}

\newcommand{\ee}      {{e^+ e^-}}
\newcommand{\mmu}      {{\mu^+ \mu^-}}

\newcommand{\ttau}      {{\tau^+\tau^-}} 

\newcommand{\ttop}      {{t\bar{t}}}

\newcommand{\bb}      {{b \bar{b}}}

\newcommand{\qq}      {{q \bar{q}}}

\definecolor{mint}{rgb}{0.24, 0.71, 0.54}

\begin{document}

\title{\color{verde} Type-X two Higgs doublet model in light of the muon $g-2$: confronting Higgs and collider data}

\author{Adil Jueid}
\email{adiljueid@konkuk.ac.kr}
\address{Department of Physics, Konkuk University, Seoul 05029, Republic of Korea}
\author{Jinheung Kim}
\email{jinheung.kim1216@gmail.com}
\address{Department of Physics, Konkuk University, Seoul 05029, Republic of Korea}
\author{Soojin Lee}
\email{soojinlee957@gmail.com}
\address{Department of Physics, Konkuk University, Seoul 05029, Republic of Korea}
\author{Jeonghyeon Song}
\email{jhsong@konkuk.ac.kr}
\address{Department of Physics, Konkuk University, Seoul 05029, Republic of Korea}

\begin{abstract}
The recent Fermilab measurement of the muon anomalous magnetic moment
yields $4.2 \sigma$ deviations from the SM prediction when combined with the BNL E821 experiment results.
In the Type-X two Higgs doublet model,
we study the consequence of 
imposing the observed muon $g-2$, along with 
the constraints from theoretical stabilities, electroweak oblique parameters, Higgs precision data, and direct searches.
For a comprehensive study,
we scan the whole parameter space in two scenarios,
the normal scenario where $h_{\rm SM} = h$ and the inverted scenario where $h_{\rm SM}=H$,
where $h$ ($H$) is the light (heavy) \textit{CP}-even Higgs boson.
We found that large $\tan\beta$ (above 100) and light pseudoscalar mass $M_A$
are required to explain the muon $g-2$ anomaly. 
This breaks the theoretical stability unless the scalar masses satisfy $M_A^2 \simeq M_{H^\pm}^2 \simeq  m_{12}^2 \tan\beta \approx M_{H/h}^2$.
The direct search bounds at the LEP and LHC exclude the light $A$ window with $M_A \lesssim 62.5~$GeV.
We also show that the observed electron anomalous magnetic moment is consistent with the model prediction, but the lepton flavor universality data in the $\tau$ and $Z$ decays are not.
For a separate exploration of the model,
we propose the golden mode $pp \to A h/AH \to 4 \tau$ at the HL-LHC.
\end{abstract}


\maketitle
\tableofcontents

\section{Intoduction}
\label{sec:intro}

The recent measurement of the muon anomalous magnetic moment by the Fermilab National Accelerator Laboratory (FNAL) Muon $g - 2$ experiment~\cite{Abi:2021gix,Albahri:2021ixb}  
achieved unprecedented precision. 
When combined with the old result of the Brookhaven National Laboratory (BNL) E821 measurement~\cite{Bennett:2006fi}, it reads as
\bea
\label{eq:Damu:exp}
a_\mu^\text{exp}&=&116\,592\,061(41)\times 10^{-11},
\eea
where $a_\mu = (g - 2)_\mu/2$.
As the experimental error is becoming comparable with the theoretical error,\footnote{The experimental measurement will be improved in a short time scale. For instance, FNAL will provide a new measurement of $a_\mu$ in the summer of 2022 after including more datasets.} reliable and accurate calculation of the SM prediction is more important than ever.
The recent progress includes five loops in QED \cite{Aoyama:2012wk} and two loops in electroweak interactions \cite{Czarnecki:2002nt,Gnendiger:2013pva}. 
Nevertheless, the most dominant contribution is from the strong interaction dynamics 
at $\mathcal{O}(1)~{\rm GeV}$,
which is categorized into the hadronic vacuum polarization (HVP)~\cite{Kurz:2014wya, Davier:2017zfy, Colangelo:2018mtw, Keshavarzi:2018mgv, Keshavarzi:2019abf, Davier:2019can, Hoid:2020xjs, Colangelo:2020lcg} 
and the hadronic light-by-light (HLbL) scattering~\cite{Melnikov:2003xd, Colangelo:2014pva, Colangelo:2014qya, Colangelo:2015ama, Colangelo:2017fiz, Masjuan:2017tvw, Colangelo:2017qdm, Hoferichter:2018dmo, Hoferichter:2018kwz, Colangelo:2019lpu, Bijnens:2019ghy, Blum:2019ugy, Bijnens:2020xnl}. 
These QCD corrections cannot be computed using perturbation theory.
We have to resort to non-perturbative methods, either Lattice QCD or data-driven methods. 
On the Lattice side,
the recent calculation of the leading order HVP (LO-HVP) contributions to $a_\mu$ 
by the Budapest-Marseille-Wuppertal collaboration~\cite{Borsanyi:2020mff}
yields $a_\mu|_{{\rm LO-HVP}} = 707.5(5.3) \times 10^{-10}$.
If we take this result at face value, 
the Fermilab measurement of $a_\mu$ is consistent with the SM prediction at $\sim 2\sg$.
On the data-driven method side, however,
the calculation of the HVP contribution~\cite{Colangelo:2018mtw, Davier:2019can, Keshavarzi:2019abf}
supports the long-standing discrepancy between the muon $g-2$ experiment and the SM prediction, as
\bea
\label{eq:Damu}
\damuo&=&a_\mu^\text{exp}-a_\mu^\text{SM}=251(59)\times 10^{-11}.
\eea

Which method is more appropriate needs further investigation.
One checking point is the connection of the QCD corrections to electroweak precision fits~\cite{Lehner:2020crt, Crivellin:2020zul, Keshavarzi:2020bfy, Malaescu:2020zuc},
since some of the most important inputs to HVP and HLbL contributions come from measuring the $R(s)$-ratio in $e^+ e^-$ collisions.
Lately, some tension was reported between the Lattice result and the electroweak data~\cite{Crivellin:2020zul, Keshavarzi:2020bfy}.
Another critical topic is how to combine the probability distribution functions with different errors.

In this paper, we take the $4.2\sg$ deviation in \eq{eq:Damu},
which calls for new physics (NP) explanation.
In a short time, 
various NP models have been vigorously studied for the muon $g-2$,
focusing on a supersymmetric theory~\cite{Czarnecki:2001pv,Baer:2021aax,Aboubrahim:2021rwz,Cao:2021tuh,Wang:2021bcx,VanBeekveld:2021tgn,Abdughani:2021pdc,Baum:2021qzx,Ahmed:2021htr,Zhang:2021gun,Chakraborti:2021bmv,Athron:2021iuf,Yin:2021mls},
leptophilic boson model~\cite{Buras:2021btx,Chun:2021dwx},
singlet scalar model~\cite{Liu:2018xkx},
three Higgs doublet model~\cite{CarcamoHernandez:2021qhf},
leptoquark model~\cite{Ban:2021tos,Du:2021zkq},
$L_\mu - L_\tau$ model~\cite{Borah:2021jzu,Zu:2021odn},
$B-L$ or $B-3L$ gauge model~\cite{Yang:2021duj,Greljo:2021xmg},
flavorful scalar model~\cite{Zhu:2021vlz},
seesaw model~\cite{Escribano:2021css},
simplified model with minimal field contents~\cite{Arcadi:2021cwg},
effective field theory~\cite{Crivellin:2021rbq},
axion model~\cite{Buen-Abad:2021fwq,Ge:2021cjz}, 
two Higgs doublet model (2HDM)~\cite{Ferreira:2021gke,Han:2021gfu,Chen:2021jok,Ghosh:2020tfq,Ghosh:2021jeg,Li:2020dbg,Botella:2020xzf,Jana:2020pxx,Jana:2020joi,Anselmi:2021chp},
or
2HDM with a singlet scalar model~\cite{Keus:2017ioh,Sabatta:2019nfg}.
These efforts shall continue because each NP model as a solution for the observed $\damu$ should simultaneously explain a vast amount of experimental data in particle physics.

From this motivation,
we study the \textit{CP} invariant Type-X (lepton-specific) 2HDM in light of the muon $g-2$.
In Type-X,
the couplings of the new scalar bosons to the SM quarks 
are inversely proportional to $\tan\bt$, the ratio of two vacuum expectation values of two Higgs doublet fields,
but those to the charged leptons are linearly proportional to $\tan\bt$.
Large $\tan\bt$ can enhance the
new contributions of extra Higgs bosons to $\damu$, while suppressing the contributions to
the hadron-related data such as $B \to K\mmu$ and $B_s \to \mmu$~\cite{Schmidt-Hoberg:2013hba}.
Since the other three types (Type-I, Type-II, and Type-Y) cannot accommodate this feature,
there have been extensive studies of Type-X for the muon anomalous magnetic moment~\cite{Cao:2009as,Broggio:2014mna,Wang:2014sda,Abe:2015oca,Chun:2017yob,Chun:2016hzs,Cherchiglia:2017uwv,Wang:2018hnw}.

A comprehensive study of Type-X for $\damu$,
including the LHC Run-1 and LEP results as well as the lepton flavor universality (LFU) data in the $Z$ and $\tau$ decays,
was first conducted in Ref.~\cite{Abe:2015oca}.
Partial updates have followed, focusing on the LFU data~\cite{Chun:2017yob}
or the LHC data~\cite{Chun:2016hzs,Cherchiglia:2017uwv,Wang:2018hnw}.
We generalize the previous studies of the Type-X 2HDM
both in theoretical setup and in data analysis.
First, we take the general setting in the Higgs sector,
by considering two scenarios, the ``normal" scenario
where the observed Higgs boson is the lighter \textit{CP}-even scalar $h$
and the ``inverted" scenario where 
the heavier \textit{CP}-even scalar $H$ is the observed one.
The inverted scenario with a new light \textit{CP}-even scalar
has recently drawn a lot of interest because of the $3\sg$ excess
in the diphoton invariant mass distribution at around $96~{\rm GeV}$~\cite{CMS:2018cyk},
but has not been analyzed in the context of the muon $g-2$.
This scenario seems incompatible with the recent measurement of \textit{positive} $\Dt a_\mu$, 
because the dominant Barr-Zee contributions of the $\tau^\pm$ loop mediated by the light \textit{CP}-even $h$ 
are \textit{negative}~\cite{Barr:1990vd,Ilisie:2015tra}. 
We need to answer whether the inverted scenario remains viable. 

For the general data analysis,
we will investigate all the latest data of the LHC Run-2, 
the electron anomalous magnetic moment~\cite{Parker:2018vye,Morel:2020dww},
the LFU data (adopting the updated HFLAV global fit results~\cite{HFLAV:2019otj}
and Michel parameters~\cite{ALEPH:2001gaj} in the $\tau$ decay),
as well as theoretical stabilities
and electroweak oblique parameters.
We will also include the correlations among the observables.
The correlations are often neglected in the literature, but they play a vital role in constraining new physics models.
To draw a general conclusion on the Type-X 2HDM,
we will scan the whole parameter space without any extra assumption on the masses or the couplings.
Furthermore,
the tension in the Type-X when simultaneously explaining $\damu$ and LFU data
shall be quantified through the global $\chi^2$ fit.
Finally, the customized search strategy for the viable parameter space at the HL-LHC is to be studied.
These are our contributions to the phenomenology of the Type-X 2HDM in light of the new Fermilab measurement of
$\damu$.

The paper is organized in the following way.
In Sec.~\ref{sec:review},
we briefly review the Type-X 2HDM and describe the characteristics of the normal and inverted scenarios
in the Higgs alignment limit.
In Sec.~\ref{sec:damu},
we discuss the new contributions of the Type-X 2HDM to $\damu$.
Section \ref{sec:scan}
describes our scanning strategies in three steps
and shows the results of the allowed parameter space at each step.
Section \ref{sec:implications} deals with the electron anomalous magnetic moment
and the LHC signatures.
In Sec.~\ref{sec:LFU},
we check the consistency of the model with the LFU data in the $\tau$ and $Z$ decays.
Conclusions are given in Sec.~\ref{sec:conclusions}.

\section{Type-X 2HDM}
\label{sec:review}

The 2HDM accommodates two complex $SU(2)_L$ Higgs doublet scalar fields, $\Phi_1$ and $\Phi_2$~\cite{Branco:2011iw}:
\bea
\label{eq:phi:fields}
\Phi_i = \left( \begin{array}{c} w_i^+ \\[3pt]
\dfrac{v_i +  h_i + i \eta_i }{ \sqrt{2}}
\end{array} \right), \quad i=1,2,
\eea
where $v =\sqrt{v_1^2+v_2^2}=246\gev$.
Using the simplified notation of $s_x=\sin x$, $c_x = \cos x$, and $t_x = \tan x$,
we define $\tb =v_2/v_1$.
To prevent the tree-level flavor changing neutral currents,
a discrete $Z_2$ symmetry
is imposed as $\Phi_1 \to \Phi_1$
and $\Phi_2 \to -\Phi_2$~\cite{Glashow:1976nt,Paschos:1976ay}.
The most general, renormalizable, and \textit{CP} conserving scalar potential with softly broken $Z_2$ symmetry is
\bea
\label{eq:VH}
V_\Phi = && m^2 _{11} \Phi^\dagger _1 \Phi_1 + m^2 _{22} \Phi^\dagger _2 \Phi_2
-m^2 _{12} ( \Phi^\dagger _1 \Phi_2 + \hc) \\ \nn
&& + \frac{1}{2}\lambda_1 (\Phi^\dagger _1 \Phi_1)^2
+ \frac{1}{2}\lambda_2 (\Phi^\dagger _2 \Phi_2 )^2
+ \lambda_3 (\Phi^\dagger _1 \Phi_1) (\Phi^\dagger _2 \Phi_2)
+ \lambda_4 (\Phi^\dagger_1 \Phi_2 ) (\Phi^\dagger _2 \Phi_1) \\ \nn
&& + \frac{1}{2} \lambda_5
\left[
(\Phi^\dagger _1 \Phi_2 )^2 +  \hc
\right],
\eea
where the $m^2 _{12}$ term softly breaks the $Z_2$ parity.
There are five physical Higgs bosons, the light \textit{CP}-even scalar $h$,
the heavy \textit{CP}-even scalar $H$, the \textit{CP}-odd pseudoscalar $A$,
and two charged Higgs bosons $H^\pm$.
The relations of the physical Higgs bosons with the weak eigenstates in Eq.~(\ref{eq:phi:fields})
via two mixing angles $\al$ and $\bt$ 
are referred to Ref.~\cite{Aoki:2009ha,Song:2019aav}.
Note that the SM Higgs boson is a linear combination of $h$ and $H$, as
\bea
\label{eq:hsm}
\hsm = \sba h + \cba H.
\eea
The Yukawa couplings to the SM fermions are written by
\bea
\mathcal{L}_{\rm Yuk} &=&
- \sum_f 
\lf 
\frac{m_f}{v} y_f^h \bar{f} f h + \frac{m_f}{v} y^H_f \bar{f} f H
-i \frac{m_f}{v} y^A_f \bar{f} \gm_5 f A
\ri
\\ \nn &&
- 
\left\{
\dfrac{\sqrt{2}}{v } \overline{t}
\left(m_t y^A_t {P}_L +  m_b y^A_b {P}_R\right)b H^+
+\dfrac{\sqt m_\ell}{v}y^A_\ell \overline{\nu}_\ell P_R \ell^{}H^+
+\hc
\right\},
\eea
where $P_{R,L}=(1\pm \gm^5)/2$ and $\ell=\mu,\tau$.

The observed Higgs boson at a mass of $125\gev$ is similar to the SM Higgs boson,
more strongly in Type-X with large $\tb$~\cite{ATLAS:2020qdt}.
Therefore, we take the Higgs alignment limit where one of the \textit{CP}-even neutral Higgs bosons
is the SM Higgs boson $\hsm$~\cite{Carena:2013ooa,Celis:2013rcs,Bernon:2015qea,Chang:2015goa,Das:2015mwa}.
There are two ways to realize the Higgs alignment limit, the ``normal" and 
``inverted" scenarios.
In the normal scenario, the observed Higgs boson is the lighter \textit{CP}-even scalar $h$,
i.e., $ \sba=1$.
In the inverted scenario, $ \cba=1$ so that the heavier \textit{CP}-even scalar $H$ is observed
while the lighter one is hidden~\cite{Chang:2015goa,Bernon:2015wef}.
The model has five independent parameters in the physical basis,
\bea
\label{eq:model:parameters}
\left\{ m_{\varphi^0},~\ma,~\mch,~ M^2,~\tb \right\},
\eea
where $M^2 = m_{12}^2/(\sb\cb)$ and $\varphi^0$ is 
the new \textit{CP}-even neutral Higgs boson, i.e.,
$\varphi^0=H$ in the normal scenario and $\varphi^0=h$ in the inverted scenario.
The two scenarios are summarized as follows:
\bea
\label{eq:NS:IS}
  {\renewcommand{\arraystretch}{1.5} 
\begin{array}{c|c}
\hbox{normal scenario (NS)} & \hbox{inverted scenario (IS)} \\ \hline
\hsm = h,\quad \varphi^0 = H & \hsm = H,\quad \varphi^0 = h \\
y_f^\hsm = 1,\quad \sba=1 & y_f^\hsm =1,\quad \cba=1 \\
y_t^A = - y_t^{\varphi^0} = \tbi,\quad y_\ell^A =y_\ell^{\varphi^0} = \tb ~~& 
~~y_t^A = y_t^{\varphi^0} = \tbi,\quad y_\ell^A =- y_\ell^{\varphi^0} = \tb \\ 
\end{array}
}
\eea

In the Higgs alignment limit, the quartic couplings 
are~\cite{Das:2015mwa}
\bea
\label{eq:lm1}
\lm_1 &=& \frac{1}{v^2}
\left[
m_{125}^2 + \tb^2 \lf m_{\varphi^0}^2 - M^2 \ri
\right],
\\ \nn
\lm_2 &=& \frac{1}{v^2}
\left[
m_{125}^2 + \frac{1}{\tb^2} \lf m_{\varphi^0}^2 - M^2 \ri
\right],
\\ \nn
\lm_3 &=& \frac{1}{v^2}
\left[
m_{125}^2 - m_{\varphi^0}^2 -M^2 +2 \mch^2 
\right],
\\ \nn
\lm_4 &=& \frac{1}{v^2}
\left[
M^2+\ma^2-2 \mch^2
\right],
\\ \nn
\lm_5 &=& \frac{1}{v^2}
\left[
M^2-\ma^2
\right],
\eea
where $m_{125}=125\gev$.
As shall be shown in the next section,
the observed $\damu$ requires large $\tb$.
Then, the $\tb^2$ terms in $\lm_1$ easily break
the perturbativity of $\lm_1$ unless $m_{\varphi^0}^2$ is extremely close to $M^2$,
which is to be denoted by $m_{\varphi^0}^2\approx M^2$.
When applying this approximate equality to the perturbativity of $\lm_3$,
we should accommodate quasi-degeneracy between $M^2$ and $\mch^2$.
The mass degeneracy is weaker because of the absence of $\tb^2$ terms in $\lm_3$.
We use the notation of $M^2 \simeq \mch^2$ for the weak equality.
The perturbativity of $\lm_4$ and $\lm_5$
finally yields $\ma \simeq \mch$.
In summary,
the perturbativity of the quartic couplings for large $\tb$
limits the masses as
\bea
\label{eq:theory:mass}
\ma \simeq \mch \simeq M \approx m_{\varphi^0}.
\eea

For light $\ma$, the exotic Higgs decay of $\hsm \to AA$ severely restricts the model.
When writing $\mathcal{L} = (1/2) \lm^{\hsm}_{ A A}  \hsm A A$, the vertex is
\bea
\label{eq:lm:hAA}
 \lm^{\hsm}_{ A A} = \frac{1}{v}
\lf -m_{125}^2 - 2 \ma^2 +  2M^2 \ri.
\eea
Because of the condition in \eq{eq:theory:mass},
it is difficult to accommodate $ \lm^{\hsm}_{ A A}=0$.\footnote{If
the Higgs alignment is broken and $t_{\beta-\al} = (\tb-1/\tb)(M^2 - m_{125}^2)/(2 M^2-2 \ma^2-m_{125}^2)$,
$\lm^{\hsm}_{ A A}$ vanishes and the constraint from $\hsm\to AA$ can be evaded.
However, the equality involves five independent parameters of $\al$, $\bt$, $m_{125}$, $M^2$, and $\ma$,
which is an unnatural fine-tuning without underlying symmetries.
}
Since the Higgs precision measurement puts a strong bound on the exotic Higgs decay as
$\br(\hsm \to XX) \lsim \mco(0.1) $~\cite{Aad:2019mbh}, 
the parameter region with $\ma \leq m_{125}/2$ is highly disfavored.

\section{$\Dt a_\mu$ in the Type-X 2HDM}
\label{sec:damu}

The Type-X 2HDM makes two kinds of new contributions to $\damu$,
one-loop contributions 
and two-loop Barr-Zee contributions~\cite{Barr:1990vd,Ilisie:2015tra}.
The one-loop contributions are mediated by $\varphi^0$, $A$, and $\ch$, as~\cite{Chun:2016hzs}
\bea
\label{eq:amu:oneloop}
	\Delta a_\mu^{\rm 1-loop} &=&
	\frac{G_F \, m_{\mu}^2}{4 \pi^2 \sqrt{2}}  \, 
	\sum_\phi  \left( y_{\mu}^\phi \right)^2  \rho^\mu_\phi \, 
	f_\phi(\rho^\mu_\phi)
	\\ \nn
	&\simeq& 
	2.6 \times 10^{-15}\, 
	\sum_\phi  \left( y_{\mu}^\phi \right)^2  \lf \frac{100\gev}{M_\phi} \ri^2 \, 
	f_\phi(\rho^\mu_\phi),
\eea
where $\phi =  \{\varphi^0, A , H^\pm\}$, 
$\rho^i_j=  m_i^2/m_j^2$, and the expressions for the loop function
$f_{\phi}$ are referred to Ref.~\cite{Chun:2016hzs}.
The numerical factor in the second equality of \eq{eq:amu:oneloop}
implies that the observed $\Dt a_\mu$ requires light $M_\phi$ and large $y_\mu^\phi$.
Because  $\rho^{\mu}_\phi \ll 1$, the loop functions show the following asymptotic behaviors:
\begin{eqnarray}
\label{eq:f:1loop}
	f_{\varphi^0}(\rho) &=&- \ln \rho - {7}/{6} + \mco(\rho),
 \\ \nn
	f_A (\rho) &=& +\ln \rho +11/6 + \mco(\rho),
 \\ \nn
	f_{H^\pm} (\rho) &=& -1/6 + \mco(\rho).	
\end{eqnarray}
It is clear to see that the one-loop contributions of the \textit{CP}-even scalar $\varphi^0$ 
are positive while those of $A$ and $\ch$ are negative: $\Delta a_\mu^{\rm 1-loop}$ is proportional to the \textit{square} of $y_\mu^\phi$.

More significant contributions to $\damu$ are from the two-loop Barr-Zee type diagrams with heavy fermions 
in the loop~\cite{Barr:1990vd}:
\bea
\label{eq:BZ}
	\Delta a_\mu^{\rm BZ} 
	&=& \frac{G_F \, m_{\mu}^2}{4 \pi^2 \sqrt{2}} \, \frac{\alpha_{\rm em}}{\pi}
	\, \sum_{f,\phi^0}  N^c_f  \, Q_f^2  \,  y_{\mu}^{\phi^0}  \, y_{f}^{\phi^0} \,  \rho^{f}_{\phi^0}\,  
	g_{\phi^0}(\rho^{f}_{\phi^0})
	,
\eea
where $f=t,b,\tau$, $\phi^0 =  \{\varphi^0, A \}$,  $m_f$, $Q_f$ and $N^c_f$ are 
the mass, electric charge and color factor of the fermion $f$,
and the loop functions are
\bea
\label{eq:2loop:ft}
	g_{\varphi^0}(\rho) &=& \int_0^1 \! dx \, \frac{2x (1-x)-1}{x(1-x)-\rho} \ln \frac{x(1-x)}{\rho},
	\\[3pt] \nn
	g_A(\rho) &=& \int_0^1 \! dx \, \frac{1}{x(1-x)-\rho} \ln \frac{x(1-x)}{\rho}.
\eea
For the top quark and $\tau^\pm$ loops,
the factor $\rho^{f}_{\phi^0} $ in \eq{eq:BZ} 
significantly enhances $\Delta a_\mu^{\rm BZ}$
with respect to $\Delta a_\mu^{\rm 1-loop}$ in \eq{eq:amu:oneloop}.
The usual conclusion that a \textit{CP}-even 
scalar boson makes a negative contribution 
to $\Delta a_\mu^{\rm BZ} $ holds true when $y_{\mu}^{\varphi^0}  \, y_{f}^{\varphi^0} > 0$.
As shown in \eq{eq:NS:IS},
the top quark 
 incorporates $y_\mu^{\varphi^0} y_t^{\varphi^0} <0$ in both scenarios
and thus generates \textit{positive} two-loop Barr-Zee contributions.

\begin{figure}[h] \centering
\begin{center}
\includegraphics[width=0.49\textwidth]{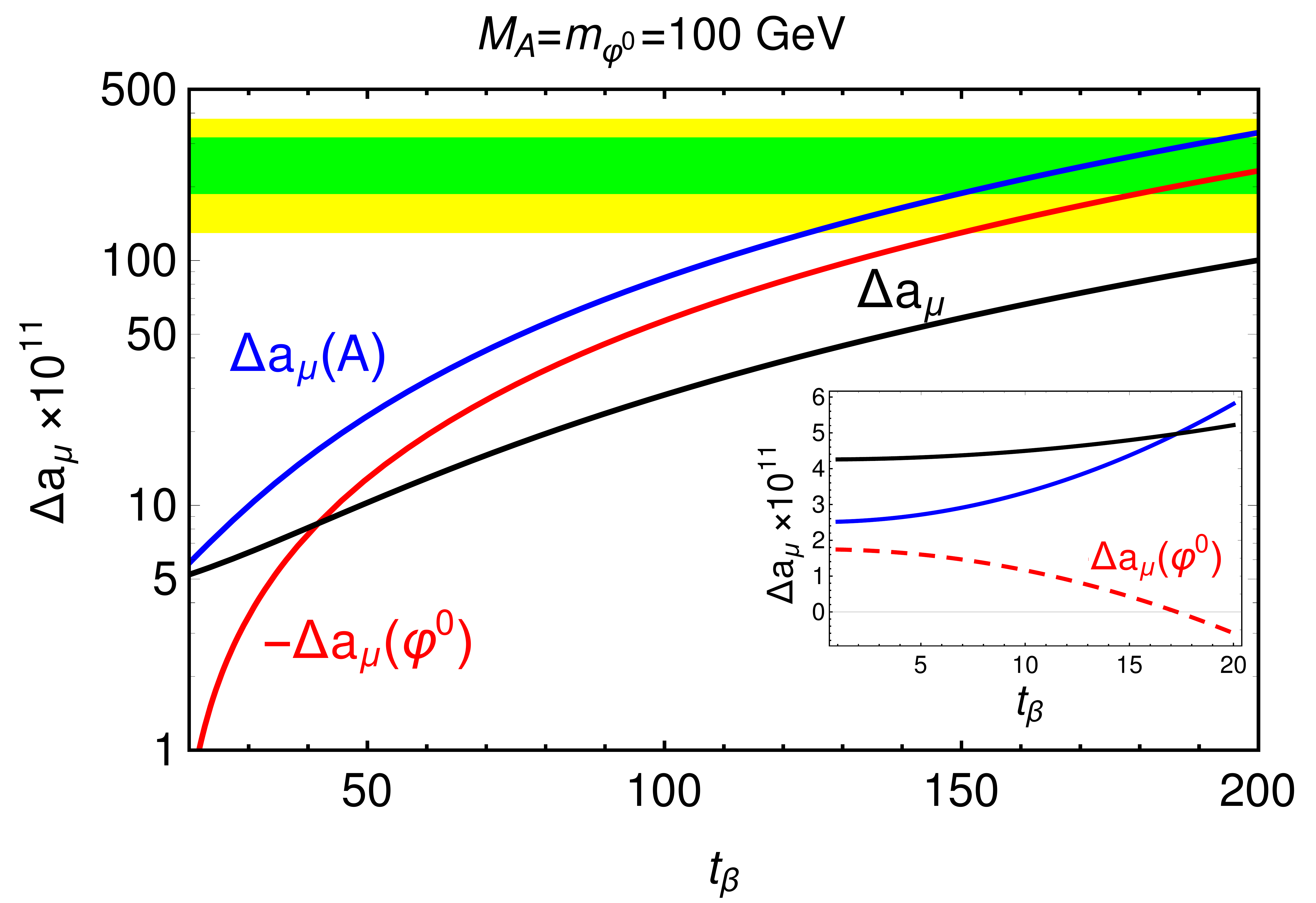}~
\includegraphics[width=0.48\textwidth]{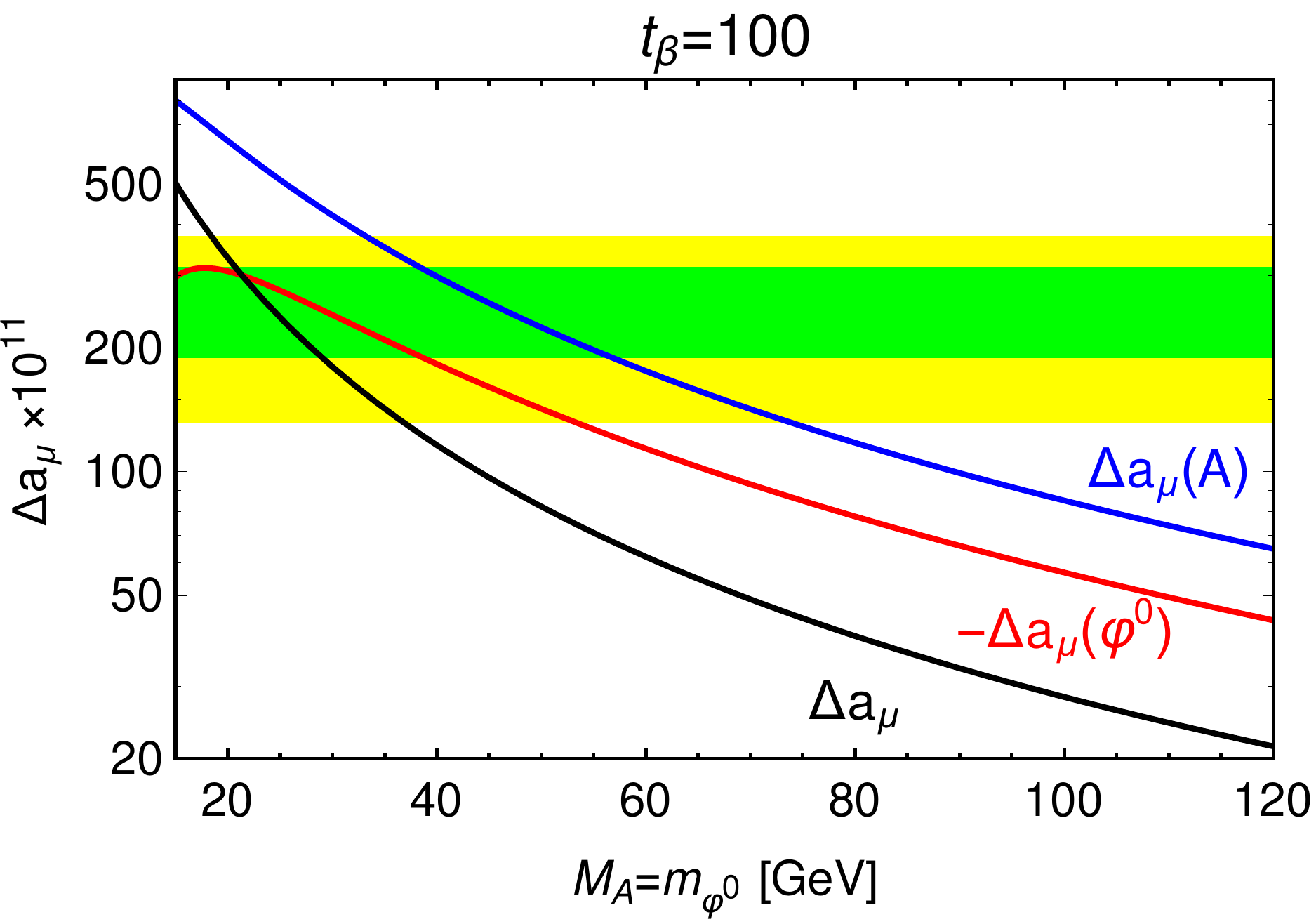}
\end{center}
\caption{\label{fig-damu-tb}
$\Dt a_\mu$ from $A$, $-\Dt a_\mu$ from $\varphi^0$, and $\damu(A)+\damu(\varphi^0)$ 
as a function of $\tb$ with $\ma=m_{\varphi^0}=100\gev$ (left panel)
and as a function of $\ma=m_{\varphi^0}$ for $\tb=100$ (right panel).
}
\end{figure}

In Fig.~\ref{fig-damu-tb},
we show  $\Dt a_\mu(A)$ (blue line), 
$-\Dt a_\mu(\varphi^0)$ (red line), and $\Dt a_\mu(A)+ \Dt a_\mu(\varphi^0)$ (black line)
as a function of $\tb$ with $\ma=m_{\varphi^0}=100\gev$ in the left panel
and as a function of $\ma=m_{\varphi^0}$ with $\tb=100$ in the right panel.
To show negative $\Dt a_\mu(\varphi^0)$ in the logarithmic scale,
we present $-\Dt a_\mu(\varphi^0)$.
The horizontal green (yellow) area denotes the allowed region of $\Dt a_\mu$ at $1\sg$ ($2\sg$).
The dominant contribution of $A$ is from two-loop Barr-Zee diagrams,
which is always positive.
The sign of the $\varphi^0$ contribution depends on the value of $\tb$.
For very large $\tb$, $\Dt a_\mu(\varphi^0)$ is negative since the contribution of the $\tau^\pm$ loop
in the two-loop Barr-Zee diagram is dominant.
If $\tb \lsim 17$, however,
$\Dt a_\mu(\varphi^0)$ becomes positive (see the small figure inside the left panel)
because dominant is the top quark loop in the two-loop Barr-Zee diagram.
Although the contributions from both $A$ and $\varphi^0$ are positive for $\tb \lsim 17$,
the absolute value of $\Dt a_\mu$ is not large enough to explain $\damuo$.
In the right panel,
we show $\Dt a_\mu (A)$ and $-\Dt a_\mu(\varphi^0)$ as a function of $\ma=m_{\varphi^0}$
by fixing $\tb=100$.
$\Dt a_\mu$ increases rapidly with decreasing scalar masses.
Since the negative contributions of the \textit{CP}-even $\varphi^0$ become severe with decreasing $m_{\varphi^0}$, the inverted scenario receives a stronger constraint.

\section{Theoretical and experimental constraints on the Type-X 2HDM}
\label{sec:scan}
\subsection{Scanning strategies in three steps}
\label{subsec:steps}

For the comprehensively study of the Type-X 2HDM in light of the muon $g-2$, 
we perform the successive and cumulative scan of the model parameters in three steps.
\begin{description}
\item[Step I:] We demand that the model explains $\damuo$ at $2\sg$.
\item[Step II:] Among the parameters that survive Step I, we impose the constraints from theoretical stabilities and electroweak precision data, as detailed below.
	\ben
	\item Theoretical stabilities~\cite{Ivanov:2006yq,Barroso:2013awa,Chang:2015goa}
		\bit
		\item Higgs potential being bounded from below~\cite{Ivanov:2006yq};
		\item Unitarity of scalar-scalar scatterings~\cite{Arhrib:2000is,Branco:2011iw};
		\item Perturbativity~\cite{Chang:2015goa};
		\item Vacuum stability~\cite{Barroso:2013awa}.
		\eit
	\item Peskin-Takeuchi electroweak oblique parameters~\cite{Peskin:1991sw}\\	
	We take the current best-fit results of~\cite{PDG2020}
	\bea
	S &=& -0.01 \pm 0.10,
	\quad
	T = 0.03 \pm 0.12, \quad
	U=0.02 \pm 0.11, 
	\\ \nn
	 \rho_{ST} &=& 0.92, \quad \rho_{SU}=-0.80,\quad \rho_{TU}=-0.93,
	\eea
	where $\rho_{ij}$ is the correlation matrix.	
	The expressions of the contributions from the scalar boson loops to $S$, $T$, and $U$ are referred to Ref.~\cite{He:2001tp,Grimus:2008nb}.
	We require $\Dt \chi^2 \lf = \chi^2 - \chi^2_{\rm min} \ri< 7.81$. 
	\een
\item[Step III:] For the parameters that survive Step II, we demand to satisfy the collider bounds.
	\ben
	\item  Higgs precision data by using \texttt{HiggsSignals}~\cite{Bechtle:2013xfa,Bechtle:2020uwn}:\\
	The \texttt{HiggsSignals}-v2.2.0~\cite{Bechtle:2020uwn} provides the $\chi^2$ value for 107 Higgs observables.
	Since our model has five parameters,
	the number of degrees of freedom for the $\chi^2$ analysis is 102. 
	We require that the calculated Higgs signal strengths be consistent 
	with the experimental measurements at $2\sg$.
	\item Direct searches for new scalars at the LEP, Tevatron, and LHC:\\
	We use the public code \texttt{HiggsBounds}~\cite{Bechtle:2020pkv}. 
	Main search channels which affect the Type-X 2HDM are 
		\ben
		\item LEP experiments:
			\bit
			\item $\ee\to Z \to A h \to \ttau\ttau$~\cite{Schael:2006cr}.
			\eit
		\item LHC experiments:
			\bit
			\item $h \to AA$~\cite{Aaboud:2018esj,Aaboud:2018iil,Sirunyan:2018mbx,Sirunyan:2018mot,Sirunyan:2018pzn,Sirunyan:2019gou};
			\item  $H \to ZZ/W^+ W^-$~\cite{Aaboud:2018bun,Sirunyan:2018qlb,Sirunyan:2019pqw};
			\item $H \to hh$~\cite{Sirunyan:2018ayu,Aaboud:2018bun,Aaboud:2018ewm,Aaboud:2018knk,Aaboud:2018ksn,Aaboud:2018sfw,Aad:2019uzh,Aad:2020kub};
						\item $H/A \to \rr$~\cite{Aaboud:2017yyg,Sirunyan:2018aui},
						$\ttau$~\cite{Sirunyan:2018zut,Aad:2020zxo},
						$ \mmu$~\cite{Aad:2019fac,Aaboud:2019sgt,Sirunyan:2019tkw},
						$ \bb$~\cite{Sirunyan:2018ikr,Sirunyan:2018taj,Aad:2019zwb},
						$\ttop$~\cite{Sirunyan:2019wph};
			\item $ A \to Z h$~\cite{Aad:2015wra,Sirunyan:2019xls};
			\item $A(H)\to Z H(A)$~\cite{Aaboud:2018eoy,Sirunyan:2019wrn};
			\item $H^\pm \to t b$~\cite{Aaboud:2018cwk,Sirunyan:2020hwv},
			$ \tau^\pm\nu$~\cite{Aaboud:2018gjj,Sirunyan:2019hkq}.
			\eit			
		\een
		For each scattering process, we compute the $r_{95 \%}$ defined by 
			\bea
			\label{eq:r95}
			r_{95\%} = \frac{S_{\rm 2HDM}}{S_{\rm obs}^{95\%}},
			\eea
			where $S_{\rm 2HDM}~(S_{\rm obs}^{95\%})$ is the predicted~(observed) cross section. A point in the parameter space is excluded at the $95\%$ confidence level if $r_{95\%} > 1$.
	\een
\end{description}
In the normal scenario, we obtained $5 \times 10^5$ parameter sets that satisfy Step II.
Step III excludes about 80\% of the parameter sets that survived Step II. 
The exclusion is more severe in the inverted scenario,
for which we separately collected $5 \times 10^5$ parameter sets that pass Step II.
Only $\sim 1.8\%$ parameter sets survive at Step III.

\subsection{Results in the normal scenario}

\begin{figure}[h] \centering
\begin{center}
\includegraphics[width=\textwidth]{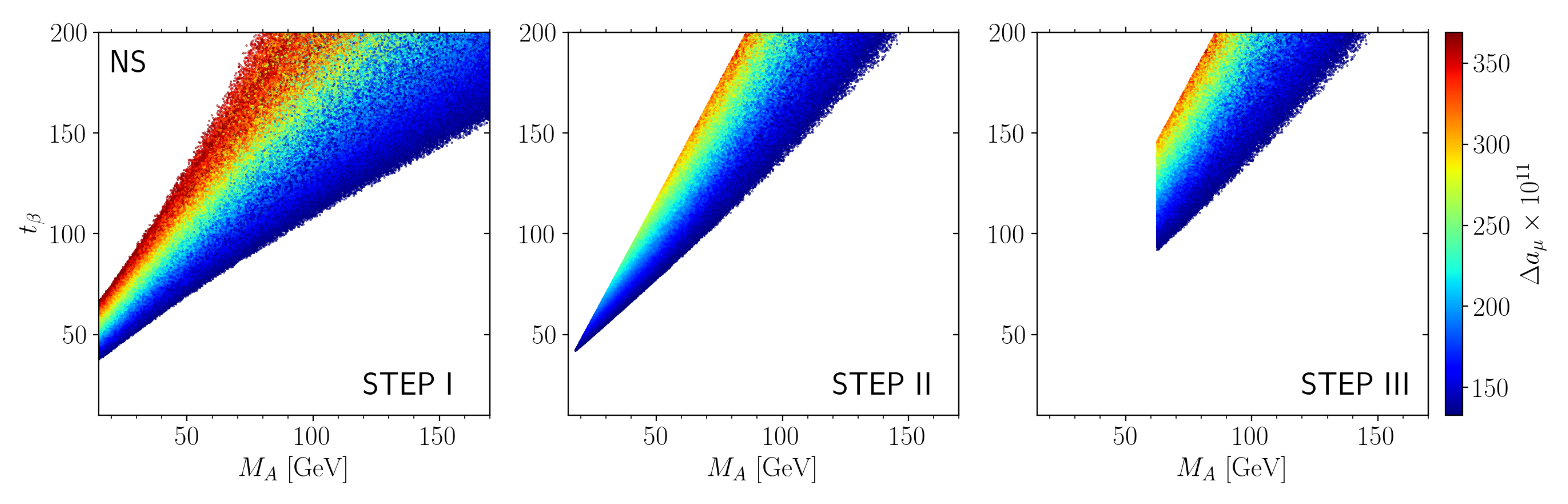}
\end{center}
\caption{\label{fig-tb-MA-damu-NS}
In the normal scenario,
the allowed parameter space of $(\ma,\tb)$
after Step I ($\damuo$), Step II (Step I+Theory+EWPD), and Step III (Step II+Collider),
with the color code indicating the value of $\Dt a_\mu$.
}
\end{figure}

\begin{figure}[h] \centering
\begin{center}
\includegraphics[width=0.55\textwidth]{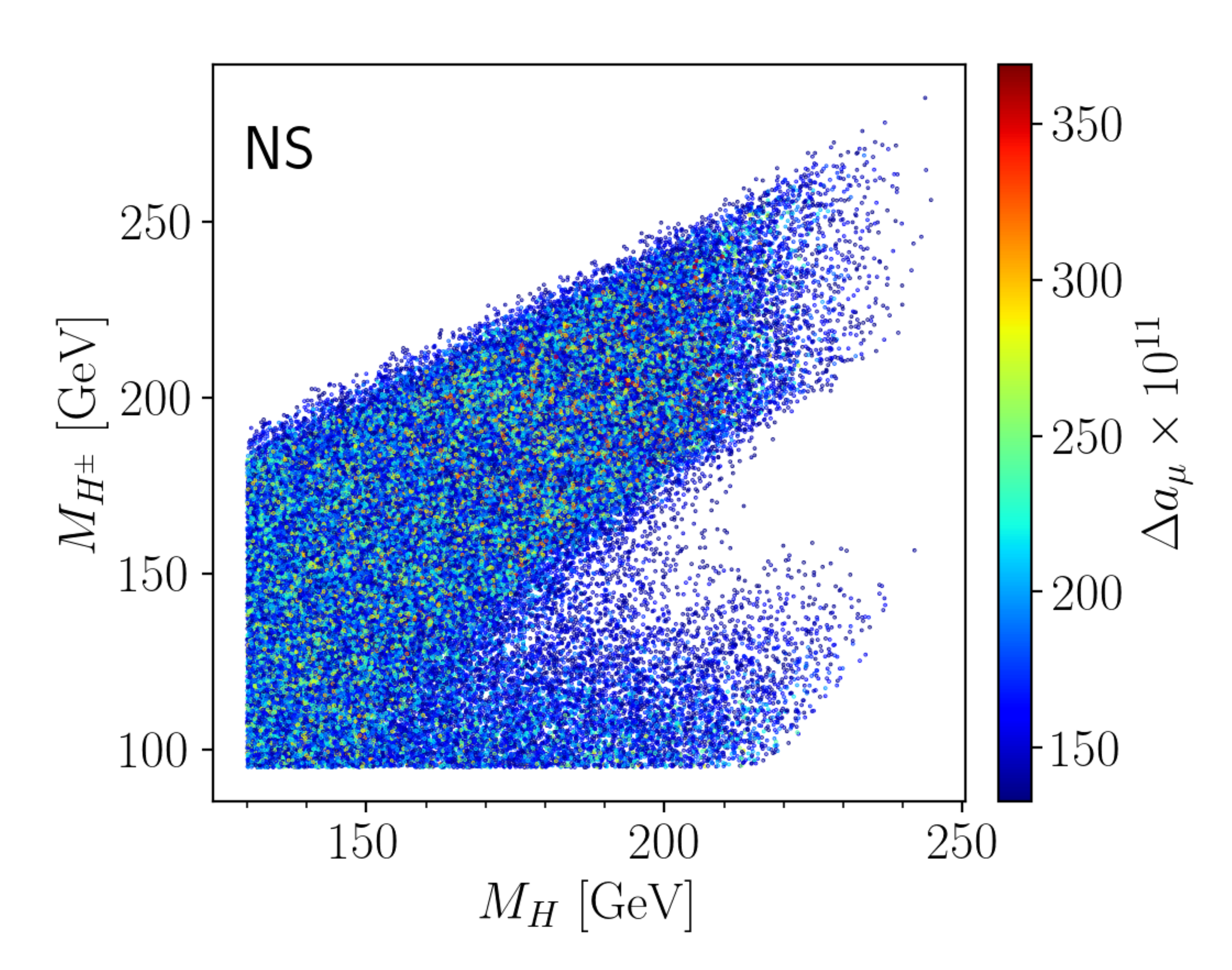}
\end{center}
\caption{\label{fig-mch-mhh-NS}
The allowed $(\mhh,\mch)$ in the normal scenario at Step III.
The color code indicates the value of $\damu$.
}
\end{figure}

In Fig.~\ref{fig-tb-MA-damu-NS},
we show the allowed $(\ma,\tb)$ at each step.
The observed $\damu$ at Step I (left panel)
demands $\tb\gsim 30$ and $\ma \lsim 200\gev$,
but does not limit the masses of $H$ and $\ch$.
Both $\mhh$ and $\mch$ can reach about $1\tev$.
At Step II (middle panel),
a large cut on the parameter space is made, mainly on $\mhh$ and $\mch$.
It is because the combination of the theoretical stability condition ($\ma \simeq \mch \simeq \mhh \approx M $)
and the intermediate $\ma$ at Step I lowers $\mhh$ and $\mch$.
This feature is shown in Fig.~\ref{fig-mch-mhh-NS}
by the allowed $(\mhh,\mch)$ at Step III:
Step II and Step III have similar results for $\mhh$ and $\mch$.
There exist upper bounds of $\mhh \lesssim 245\gev$ and $\mch\lesssim 285\gev$.
Besides, the correlation of $\damu$ with $\mhh$ or $\mch$ is weak,
as indicated by the mixed colors of $\damu$.
As only the intermediate $\mhh$ survives,
the negative contribution of the \textit{CP}-even $H$ to $\damu$ becomes significant.

\begin{figure}[h] \centering
\begin{center}
\includegraphics[width=0.45\textwidth]{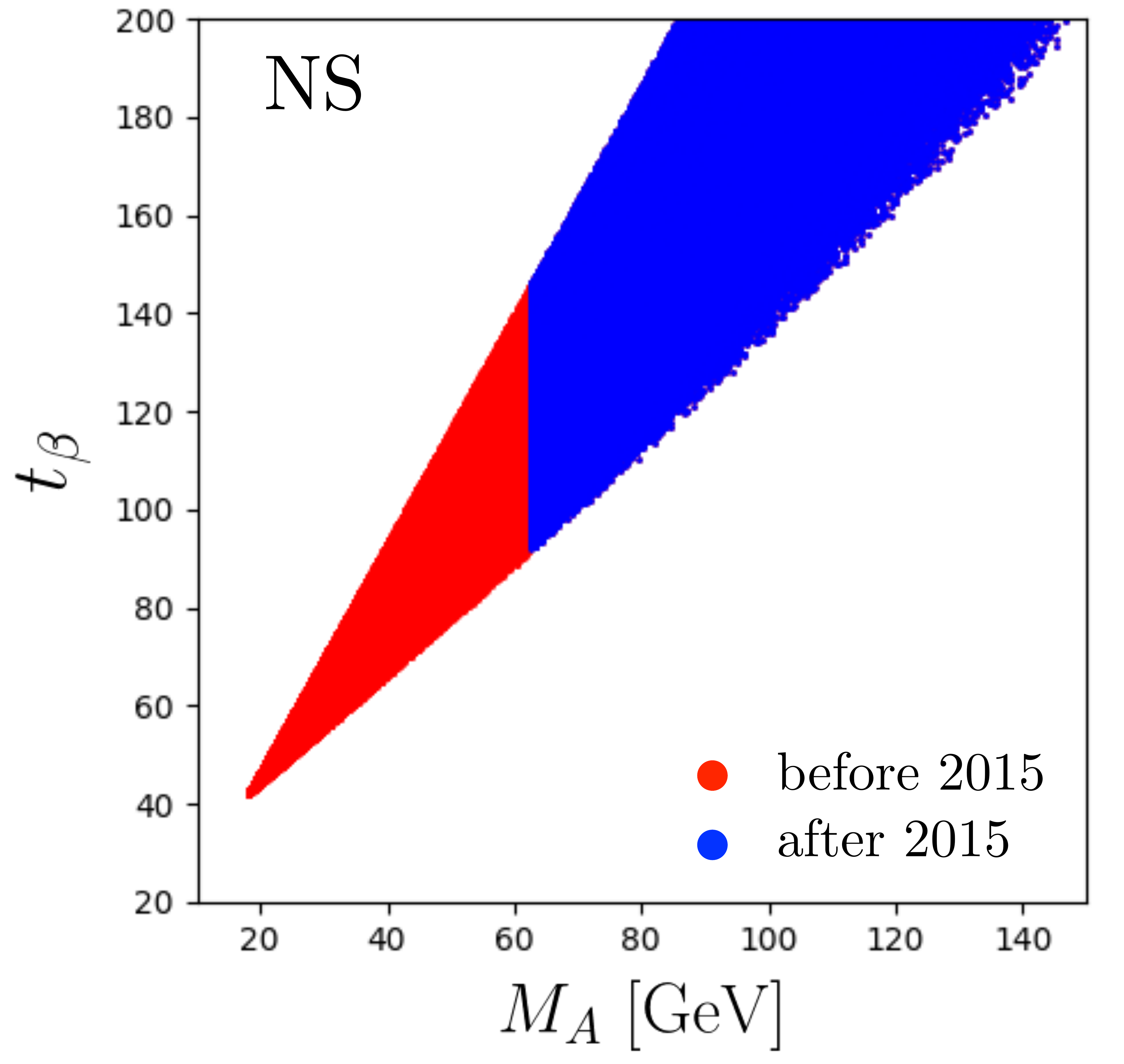}
\end{center}
\caption{\label{fig-compare}
In the normal scenario, the comparison of the allowed $(\ma,\tb)$
by using the LHC data before 2015 (red) and after 2015 (blue).
}
\end{figure}

Let us go back to discussing the allowed $(\ma,\tb)$.
At Step III, which additionally imposes the constraints from the collider data at the LEP, Tevatron, and LHC,
a large portion of the parameter space is removed:
see the right panel in Fig.~\ref{fig-tb-MA-damu-NS}.
We found that the \emph{recent} LHC data plays a crucial role in the curtailment.
To demonstrate the role,
we present the allowed parameter points in $(\ma,\tb)$ by the LHC data before 2015 (red) 
and those after 2015 (blue) in Fig.~\ref{fig-compare}:
2015 is taken as the reference point in consideration of Ref.~\cite{Abe:2015oca}.
The new LHC data exclude the whole parameter space of $\ma < m_{125}/2$ and $\tb \lsim 90$.
The accumulation of the LHC null results in the NP searches
gives a significant implication on the Type-X 2HDM in the context of muon $g-2$.

\begin{figure}[h] \centering
\begin{center}
\includegraphics[width=0.45\textwidth]{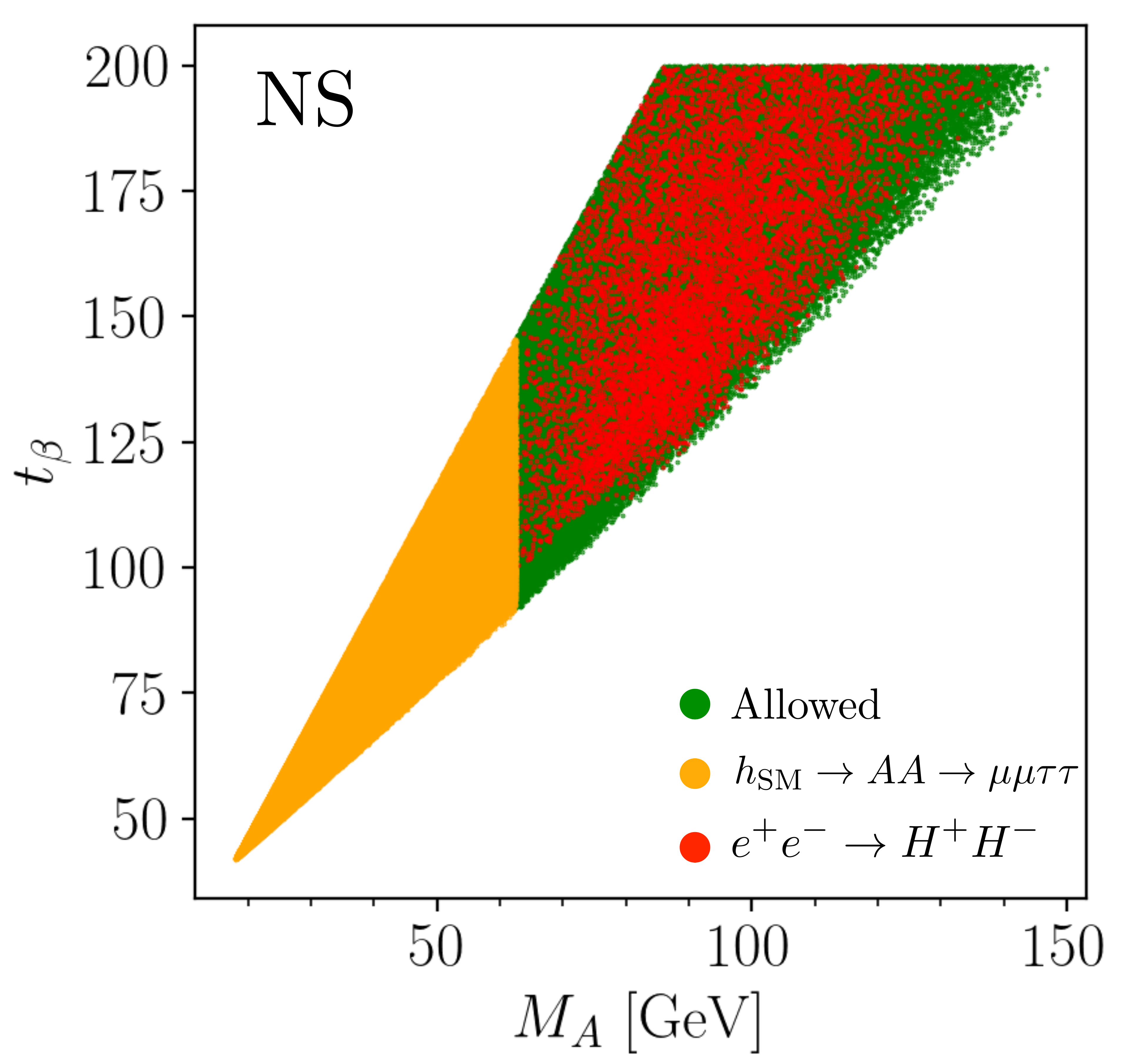}
\end{center}
\caption{\label{fig-HB-NS}
In the normal scenario at Step III,
the allowed points (green) and the excluded points (orange and red) in the parameter space of $(\ma,\tb)$.
The orange points are excluded by $\hsm\to AA\to \mmu\ttau$ at the LHC,
and the red points are excluded by $\ee\to H^+ H^-$ at the LEP.
}
\end{figure}

The question that follows is which LHC processes exclude the region of $\ma < m_{125}/2$. 
In principle, multiple processes exclude one parameter set simultaneously.
For efficient illustration,
we present in Fig.~\ref{fig-HB-NS} the smoking-gun process that has the largest deviation of the model prediction from the observation,
$r_{95\%}$ in \eq{eq:r95}.
The green points pass all the constraints.
The orange points are rejected by the LHC bounds on
$\hsm \to AA\to \mmu\ttau$~\cite{Khachatryan:2017mnf,Sirunyan:2018mbx}.
The red points are excluded by the combined LEP results of $\ee\to H^+ H^-$ including the decays of
$H^+ H^-$ into $c\bar{s}c\bar{s}$, $c\bar{s}\tau\nu$, $\tau\nu\tau\nu$, $W^{*} A \tau\nu$, 
and $W^{*} AW^{*} A$~\cite{Abbiendi:2013hk}.
The overlap of the allowed (green) and excluded (red) points
is attributed to the projection of the five-dimensional hypervolume onto the two-dimensional $(\ma,\tb)$ plane.
In summary,
the normal scenario of the Type-X 2HDM in light of the muon $g-2$ is phenomenologically viable for $\tb\gsim 90$, $\ma \in [62.5, 145]\gev$, $\mhh \in [130,245]\gev$, and $\mch \in [95,285]\gev$.

\subsection{Results in the inverted scenario}

\begin{figure}[h] \centering
\begin{center}
\includegraphics[width=\textwidth]{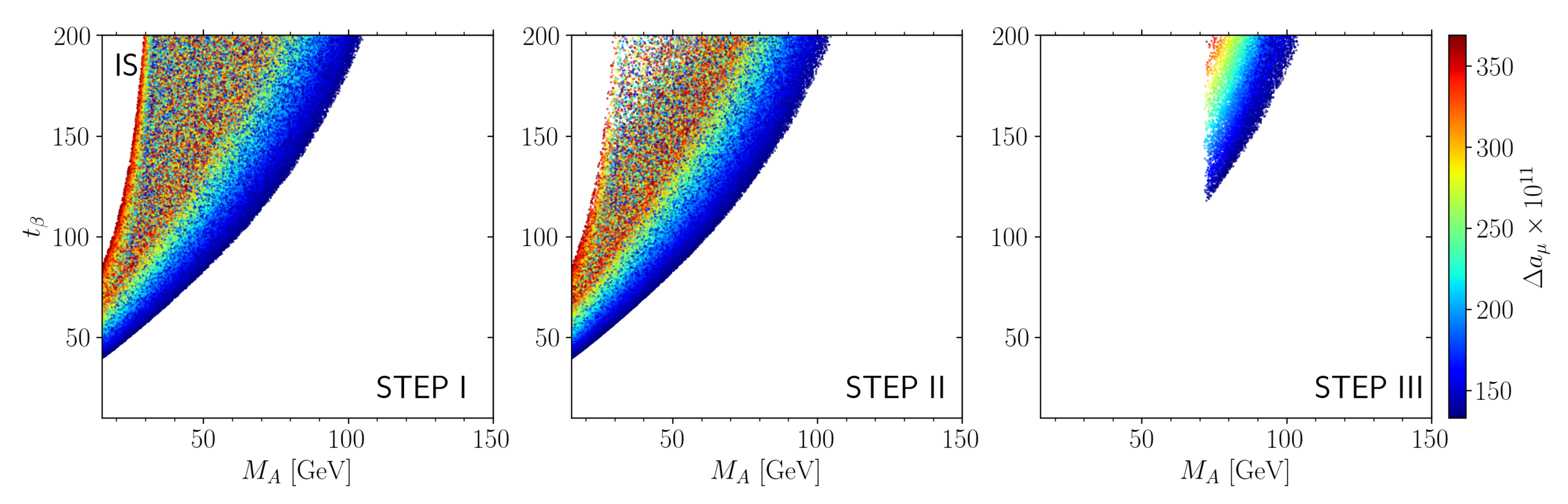}
\end{center}
\caption{\label{fig-tb-MA-damu-IS}
In the inverted scenario,
the allowed parameter space of $(\ma,\tb)$
at Step I, Step II, and Step III,
with the color code indicating the value of $\Dt a_\mu$.}
\end{figure}

In the inverted scenario,
the pattern of the exclusion at Step I, Step II, and Step III is similar to that in the normal scenario: see Fig.~\ref{fig-tb-MA-damu-IS}.
In the quantitative aspect, however,
there are some differences.
At Step I,
the observed $\Dt a_\mu$ prefers lighter $\ma$ than in the normal scenario,
as the light \textit{CP}-even $h$ makes a sizably negative contribution.
The constraints at Step II are weaker than in the normal scenario.
The perturbativity of $\lm_1$, the most critical factor for the theoretical stability,
is easier to satisfy with light $m_{h}$.  
At Step III (right panel), the collider constraints in the inverted scenario are stronger than in the normal scenario,
leading to larger $\tb$ as $\tb \gsim 120$.

\begin{figure}[h] \centering
\begin{center}
\includegraphics[width=0.5\textwidth]{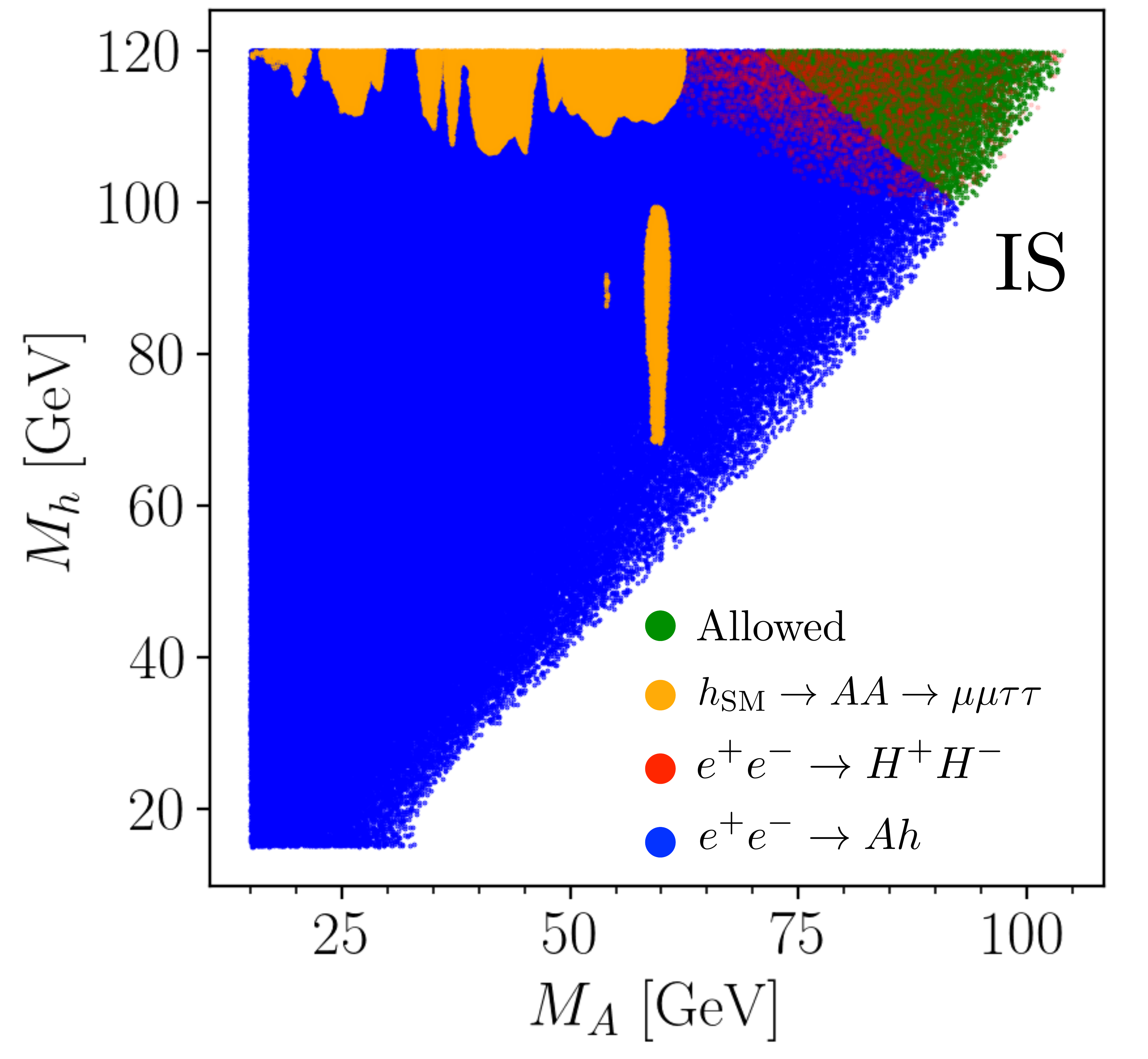}
\end{center}
\caption{\label{fig-HB-IS}
In the inverted scenario at the final Step III,
the allowed points (green)
and the excluded points (orange, red, and blue).
The orange points are excluded by $\hsm\to AA\to \mmu\ttau$ at the LHC,
the red points by $\ee\to H^+ H^-$ at the LEP,
and the blue points by $\ee\to Ah$ at the LEP.
}
\end{figure}

Figure \ref{fig-HB-IS} presents the collider smoking-gun processes in the inverted scenario.
The green points are finally allowed.
The orange and red points are excluded 
by $\hsm\to AA$~\cite{Khachatryan:2017mnf,Sirunyan:2018mbx}
and the LEP process $\ee\to H^+ H^-$~\cite{Abbiendi:2013hk}, respectively.
But the most vital role is played by the LEP process $\ee\to Z^* \to A h$~\cite{Schael:2006cr} (blue points)
because the $Z$-$A$-$h$ vertex, proportional to $\cba$,
is maximal in the alignment limit of the inverted scenario.
We found that the constraint from $\ee\to A h$ is so strong that only the kinematic ban of $\sqrt{s_{ee}}< \ma+\mh$
saves the parameter point.

\begin{figure}[h] \centering
\begin{center}
\includegraphics[width=0.6\textwidth]{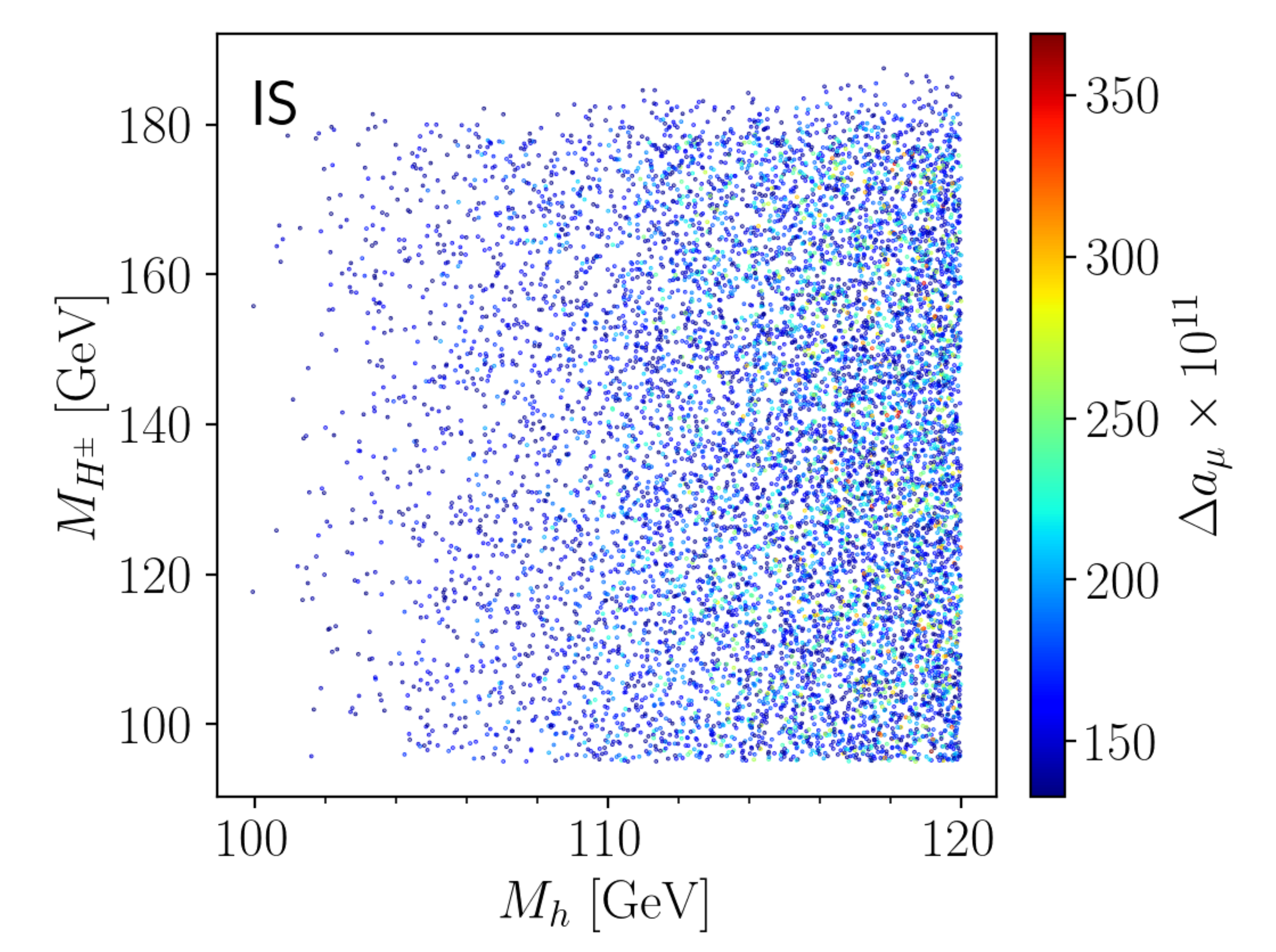}
\end{center}
\caption{\label{fig-mch-mh-IS}
The finally allowed $(\mh,\mch)$ in the inverted scenario 
with the color code indicating the value of $\damu$.
}
\end{figure}

In Fig.~\ref{fig-mch-mh-IS},
we present the finally allowed $(\mh,\mch)$ in the inverted scenario
with the color code indicating the value of $\damu$.
As in the normal scenario,
there is no correlation of $\damu$ with $\mh$ or $\mch$: see the mixed color distribution.
In summary, the inverted scenario survives for $\tb \gsim 120$, $\ma \in [70,105]\gev$, $\mhh \in [100,120]\gev$, and $\mch \in [95,185]\gev$.

\section{Implications on the electron $g-2$ and the LHC collider signatures}
\label{sec:implications}

Upon obtaining the finally allowed parameter points of the Type-X 2HDM
in light of the new muon $g-2$,
we investigate the phenomenological implications of the surviving parameters.
First, we study the electron anomalous magnetic moment.
For $\dae$,
there is controversy over the value of the fine structure constant $\alpha$.
Therefore, we check the consistency of the surviving parameters with $\dae$
rather than accept $\dae$ as an observable.
Second, we study the LHC phenomenology
so to suggest the golden mode for the hadro-phobic scalar bosons.
Since direct searches at high energy colliders provide independent information, the LHC exploration should continue.

\subsection{Electron anomalous magnetic moment}
\label{subsec:e:g-2}

As a flavor universal theory, Type-X 2HDM has the same contributions to $\dae$ and $\damu$
except the differences of the electron and muon masses. 
Positive $\damu$ demands positive $\dae$.
In the measurement, however, $\dae$ has not been settled yet
because of the discrepancy in the recent two experiments for the fine structure constant $\alpha$, the most sensitive input to $\dae$.
Depending on whether we take the data from $^{133}$Cs~\cite{Parker:2018vye}
or from $^{87}$Rb~\cite{Morel:2020dww},
the deviations of the electron $g-2$ from the SM prediction~\cite{Aoyama:2017uqe,Laporta:2017okg,Buras:2021btx} are substantially different as
\bea
\label{eq:dae}
\dae^{\rm Cs} &=& -8.8(3.6) \times 10^{-13},\\ \nn
\dae^{\rm Rb} &=& \phantom{-}4.8(3.0) \times 10^{-13}.
\eea
At $2\sg$ level,
$\dae^{\rm Cs} $
is negative while $\dae^{\rm Rb}$
can be positive.

\begin{figure}[h] \centering
\begin{center}
\includegraphics[width=0.48\textwidth]{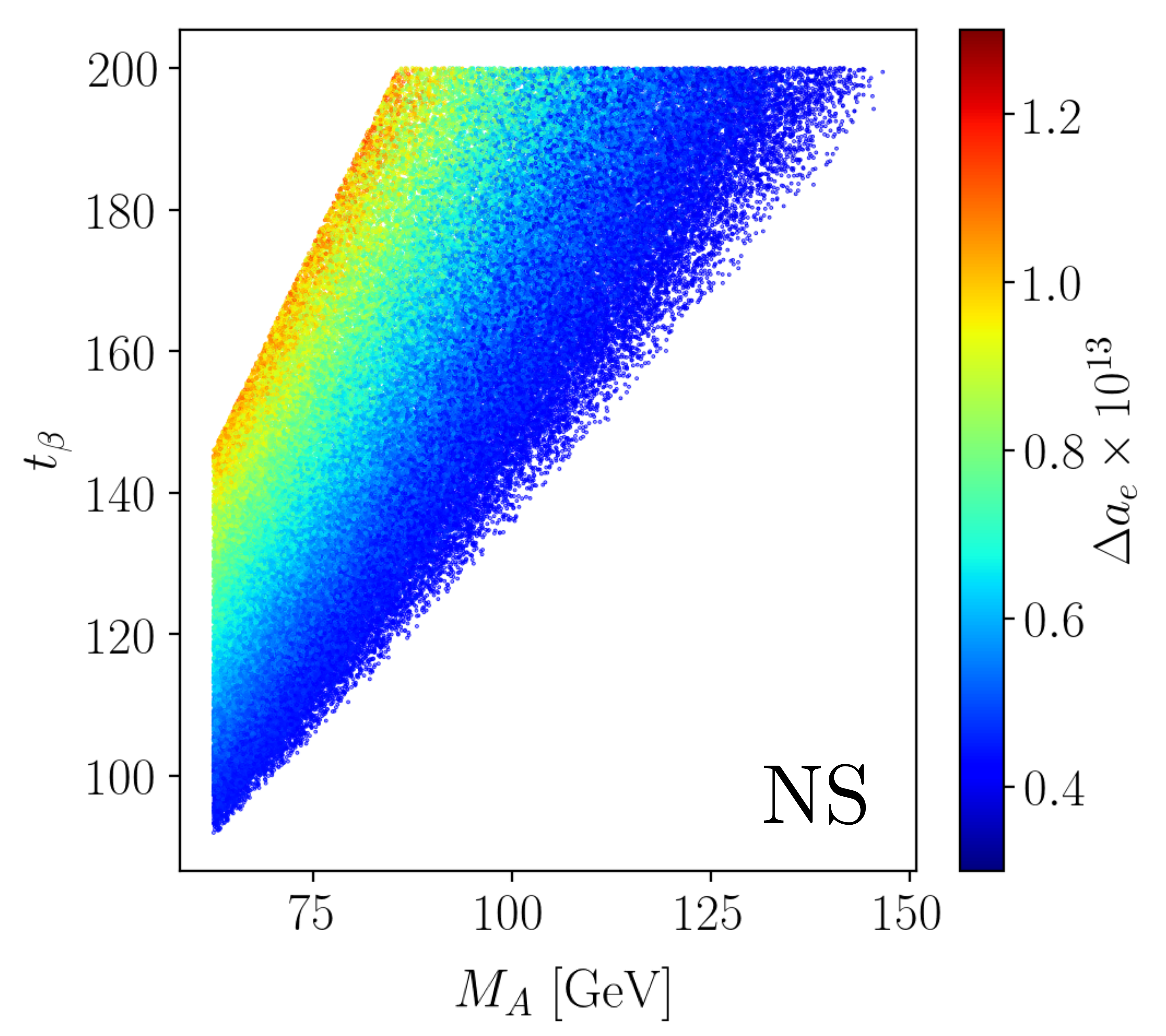}~
\includegraphics[width=0.48\textwidth]{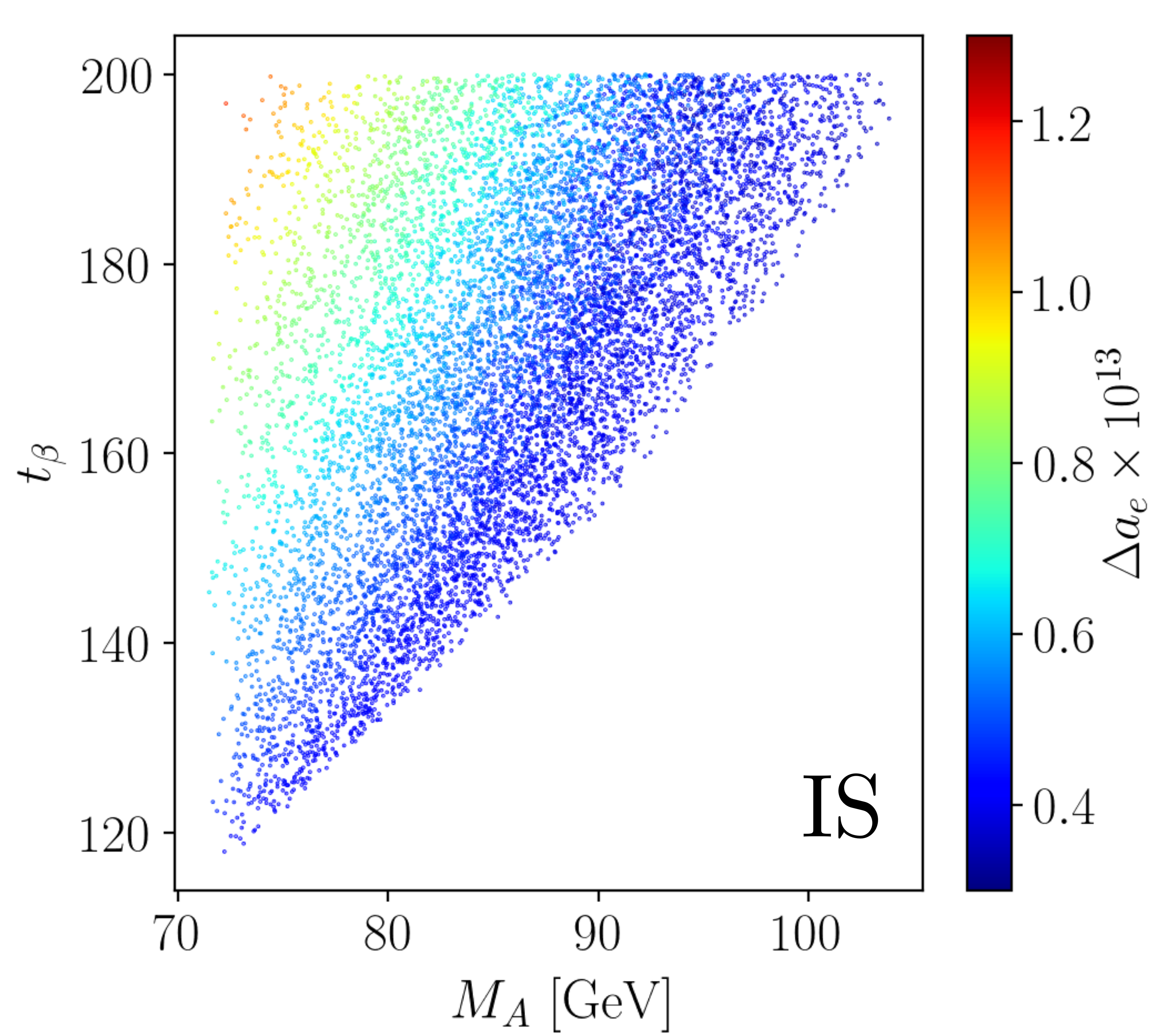}
\end{center}
\caption{\label{fig-dae}
$\dae$ of the finally allowed parameter points that satisfy all the theoretical and experimental constraints
including $\damu$, projected on $(\ma,\tb)$.
The left (right) panel corresponds to the normal (inverted) scenario.
}
\end{figure}

In Fig.~\ref{fig-dae},
we present the $\dae$ over the finally allowed $(\ma,\tb)$.
The left (right) panel corresponds to the normal (inverted) scenario.
The viable parameters predict
$\dae \in [0.36,1.17]\times 10^{-13}$ in the normal scenario and 
$\dae \in [0.33,1.15]\times 10^{-13}$ in the inverted scenario.
The $\dae^{\rm Rb}$ is explained at $2\sg$,
except for the points along the upper-left boundary of the allowed $(\ma,\tb)$ space.
The $\dae^{\rm Cs}$ is negative at $2\sigma$, which is contradictory to the prediction of the model.
At $3\sigma$, however, it is consistent with
the model.

\subsection{Production of the hadro-phobic new scalars at the LHC}
\label{subsec:production:LHC}
For the LHC phenomenology, we point out
two characteristics of the finally allowed parameters:
(i) $\tb$ is extremely large;
(ii) the masses of new scalar bosons are below about $ 300\gev$.
The new scalar bosons with intermediate mass have escaped the LHC searches because of their hadro-phobic nature due to large $\tb$.
For the intermediate-mass $\ch$, 
the current LHC search depends on its production via the decay of a top quark into $b \ch$,
followed by $\ch\to\tau\nu$~\cite{Aaboud:2018gjj,Sirunyan:2019hkq}.
When $\tb \gsim 100$, however, the $\ch$-$t$-$b$ vertex is extremely small, suppressing the production
of the charged Higgs boson.
For the intermediate-mass $A$ and $\varphi^0$, the LHC searches resort to the gluon fusion production via top quark loops,
which is also suppressed.
These hadro-phobic new scalar bosons need different search strategies.

\begin{figure}[h] \centering
\begin{center}
\includegraphics[width=0.45\textwidth]{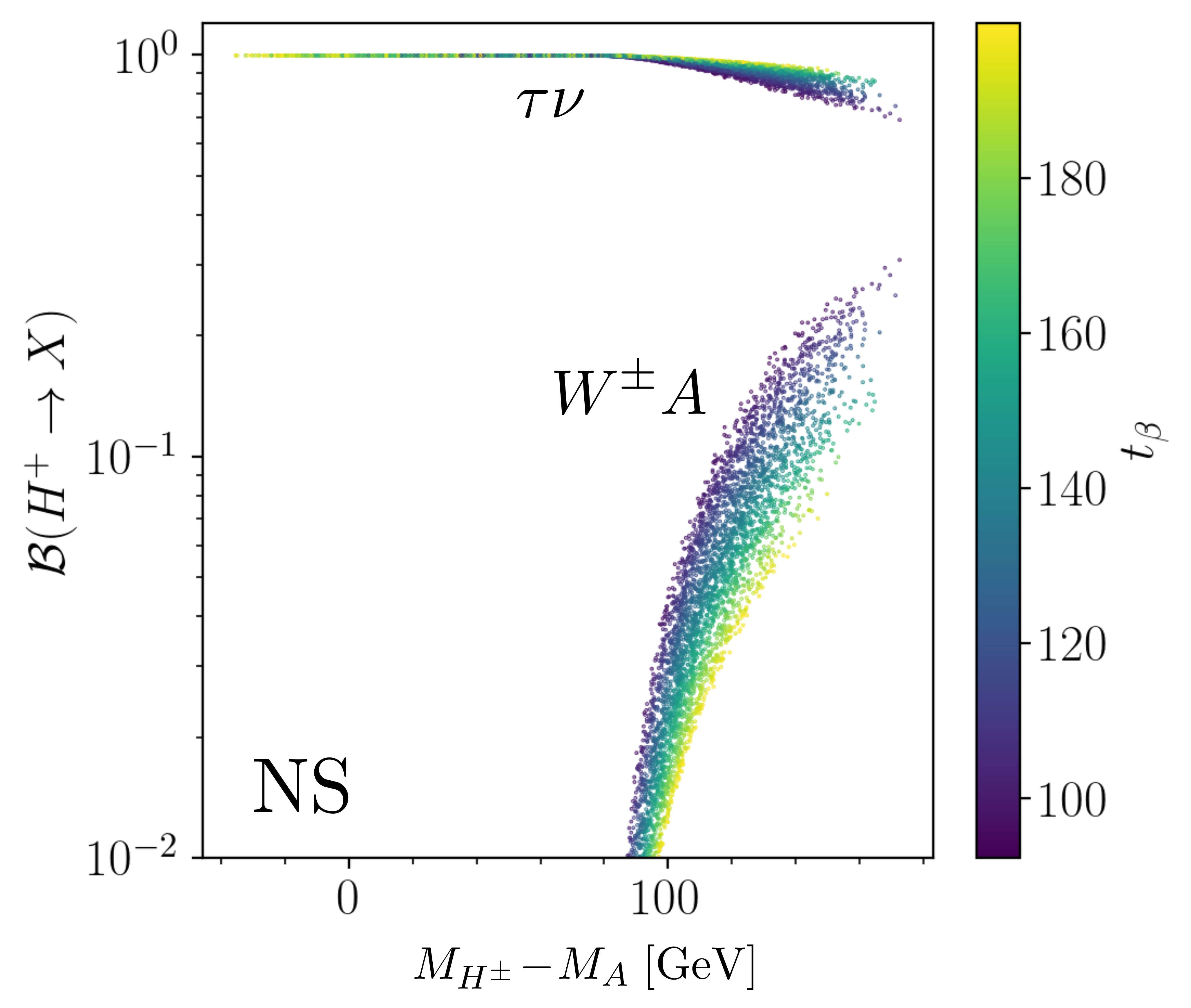}~
\includegraphics[width=0.45\textwidth]{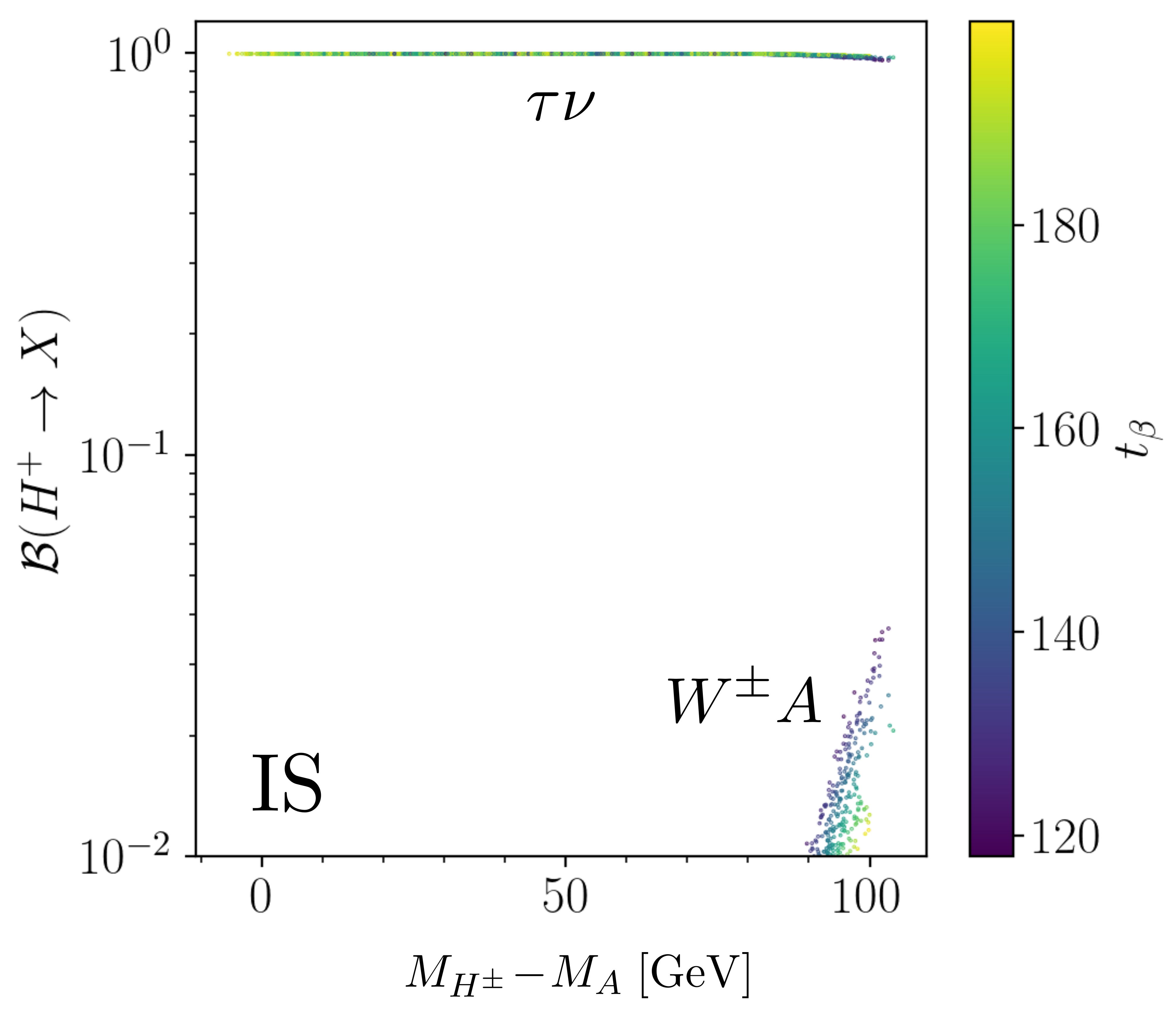}
\end{center}
\caption{\label{fig-BRcH}
The branching ratios of the charged Higgs boson 
as a function of $(\mch-\ma)$ for the finally allowed parameter points.
The left (right) panel corresponds to the normal (inverted) scenario.
}
\end{figure}

We study the branching ratios of $A$, $\varphi^0$, and $\ch$
in the viable parameter space.
Both $A$ and $\varphi^0$ dominantly decay into $\ttau$.
The branching ratios of $H \to ZA$ and $H \to H^\pm W^\mp$ are below $10\%$.
For the $\ch$ decays, Fig.~\ref{fig-BRcH} presents the scatter plot of the branching ratios as a function of $(\mch-\ma)$ in the normal (left panel) and inverted (right panel) scenario.
The color code indicates the value of $\tb$.
The primary decay channel of $\ch$ is into $\tau^\pm\nu$.
The second important mode is $\ch \to W^\pm A$,
which is sizable for larger $(\mch-\ma)$ and smaller $\tb$.
In the normal scenario, $\br(\ch \to W^\pm A)$ can reach up to about 30\%.
In the inverted scenario, its maximum is only about 3\%.

Based on these characteristics, 
we consider the following two channels:
\bea
\label{eq:Ah}
&& \qq \to Z^* \to A \varphi^0 \to \ttau\ttau,
\\ \label{eq:H+H-}
&& pp \to H^+ H^-\to \tau^+\nu \tau^-\nu.
\eea
The process in \eq{eq:Ah} is efficient since the $Z$-$A$-$\varphi^0$ vertex
has the maximal value in the alignment limit of both scenarios.
In addition,
$m_{\varphi^0}$
can be measured through the $\ttau$ invariant mass distribution,
differentiating the normal scenario from the inverted scenario.
The pair production of charged Higgs bosons in \eq{eq:H+H-}
is almost uniquely determined by $\mch$
since the production is via the gauge couplings to $\gm$ and $Z$.

\begin{figure}[h] \centering
\begin{center}
\includegraphics[width=0.48\textwidth]{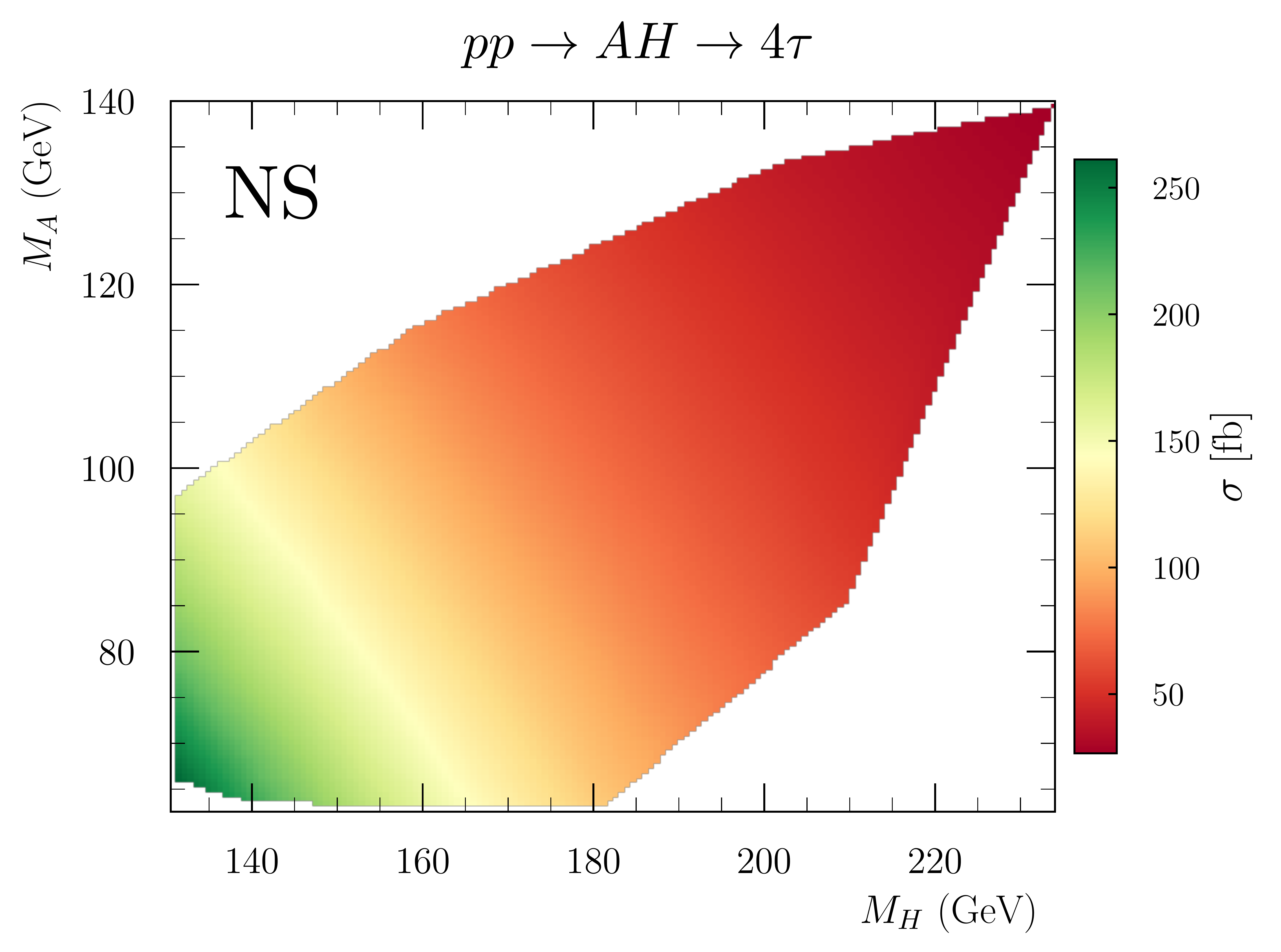}~
\includegraphics[width=0.48\textwidth]{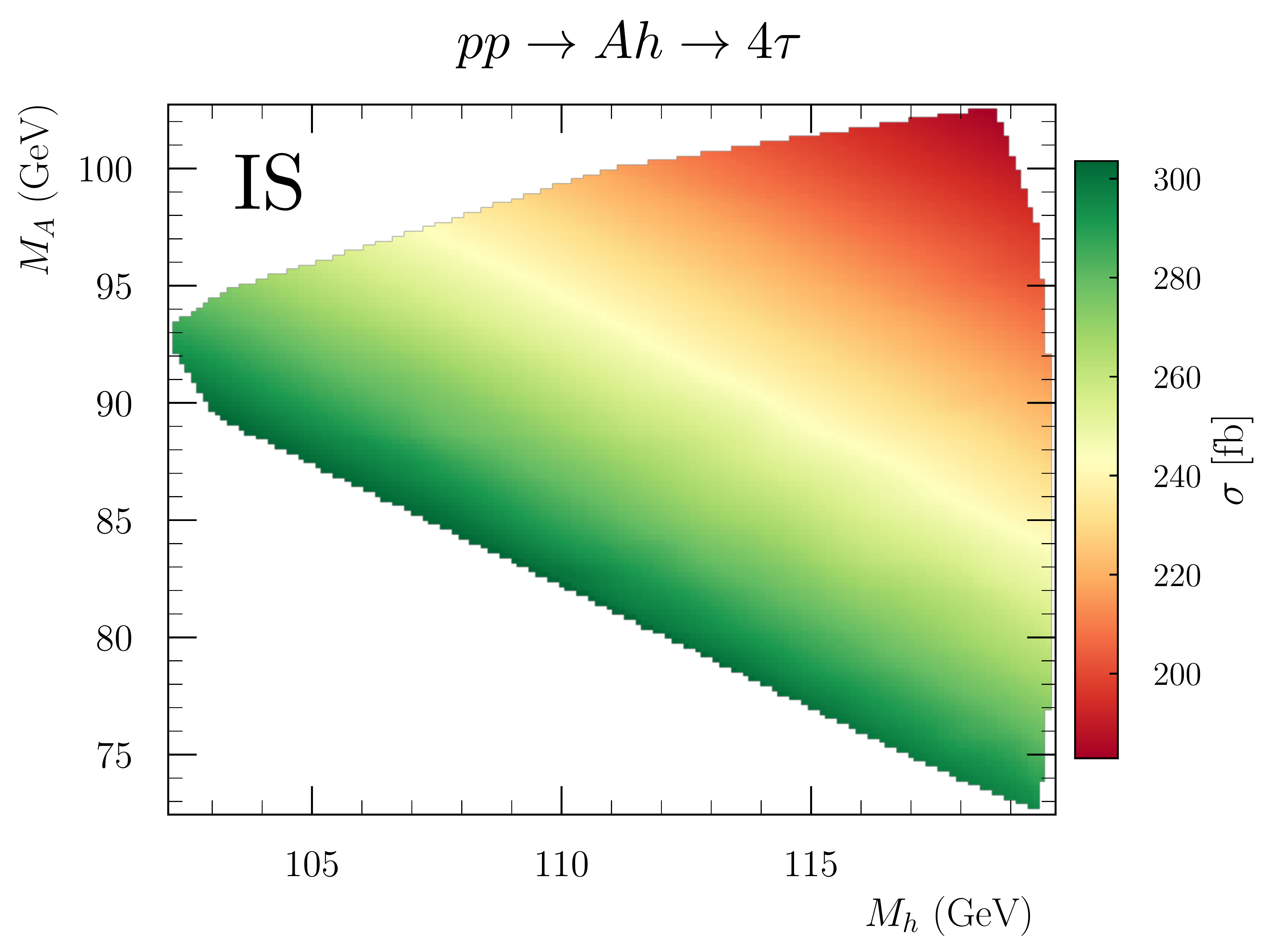}
\end{center}
\caption{\label{fig-sig-pp2AH}
The total cross section at the 14 TeV LHC for the process $pp \to Z^* \to A H/ Ah \to 4\tau $
of the finally allowed parameter points, projected on $(\mhh/\mh,\tb)$.
The left (right) panel corresponds to the normal (inverted) scenario.
}
\end{figure}

Figure \ref{fig-sig-pp2AH} shows the parton-level total cross sections 
of $pp \to Z^* \to A H/ Ah \to 4\tau $ at the 14 TeV LHC,
by scanning all the viable parameter points.
In both scenarios, we see a strong anti-correlation of $\sg_{\rm tot}$ with $\ma+\mhh$.
In the normal scenario (left panel),
the total cross section lies between $\sim 25\fb$ and $\sim 260 \fb$.
In the inverted scenario,
$\sg_{\rm tot}$ goes up to about $300\fb$, larger than in the normal scenario.
Considering the observed $\sg(pp \to ZZ \to 4\tau) \simeq 17\fb$ at the 13 TeV LHC~\cite{Grazzini:2015hta,Aad:2015zqe},
the process $pp \to A \varphi^0 \to 4\tau$ 
has a high potential to probe the model.

\begin{figure}[h] \centering
\begin{center}
\includegraphics[width=0.48\textwidth]{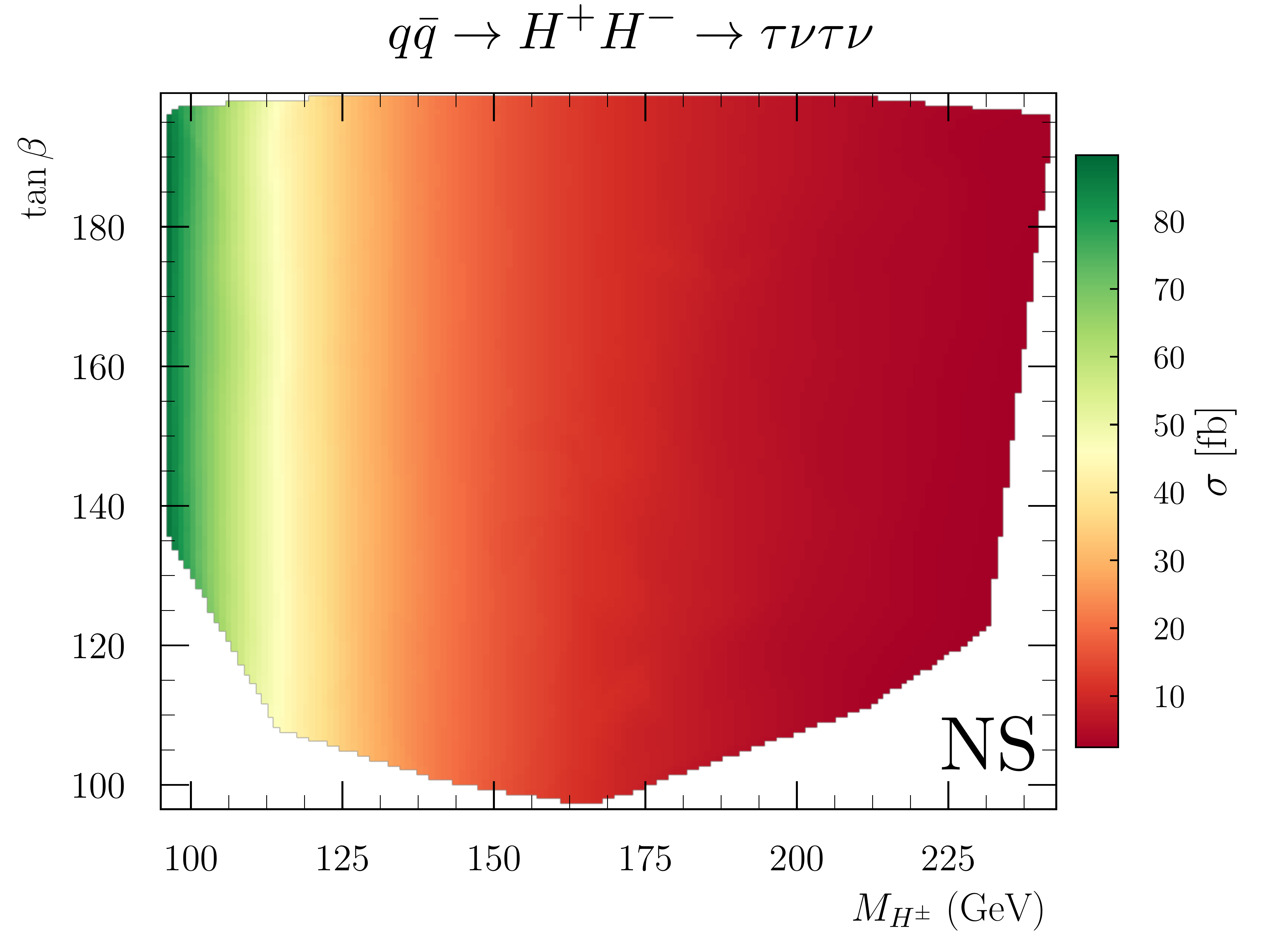}~
\includegraphics[width=0.48\textwidth]{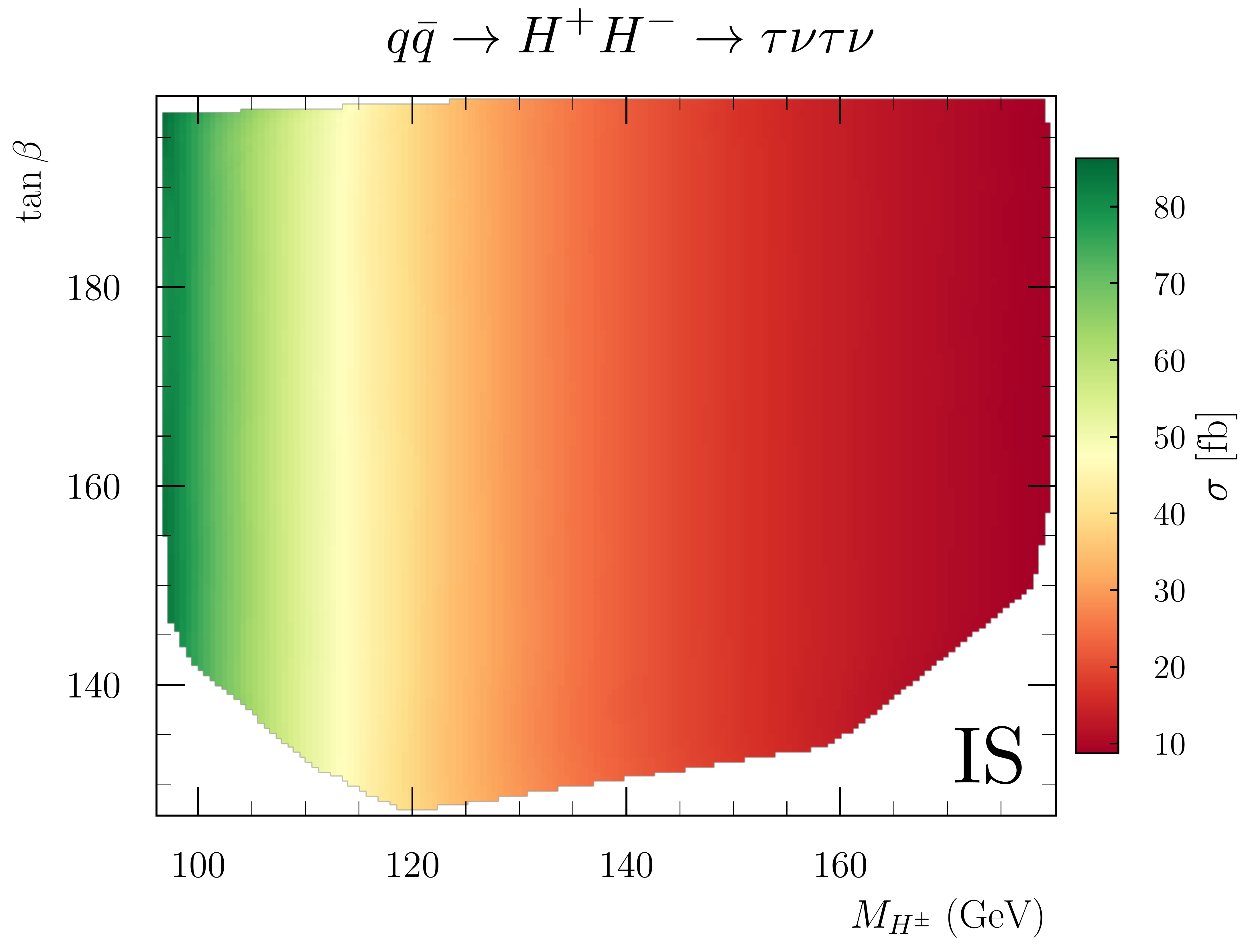}\\[3pt]
\includegraphics[width=0.48\textwidth]{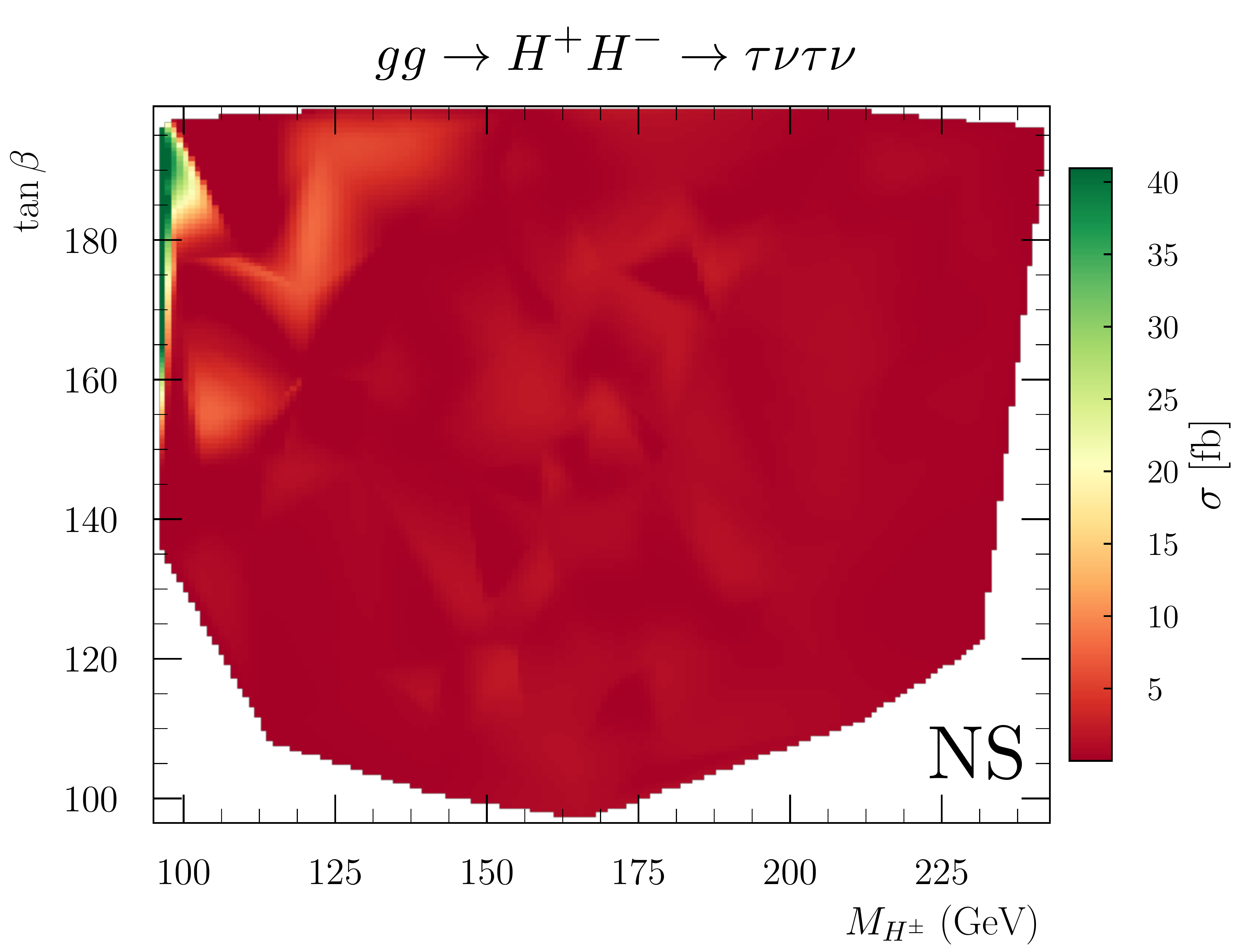}~
\includegraphics[width=0.48\textwidth]{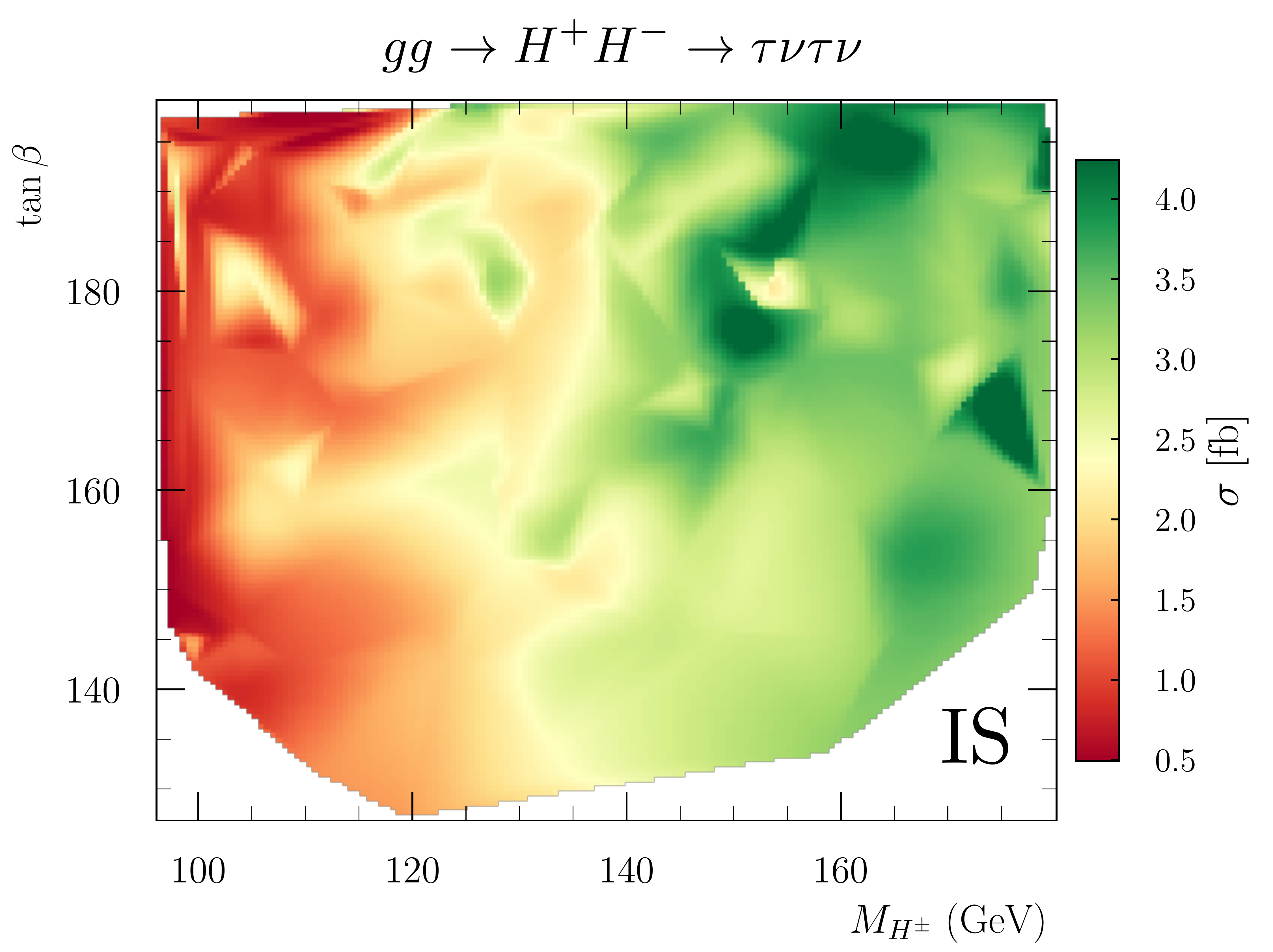}
\end{center}
\caption{\label{fig-sig-cHcH}
The total cross section at the 14 TeV LHC for the process $pp \to H^+ H^- \to \tau\nu\tau\nu $
of the finally allowed parameter points, projected on $(\mch,\tb)$.
The results in the normal (inverted) scenario are in the left (right) panels.
Two upper panels correspond to $\qq\to H^+ H^-$,
and two lower panels to the gluon fusion production.
}
\end{figure}

In Fig.~\ref{fig-sig-cHcH},
we present the prediction of the viable parameters to the total cross sections of the process $pp \to H^+ H^- \to \tau\nu\tau\nu $ at the 14 TeV LHC.
The results in the normal (inverted) scenario are in the left (right) panels.
Two upper (lower) panels present the total cross sections for $\qq\to H^+ H^-$ ($gg\to H^+ H^-$).
In most parameter spaces,
the Drell-Yan production
has a much larger signal rate,
since the hadro-phobic nature of the charged Higgs boson suppresses the gluon fusion production
mediated by the top quark loop.
The irreducible backgrounds for the final state of $\tau^+\nu\tau^-\nu$ 
are $pp \to W^+ W^- \to \tau^+\nu\tau^-\nu$ and $pp \to Z Z \to \ttau\nu\nu$.
Considering  $\sg_{\rm tot}^\sm(pp \to W^+ W^- \to \tau^+\nu\tau^-\nu)\simeq 1.7 \pb$~\cite{Aaboud:2019nkz}
and $\sg_{\rm tot}^\sm(pp \to Z Z \to \ttau\nu\nu)\simeq 100\fb$~\cite{Grazzini:2015hta,Aad:2015zqe}
at the 13 TeV LHC,
there is a chance to see the process.

\section{Lepton flavor universality data in the $\tau$ and $Z$ decays}
\label{sec:LFU}
In Sec.~\ref{sec:scan}, we found that the Type-X 2HDM as a solution to the muon $g-2$ does not allow decoupling of any new Higgs boson.
The generically flavor-universal model may yield excessive violation of the LFU in the $\tau$ and $Z$ decays,
through the loop contributions mediated by new Higgs bosons.
For the rigorous analysis, we first categorize the LFU data as follows:
\begin{description}
\item[(i) HFLAV global fit results in the $\tau$ decay:] 
To parameterize the LFU in the $\tau$ decays,
we introduce the coupling ratios defined by
\bea
\label{eq:tau:LFU}
\lf \frac{g_\tau}{g_\mu} \ri^2 &\equiv& \frac{\Gm(\tau \to e\nu \nu) }{\Gm(\mu \to e\nu \nu) } \times
\frac{f(\rho^e_\mu)}{f(\rho^e_\tau)},
\\[3pt] \nn
\lf \frac{g_\tau}{g_e} \ri^2 &\equiv& \frac{\Gm(\tau \to \mu\nu \nu) }{\Gm(\mu \to e\nu \nu) } \times
\frac{f(\rho^e_\mu)}{f(\rho^\mu_\tau)},
\\[3pt] \nn
\lf \frac{g_\mu}{g_e} \ri^2 &\equiv& \frac{\Gm(\tau \to \mu\nu \nu) }{\Gm(\tau \to e\nu \nu) }\times
\frac{f(\rho^e_\tau)}{f(\rho^\mu_\tau)},
\eea
where $\rho^i_j=m_i^2/m_j^2$ and 
\bea
\label{eq:f(x)}
f(x) &=& 1-8x+8x^3-x^4 -12 x^2 \ln x.
\eea
The second factors in the right-hand sides of \eq{eq:tau:LFU} cancel the mass differences of the charged leptons.
Including the hadronic decays of $\tau \to \pi\nu/K\nu \to \mu\nu\nu$,
we consider
\bea
\label{eq:Rtau:def}
\mathcal{R}^\tau_1 &&\equiv \frac{g_\tau }{ g_\mu}, \quad
\mathcal{R}^\tau_2 \equiv   \frac{g_\tau}{ g_e}  , 
 \quad
\mathcal{R}^\tau_3 \equiv \frac{g_\mu}{g_e}  , 
 \quad
\mathcal{R}^\tau_4 \equiv\lf  g_\tau \over g_\mu \ri_\pi, 
 \quad
\mathcal{R}^\tau_5 \equiv\lf  g_\tau \over g_\mu \ri_K,
\eea
where $\lf \mathcal{R}^{\tau}_{1,\cdots 5} \ri^\sm=1$.
Since $\mathcal{R}^\tau_2/\mathcal{R}^\tau_1=\mathcal{R}^\tau_3$,
only four in \eq{eq:Rtau:def} are independent.
We should remove one redundant degree of freedom that has a zero eigenvalue in the covariance matrix.
\item[(ii) Michel parameters:] 
In the decay of $\tau^- \to \ell^- \nu\nu_\tau$,
the energy and angular distribution of $\ell^-$
provides valuable information on the LFU.
The distribution is written in terms of the Michel parameters $\rho$, $\eta$, $\xi$, and $\dt$,
as~\cite{Michel:1949qe,Logan:2009uf}
\bea
\label{eq:Michel:definition}
\frac{d^2\Gm}{d x\, d\cos\theta^*}
\propto &&
x^2 \left[
3(1-x) + \frac{2\rho}{3} (4x-3) +  3 \,\eta \,x_0 \frac{1-x}{x} \right.
\\ \nn && \left. \qquad +  P_\tau \xi\cos\theta^*
\left\{
1-x + \frac{2\dt}{3}  (4x-3)
\right\}
\right],
\eea
where $x=2 E_{\ell}/m_\tau$, $x_0=2 m_{\ell}/m_\tau$, $P_\tau$ is the $\tau^-$ polarization,
and $\theta^*$ is the angle between the $\ell^-$ momentum and the $\tau^-$ spin quantization axis.
In the SM~\cite{Stahl:1999ui}, they are\footnote{For the hadronic $\tau$ decay,
only the $\xi_h$ was measured.
In the literature, two conventions for the sign of $\xi_h$ coexist,
$\xi_h>0$ in Refs.~\cite{Stahl:1999ui,Belle:2017wxw,ALEPH:2001gaj} and $\xi_h<0$ in Refs.~\cite{Gentile:1995ue,Chun:2016hzs}.
To unify with $\xi_\ell=1$ from the leptonic $\tau$ decays,
we adopt the positive $\xi_h$ convention.
}
\bea
\label{eq:Michel:SM}
\rho^\sm = \frac{3}{4},\quad \eta^\sm =0 , \quad \xi^\sm=1, \quad \lf \xi \dt\ri^\sm = \frac{3}{4}.
\eea
Including the leptonic and hadronic decays of $\tau^-$,
we consider the Michel parameters of
\begin{eqnarray}
\label{eq:Michel:def}
\mathcal{R}^{\rm M}_1 &\equiv& \rho_e , 
\quad \mathcal{R}^{\rm M}_2\equiv \lf \xi \delta \ri_e , 
\quad 
\mathcal{R}^{\rm M}_3\equiv \xi_e , 
\\ \nn
\mathcal{R}^{\rm M}_4 &\equiv&
\eta_\mu , 
\quad 
\mathcal{R}^{\rm M}_5 \equiv\rho_\mu, 
\quad 
\mathcal{R}^{\rm M}_6 \equiv \lf \xi \delta \ri_\mu , 
\quad 
\mathcal{R}^{\rm M}_7 \equiv \xi_\mu , 
\\ \nn
\mathcal{R}^{\rm M}_8 &\equiv& \xi_\pi , 
\quad
\mathcal{R}^{\rm M}_9 \equiv \xi_\rho , 
\quad
\mathcal{R}^{\rm M}_{10}\equiv \xi_{a_1} .
\end{eqnarray}
\item[(iii) LFU in the $Z$ decay:] From the partial decay rates of the leptonic $Z$ decays, 
we take two ratios of~\cite{ALEPH:2005ab}
\bea
\label{eq:obs:Z}
\mathcal{R}^Z_1 \equiv\frac{\Gm(Z\to \mmu)}{\Gm(Z \to \ee)}, \quad
\mathcal{R}^Z_2 \equiv\frac{\Gm(Z\to \ttau)}{\Gm(Z \to \ee)}.
\eea
where $\lf\mathcal{R}^{Z}_{1,2}\ri^\sm=1$ if neglecting $m_\tau^2/m_Z^2$.
\end{description}

The Type-X 2HDM makes two sorts of contributions to the observables in the $\tau$ decays,
the tree-level contributions (mediated by the charged Higgs boson)
and the one-loop level contributions.
We parameterize them by
\bea
\label{eq:dt:loop:tree}
\dt_{\rm tree} &=& \frac{m_\mu m_\tau \tb^2}{\mch^2}
, 
\\[3pt] \nn
\dt_{\rm loop} &=& \frac{1}{16\pi^2} \frac{m_\tau^2\tb^2}{v^2}
\left[
1+\frac{1}{4} \left\{
H\lf \rho^A_{\ch} \ri + H \lf \rho^{\varphi^0}_{\ch} \ri
\right\}
\right],
\eea
where $\rho^i_j=m_i^2/m_j^2$, 
and $H(x) = (1+x)\ln x/(1-x)$.
If $\ma=\mhh=\mch$,
we have $\dt_{\rm loop}=0$ since $\lim_{x\to 1}H(x)=-2$.
Therefore, similar masses of $H$, $A$, and $\ch$ in the viable parameter space
help to suppress the loop-corrections to the $\tau$ decays.

For the HFLAV results, $\mathcal{R}^\tau_i$'s in \eq{eq:Rtau:def} receive new contributions, given by
\bea
\label{eq:tau:new}
\mathcal{R}^\tau_1 &=&\mathcal{R}^\tau_4=\mathcal{R}^\tau_5= 1+\es^{\tau}_{\rm loop},
\\ \nn
\mathcal{R}^\tau_2 &=& 1+ \dt_{\rm loop}+ \es^\tau_{\rm tree} ,
\\ \nn
\mathcal{R}^\tau_3 &=& 1+ \es^\tau_{\rm tree}.
\eea 
Here $\es^\tau_{\rm tree}$ is
\bea
\label{eq:dt:tau:tree}
\es^\tau_{\rm tree} &=& \dt_{\rm tree} 
\left[\frac{ \dt_{\rm tree}}{8}
-\frac{m_\mu}{m_\tau} \frac{g\lf \rho^\mu_\tau \ri}{f\lf \rho^\mu_\tau \ri}
\right],
\eea
where $g(x) = 1+9x -9 x^2-x^3 + 6 x (1+x) \ln x$ and $f(x)$ is in \eq{eq:f(x)}.

For the Michel parameters,
only the $\eta_\mu$, $ \lf \xi \dt \ri_\mu$, and $\xi_\mu$ are modified as
\bea
\label{eq:Michel:2HDM}
\mathcal{R}^{\rm M}_4 &&\equiv \eta_\mu = - \frac{2  \dt_{\rm tree}  (1+\dt_{\rm loop})}{4 + \dt_{\rm tree}^2},
\\[3pt] \nn
\mathcal{R}^{\rm M}_6 &&\equiv\lf \xi \dt \ri_\mu =
\frac{3}{4}\times
\frac{4(1+\dt_{\rm loop})^2-\dt_{\rm tree}^2}{4(1+\dt_{\rm loop})^2+\dt_{\rm tree}^2},
\\[3pt] \nn
\mathcal{R}^{\rm M}_7 &&\equiv \xi_\mu = \frac{4(1+\dt_{\rm loop})^2-\dt_{\rm tree}^2}{4(1+\dt_{\rm loop})^2+\dt_{\rm tree}^2}.
\eea
The corrections to $\rho_e$, $\lf \xi \delta \ri_e$, and $\xi_e$
are suppressed by the small electron mass.
For the hadronic $\tau$ decays, the corrections are independent of $\tb$,
which is much smaller in the large $\tb$ limit than the $\tb^2$ corrections in the leptonic $\tau$ decays.

For $\mathcal{R}^Z_i$'s,
new contributions are written as
\bea
\mathcal{R}^Z_i -1 = 
\frac{2 g_L^\sm {\rm Re}\lf \dt g_{L}^i \ri +2 g_R^\sm {\rm Re}\lf \dt g_{R}^i \ri}
{\lf g_L^\sm\ri^2+\lf g_L^\sm\ri^2},\quad \lf i=\mu,\tau \ri
\eea
where $g_L^\sm=s_W^2-1/2$, $g_R^\sm=s_W^2$,
and the full expressions for $\dt g_{L/R}^{\mu,\tau}$ at one-loop level are referred to Ref.~\cite{Chun:2016hzs}.

Rough estimation of new contributions is useful.
Since $\dt_{\rm tree} \gg \dt_{\rm loop}$ and $\dt_{\rm tree} \gg\es_{\rm tree}^\tau$, 
the dominant contribution is
\bea
\eta_\mu \simeq -\frac{1}{2} \dt_{\rm tree}.
\eea
While $\dt_{\rm tree}$ is positive so that $\eta_\mu<0$ in the model,
the ALEPH result is $\eta_\mu^{\rm ALEPH} = 0.160 \pm 0.150$~\cite{ALEPH:2001gaj}.
The observed $\eta_\mu$ threatens the consistency of the Type-X 2HDM with the LFU data.

Now we perform the global $\chi^2$ fit of the Type-X 2HDM to
\bea
\label{eq:chi2:observables}
\damu,\quad \mathcal{R}^\tau_{1,\cdots,5},\quad \mathcal{R}^{\rm M}_{1,\cdots ,10},\quad
\mathcal{R}^Z_{1,2}.
\eea
The experimental results of $\mathcal{R}$'s and the correlation matrices
are summarized in Appendix \ref{appendix:obs:tau:Z:decays}.
Altogether we have 17 independent observables, $N_{\rm obs}=17$,
since we removed one redundant degree of freedom in $\mathcal{R}^\tau_{1,\cdots,5}$.
In the SM where the number of degree of freedom is $N_{\rm dof}=17$, 
$\chi^2_{\rm min}$ and $p$ value are
\bea
\label{eq:SM:chi2:p}
\chi^2_{\rm min}({\rm SM})=37.3, \quad  p({\rm SM})=0.003.
\eea
The $\damuo$ with the LFU data calls for NP.

For the Type-X 2HDM, we address two issues.
The first is the number of degrees of freedom, $N_{\rm dof}=N_{\rm obs}-N_{\rm par}$,
where $N_{\rm par}$ is the number of free parameters.
We subtract $N_{\rm par}$ under the assumption that we use one \text{free} parameter to explain one observable.
But our hypothesis model is not a \textit{free} Type-X 2HDM.
It is the model severely limited by the theoretical and experimental constraints.
In favor of the Type-X 2HDM, we take $N_{\rm dof}=17$.
The second issue is the range of the model parameters in the global $\chi^2$ fit.
When finding $\chi^2_{\rm min}$,
we may scan either the whole parameter space without imposing other constraints or
only the parameter space consistent with all the constraints.
In the two cases,
$p$-values show big differences as follows:
\bea
\label{eq:min:chi2}
p(\hbox{NS: Step I} )&=&0.58,\quad p(\hbox{NS: Step III} )=0.02,
\\ \nn
p(\hbox{IS: Step I} )&=&0.059,\quad p(\hbox{IS: Step III} )=0.02.
\eea
Without the LHC data,
the Type-X 2HDM in both scenarios well explains the $\damu$ and LFU data.
With the combination of the LHC data and LFU data,
however,
the model is excluded as a solution to the new Fermilab measurement of the muon $g-2$.

\section{Conclusion}
\label{sec:conclusions}

In light of the recent measurement of the muon anomalous magnetic moment
by Fermilab Muon $g-2$ experiment,
we comprehensively study the Type-X (Lepton-specific) two Higgs doublet model (2HDM).
Beyond explaining only the observed $\damu$,
we included the theoretical stability conditions and almost all the available experimental results in the analysis.
Since the Higgs precision data
prefers the SM-like Higgs boson, more strongly for large $\tb$,
we assumed the Higgs alignment.
Two possible scenarios are studied, 
the normal scenario where the lighter \textit{CP}-even $h$ becomes $\hsm$
and 
the inverted scenario where the heavier \textit{CP}-even $H$ is $\hsm$.
The model has five parameters, 
$m_{\varphi^0}$, $\ma$, $\mch$, $M^2$, and $\tb$, where $\varphi^0=H$ in the normal scenario and $\varphi^0=h$ in the inverted scenario.

Various phenomenological conditions cause a chain reaction of constraining the model parameters. 
First, the large and positive $\damuo$ requires large $\tb$ and light $\ma$.
The dominant contribution is from the $\tau^\pm$ loop mediated by $A$ in the two-loop Barr-Zee diagram.
Unwanted is the negative contribution of $\varphi^0$ to the $\tau^\pm$ loop in the Barr-Zee diagram. 
But decoupling of $\varphi^0$ conflicts with the theoretical stability because of large $\tb$.
The Higgs quartic coupling $\lm_1$ has $\tb^2$ terms, which can easily break the perturbativity of $\lm_1$.
Requiring the $\tb^2$ terms to vanish yields $m_{\varphi^0}^2 \approx M^2$.
Perturbativity of other quartic couplings subsequently demands $\ma \simeq \mch \simeq M \approx m_{\varphi^0}$.
Decoupling of any new Higgs boson is not possible.
The direct search bounds at the LEP and LHC exclude a large portion of the parameter space:
$pp\to\hsm \to AA$ in the normal scenario and $\ee \to Z^* \to Ah$ in the inverted scenario are the smoking signals.
Only the region with $\ma > m_{\hsm}/2$ survives.
In turn, $\damuo$ demands $\tb \gsim 100$.

Through random scanning without any prior assumptions on the masses and couplings,
we obtained the parameter points consistent with the muon $g-2$, theoretical stabilities, $S/T/U$ parameters, Higgs precision data, and direct search results.
We also studied the phenomenological implications of the allowed parameter space.
The model prediction to the electron anomalous magnetic moment
is consistent with the observation,
$\dae^{\rm Cs}$ (using the fine structure constant $\al$ from $^{133}$Cs) at $3\sg$,
and $\dae^{\rm Rb}$ (using $\al$ from $^{87}$Rb) at $2\sg$.
For the HL-LHC searches, 
we calculated the total cross sections for the hadro-phobic new scalar bosons
in two processes, $pp \to A \varphi^0 \to 4 \tau$ and $pp \to H^+ H^-\to \tau\nu\tau\nu$.
In particular, $pp \to A \varphi^0 \to 4 \tau$ has the total cross section around $25\sim 260 \fb$ in the normal scenario and $180\sim300\fb$ in the inverted scenario.
The model has a high potential to be probed at the LHC.
As the final check of the model, 
we studied the lepton flavor universality in the $\tau$ and $Z$ decays.
Through a global $\chi^2$ fit to 16 LFU data and $\damuo$,
we showed that the combination of the LHC results and the LFU data
excludes the Type-X 2HDM as a solution to the muon $g-2$.

The confirmed deviation of the muon $g-2$ from the SM prediction by the recent Fermilab experiment indicates the dawn of a new physics era. The Type-X 2HDM that explains $\damu$ is consistent with the LEP and LHC data in limited parameter space, but not with the LFU data in the $\tau$ and $Z$ decays. 
The future LHC searches targeting the specific parameters shall provide a valuable and independent probe of the model,
which we strongly support.

\acknowledgments
We would like to thank Kingman Cheung and Chih-Ting Lu for useful discussions.
This work is supported 
by the National Research Foundation of Korea, Grant No. NRF-2019R1A2C1009419.

\appendix
\section{Used parameters in the $\tau^\pm$ and $Z$ decays for the global $\chi^2$ analysis}
\label{appendix:obs:tau:Z:decays}

We present the experimental data on the parameters in the $\tau^\pm$ and $Z$ decays,
which were used in the global $\chi^2$ analysis of Sec.~\ref{sec:LFU}.\\

\noindent\textbf{(i) HFLAV global fit results in the $\tau$ decay:}~\cite{HFLAV:2019otj}
\bea
\label{eq:obs:tau}
\mathcal{R}^\tau_1 &&\equiv \lf \frac{g_\tau }{ g_\mu} \ri  = 1.0010 \pm 0.0014, 
\\ \nn
\mathcal{R}^\tau_2 &&\equiv  \lf \frac{g_\tau}{ g_e} \ri = 1.0029 \pm 0.0014, 
\\ \nn
\mathcal{R}^\tau_3 &&\equiv \lf \frac{g_\mu}{g_e} \ri  = 1.0018 \pm 0.0014, 
\\ \nn
\mathcal{R}^\tau_4 &&\equiv\lf  g_\tau \over g_\mu \ri_\pi = 0.9958 \pm 0.0026, 
\\ \nn
\mathcal{R}^\tau_5 &&\equiv\lf  g_\tau \over g_\mu \ri_K = 0.9879 \pm 0.0063,
\eea
and the correlation matrix for $\lf \mathcal{R}^\tau_1,\cdots,\mathcal{R}^\tau_5\ri$ is
\bea
\label{eq:correlation:tau}
\lf \mbox{\boldmath$\rho$}_{ij}^\tau\ri = 
\lf
\begin{array}{rrrrr}
1.00 & ~~0.51 & ~-0.50 & ~~0.23 & ~~~0.11\\
0.51 & 1.00 & 0.49 & 0.25 & 0.10 \\
-0.50 & 0.49 & 1.00 & 0.02 & -0.01 \\
0.23 & 0.25 & 0.02 & 1.00 & 0.06 \\
0.11 & 0.10 & -0.01 & 0.06 & 1.00
\end{array}
\ri.
\eea

\noindent\textbf{(ii) Michel parameters in the $\tau$ decay:}~\cite{ALEPH:2001gaj}
\begin{eqnarray}
\mathcal{R}^{\rm M}_1 &\equiv& \rho_e = 0.747 \pm 0.019, 
\\ \nn
\mathcal{R}^{\rm M}_2 &\equiv& \lf \xi \delta \ri_e = 0.788 \pm 0.066, 
\\ \nn
\mathcal{R}^{\rm M}_3 &\equiv& \xi_e = 1.011 \pm 0.094, 
\\ \nn
\mathcal{R}^{\rm M}_4 &\equiv&
\eta_\mu = 0.160 \pm 0.150, 
\\ \nn
\mathcal{R}^{\rm M}_5 &\equiv& \rho_\mu = 0.776 \pm 0.045, 
\\ \nn
\mathcal{R}^{\rm M}_6 &\equiv& \lf \xi \delta \ri_\mu = 0.786 \pm 0.066, 
\\ \nn
\mathcal{R}^{\rm M}_7 &\equiv& \xi_\mu = 1.030 \pm 0.120, 
\\ \nn
\mathcal{R}^{\rm M}_8 &\equiv&
\xi_\pi = 0.994 \pm 0.020, 
\\ \nn
\mathcal{R}^{\rm M}_9 &\equiv& \xi_\rho = 0.987 \pm 0.012, 
\\ \nn
\mathcal{R}^{\rm M}_{10} &\equiv& \xi_{a_1} = 1.000 \pm 0.016.
\end{eqnarray}
and the correlation matrix for $\lf \mathcal{R}^{\rm M}_{1},\cdots,\mathcal{R}^{\rm M}_{10} \ri$ is
\begin{equation}
\lf \mbox{\boldmath$\rho$}_{ij}^{\rm M}\ri =
\left(
\begin{array}{rrrrrrrrrr}
1.00  & 0.00  & -0.23 & -0.02 & 0.02  & 0.00  & -0.02 & 0.10 & 0.00 & 0.00 \\
0.00  & 1.00  & 0.05  & 0.01  & 0.00  & 0.00  & -0.02 & -0.12 & 0.01 & 0.01 \\
-0.23 & 0.05  & 1.00  & 0.01  & 0.01  & -0.02 & -0.01 & -0.06 & -0.02 & -0.01 \\
-0.02 & 0.01  & 0.01  & 1.00  & 0.91  & 0.29  & 0.58  & 0.03 & -0.02 & -0.01 \\ 
0.02  & 0.00  & 0.01  & 0.91  & 1.00  & 0.25  & 0.45  & 0.06 & -0.02 & -0.01 \\ 
0.00  & 0.00  & -0.02 & 0.29  & 0.25  & 1.00  & 0.14  & -0.07 & -0.02 & -0.01 \\ 
-0.02 & -0.02 & -0.01 & 0.58  & 0.45  & 0.14  & 1.00  & -0.01  & -0.04 & -0.02 \\
0.10  & -0.12 & -0.06 & 0.03  & 0.06  & -0.07 & -0.01 & 1.00 & -0.28 & -0.20 \\ 
0.00  & 0.01  & -0.02 & -0.02 & -0.02 & -0.02 & -0.04 & -0.28 & 1.00 & -0.08 \\  
0.00  & 0.01  & -0.01 & -0.01 & -0.01 & -0.01 & -0.02 & -0.20 & -0.08 & 1.00 
\end{array}
\right).
\end{equation}

\noindent\textbf{(iii) LFU in the $Z$ decay:}~\cite{ALEPH:2005ab}
\bea
\label{eq:obs:Z}
\mathcal{R}^Z_1 \equiv\frac{\Gm(Z\to \mmu)}{\Gm(Z \to \ee)} &=& 1.0009 \pm 0.0028,
\\[5pt] \nn
\mathcal{R}^Z_2 \equiv\frac{\Gm(Z\to \ttau)}{\Gm(Z \to \ee)} &=& 1.0019 \pm 0.0032,
\eea
where the correlation between $\mathcal{R}^Z_1$ and $ \mathcal{R}^Z_2$ is $+0.63$.


\begin{thebibliography}{100}

\bibitem{Abi:2021gix}
{\scshape Muon g-2} collaboration, B.~Abi et~al., \textit{{Measurement of the
  Positive Muon Anomalous Magnetic Moment to 0.46 ppm}},
  \href{http://dx.doi.org/10.1103/PhysRevLett.126.141801}{\emph{Phys. Rev.
  Lett.} \textbf{ 126} (2021) 141801},
  [\href{https://arxiv.org/abs/2104.03281}{\texttt{2104.03281}}].

\bibitem{Albahri:2021ixb}
T.~Albahri et~al., \textit{{Measurement of the anomalous precession frequency
  of the muon in the Fermilab Muon g-2 experiment}},
  \href{http://dx.doi.org/10.1103/PhysRevD.103.072002}{\emph{Phys. Rev. D}
  \textbf{ 103} (2021) 072002},
  [\href{https://arxiv.org/abs/2104.03247}{\texttt{2104.03247}}].

\bibitem{Bennett:2006fi}
{\scshape Muon g-2} collaboration, G.~W. Bennett et~al., \textit{{Final Report
  of the Muon E821 Anomalous Magnetic Moment Measurement at BNL}},
  \href{http://dx.doi.org/10.1103/PhysRevD.73.072003}{\emph{Phys. Rev. D}
  \textbf{ 73} (2006) 072003},
  [\href{https://arxiv.org/abs/hep-ex/0602035}{\texttt{hep-ex/0602035}}].

\bibitem{Aoyama:2012wk}
T.~Aoyama, M.~Hayakawa, T.~Kinoshita and M.~Nio, \textit{{Complete Tenth-Order
  QED Contribution to the Muon g-2}},
  \href{http://dx.doi.org/10.1103/PhysRevLett.109.111808}{\emph{Phys. Rev.
  Lett.} \textbf{ 109} (2012) 111808},
  [\href{https://arxiv.org/abs/1205.5370}{\texttt{1205.5370}}].

\bibitem{Czarnecki:2002nt}
A.~Czarnecki, W.~J. Marciano and A.~Vainshtein, \textit{{Refinements in
  electroweak contributions to the muon anomalous magnetic moment}},
  \href{http://dx.doi.org/10.1103/PhysRevD.67.073006}{\emph{Phys. Rev. D}
  \textbf{ 67} (2003) 073006},
  [\href{https://arxiv.org/abs/hep-ph/0212229}{\texttt{hep-ph/0212229}}].

\bibitem{Gnendiger:2013pva}
C.~Gnendiger, D.~St\"ockinger and H.~St\"ockinger-Kim, \textit{{The electroweak
  contributions to $(g-2)_\mu$ after the Higgs boson mass measurement}},
  \href{http://dx.doi.org/10.1103/PhysRevD.88.053005}{\emph{Phys. Rev. D}
  \textbf{ 88} (2013) 053005},
  [\href{https://arxiv.org/abs/1306.5546}{\texttt{1306.5546}}].

\bibitem{Kurz:2014wya}
A.~Kurz, T.~Liu, P.~Marquard and M.~Steinhauser, \textit{{Hadronic contribution
  to the muon anomalous magnetic moment to next-to-next-to-leading order}},
  \href{http://dx.doi.org/10.1016/j.physletb.2014.05.043}{\emph{Phys. Lett. B}
  \textbf{ 734} (2014) 144--147},
  [\href{https://arxiv.org/abs/1403.6400}{\texttt{1403.6400}}].

\bibitem{Davier:2017zfy}
M.~Davier, A.~Hoecker, B.~Malaescu and Z.~Zhang, \textit{{Reevaluation of the
  hadronic vacuum polarisation contributions to the Standard Model predictions
  of the muon $g-2$ and ${\alpha (m_Z^2)}$ using newest hadronic cross-section
  data}}, \href{http://dx.doi.org/10.1140/epjc/s10052-017-5161-6}{\emph{Eur.
  Phys. J. C} \textbf{ 77} (2017) 827},
  [\href{https://arxiv.org/abs/1706.09436}{\texttt{1706.09436}}].

\bibitem{Colangelo:2018mtw}
G.~Colangelo, M.~Hoferichter and P.~Stoffer, \textit{{Two-pion contribution to
  hadronic vacuum polarization}},
  \href{http://dx.doi.org/10.1007/JHEP02(2019)006}{\emph{JHEP} \textbf{ 02}
  (2019) 006}, [\href{https://arxiv.org/abs/1810.00007}{\texttt{1810.00007}}].

\bibitem{Keshavarzi:2018mgv}
A.~Keshavarzi, D.~Nomura and T.~Teubner, \textit{{Muon $g-2$ and
  $\alpha(M_Z^2)$: a new data-based analysis}},
  \href{http://dx.doi.org/10.1103/PhysRevD.97.114025}{\emph{Phys. Rev. D}
  \textbf{ 97} (2018) 114025},
  [\href{https://arxiv.org/abs/1802.02995}{\texttt{1802.02995}}].

\bibitem{Keshavarzi:2019abf}
A.~Keshavarzi, D.~Nomura and T.~Teubner, \textit{{$g-2$ of charged leptons,
  $\alpha (M^2_Z)$ , and the hyperfine splitting of muonium}},
  \href{http://dx.doi.org/10.1103/PhysRevD.101.014029}{\emph{Phys. Rev. D}
  \textbf{ 101} (2020) 014029},
  [\href{https://arxiv.org/abs/1911.00367}{\texttt{1911.00367}}].

\bibitem{Davier:2019can}
M.~Davier, A.~Hoecker, B.~Malaescu and Z.~Zhang, \textit{{A new evaluation of
  the hadronic vacuum polarisation contributions to the muon anomalous magnetic
  moment and to $\mathbf{\boldsymbol\alpha(m_Z^2)}$}},
  \href{http://dx.doi.org/10.1140/epjc/s10052-020-7792-2}{\emph{Eur. Phys. J.
  C} \textbf{ 80} (2020) 241},
  [\href{https://arxiv.org/abs/1908.00921}{\texttt{1908.00921}}].

\bibitem{Hoid:2020xjs}
B.-L. Hoid, M.~Hoferichter and B.~Kubis, \textit{{Hadronic vacuum polarization
  and vector-meson resonance parameters from $e^+e^-\rightarrow \pi
  ^0\gamma$}},
  \href{http://dx.doi.org/10.1140/epjc/s10052-020-08550-2}{\emph{Eur. Phys. J.
  C} \textbf{ 80} (2020) 988},
  [\href{https://arxiv.org/abs/2007.12696}{\texttt{2007.12696}}].

\bibitem{Colangelo:2020lcg}
G.~Colangelo, M.~Hoferichter and P.~Stoffer, \textit{{Constraints on the
  two-pion contribution to hadronic vacuum polarization}},
  \href{http://dx.doi.org/10.1016/j.physletb.2021.136073}{\emph{Phys. Lett. B}
  \textbf{ 814} (2021) 136073},
  [\href{https://arxiv.org/abs/2010.07943}{\texttt{2010.07943}}].

\bibitem{Melnikov:2003xd}
K.~Melnikov and A.~Vainshtein, \textit{{Hadronic light-by-light scattering
  contribution to the muon anomalous magnetic moment revisited}},
  \href{http://dx.doi.org/10.1103/PhysRevD.70.113006}{\emph{Phys. Rev. D}
  \textbf{ 70} (2004) 113006},
  [\href{https://arxiv.org/abs/hep-ph/0312226}{\texttt{hep-ph/0312226}}].

\bibitem{Colangelo:2014pva}
G.~Colangelo, M.~Hoferichter, B.~Kubis, M.~Procura and P.~Stoffer,
  \textit{{Towards a data-driven analysis of hadronic light-by-light
  scattering}},
  \href{http://dx.doi.org/10.1016/j.physletb.2014.09.021}{\emph{Phys. Lett. B}
  \textbf{ 738} (2014) 6--12},
  [\href{https://arxiv.org/abs/1408.2517}{\texttt{1408.2517}}].

\bibitem{Colangelo:2014qya}
G.~Colangelo, M.~Hoferichter, A.~Nyffeler, M.~Passera and P.~Stoffer,
  \textit{{Remarks on higher-order hadronic corrections to the muon
  g\ensuremath{-}2}},
  \href{http://dx.doi.org/10.1016/j.physletb.2014.06.012}{\emph{Phys. Lett. B}
  \textbf{ 735} (2014) 90--91},
  [\href{https://arxiv.org/abs/1403.7512}{\texttt{1403.7512}}].

\bibitem{Colangelo:2015ama}
G.~Colangelo, M.~Hoferichter, M.~Procura and P.~Stoffer, \textit{{Dispersion
  relation for hadronic light-by-light scattering: theoretical foundations}},
  \href{http://dx.doi.org/10.1007/JHEP09(2015)074}{\emph{JHEP} \textbf{ 09}
  (2015) 074}, [\href{https://arxiv.org/abs/1506.01386}{\texttt{1506.01386}}].

\bibitem{Colangelo:2017fiz}
G.~Colangelo, M.~Hoferichter, M.~Procura and P.~Stoffer, \textit{{Dispersion
  relation for hadronic light-by-light scattering: two-pion contributions}},
  \href{http://dx.doi.org/10.1007/JHEP04(2017)161}{\emph{JHEP} \textbf{ 04}
  (2017) 161}, [\href{https://arxiv.org/abs/1702.07347}{\texttt{1702.07347}}].

\bibitem{Masjuan:2017tvw}
P.~Masjuan and P.~Sanchez-Puertas, \textit{{Pseudoscalar-pole contribution to
  the $(g_{\mu}-2)$: a rational approach}},
  \href{http://dx.doi.org/10.1103/PhysRevD.95.054026}{\emph{Phys. Rev. D}
  \textbf{ 95} (2017) 054026},
  [\href{https://arxiv.org/abs/1701.05829}{\texttt{1701.05829}}].

\bibitem{Colangelo:2017qdm}
G.~Colangelo, M.~Hoferichter, M.~Procura and P.~Stoffer, \textit{{Rescattering
  effects in the hadronic-light-by-light contribution to the anomalous magnetic
  moment of the muon}},
  \href{http://dx.doi.org/10.1103/PhysRevLett.118.232001}{\emph{Phys. Rev.
  Lett.} \textbf{ 118} (2017) 232001},
  [\href{https://arxiv.org/abs/1701.06554}{\texttt{1701.06554}}].

\bibitem{Hoferichter:2018dmo}
M.~Hoferichter, B.-L. Hoid, B.~Kubis, S.~Leupold and S.~P. Schneider,
  \textit{{Pion-pole contribution to hadronic light-by-light scattering in the
  anomalous magnetic moment of the muon}},
  \href{http://dx.doi.org/10.1103/PhysRevLett.121.112002}{\emph{Phys. Rev.
  Lett.} \textbf{ 121} (2018) 112002},
  [\href{https://arxiv.org/abs/1805.01471}{\texttt{1805.01471}}].

\bibitem{Hoferichter:2018kwz}
M.~Hoferichter, B.-L. Hoid, B.~Kubis, S.~Leupold and S.~P. Schneider,
  \textit{{Dispersion relation for hadronic light-by-light scattering: pion
  pole}}, \href{http://dx.doi.org/10.1007/JHEP10(2018)141}{\emph{JHEP} \textbf{
  10} (2018) 141},
  [\href{https://arxiv.org/abs/1808.04823}{\texttt{1808.04823}}].

\bibitem{Colangelo:2019lpu}
G.~Colangelo, F.~Hagelstein, M.~Hoferichter, L.~Laub and P.~Stoffer,
  \textit{{Short-distance constraints on hadronic light-by-light scattering in
  the anomalous magnetic moment of the muon}},
  \href{http://dx.doi.org/10.1103/PhysRevD.101.051501}{\emph{Phys. Rev. D}
  \textbf{ 101} (2020) 051501},
  [\href{https://arxiv.org/abs/1910.11881}{\texttt{1910.11881}}].

\bibitem{Bijnens:2019ghy}
J.~Bijnens, N.~Hermansson-Truedsson and A.~Rodr\'\i{}guez-S\'anchez,
  \textit{{Short-distance constraints for the HLbL contribution to the muon
  anomalous magnetic moment}},
  \href{http://dx.doi.org/10.1016/j.physletb.2019.134994}{\emph{Phys. Lett. B}
  \textbf{ 798} (2019) 134994},
  [\href{https://arxiv.org/abs/1908.03331}{\texttt{1908.03331}}].

\bibitem{Blum:2019ugy}
T.~Blum, N.~Christ, M.~Hayakawa, T.~Izubuchi, L.~Jin, C.~Jung et~al.,
  \textit{{Hadronic Light-by-Light Scattering Contribution to the Muon
  Anomalous Magnetic Moment from Lattice QCD}},
  \href{http://dx.doi.org/10.1103/PhysRevLett.124.132002}{\emph{Phys. Rev.
  Lett.} \textbf{ 124} (2020) 132002},
  [\href{https://arxiv.org/abs/1911.08123}{\texttt{1911.08123}}].

\bibitem{Bijnens:2020xnl}
J.~Bijnens, N.~Hermansson-Truedsson, L.~Laub and A.~Rodr\'\i{}guez-S\'anchez,
  \textit{{Short-distance HLbL contributions to the muon anomalous magnetic
  moment beyond perturbation theory}},
  \href{http://dx.doi.org/10.1007/JHEP10(2020)203}{\emph{JHEP} \textbf{ 10}
  (2020) 203}, [\href{https://arxiv.org/abs/2008.13487}{\texttt{2008.13487}}].

\bibitem{Borsanyi:2020mff}
S.~Borsanyi et~al., \textit{{Leading hadronic contribution to the muon 2
  magnetic moment from lattice QCD}},
  \href{https://arxiv.org/abs/2002.12347}{\texttt{2002.12347}}.

\bibitem{Lehner:2020crt}
C.~Lehner and A.~S. Meyer, \textit{{Consistency of hadronic vacuum polarization
  between lattice QCD and the R-ratio}},
  \href{http://dx.doi.org/10.1103/PhysRevD.101.074515}{\emph{Phys. Rev. D}
  \textbf{ 101} (2020) 074515},
  [\href{https://arxiv.org/abs/2003.04177}{\texttt{2003.04177}}].

\bibitem{Crivellin:2020zul}
A.~Crivellin, M.~Hoferichter, C.~A. Manzari and M.~Montull, \textit{{Hadronic
  Vacuum Polarization: $(g-2)_\mu$ versus Global Electroweak Fits}},
  \href{http://dx.doi.org/10.1103/PhysRevLett.125.091801}{\emph{Phys. Rev.
  Lett.} \textbf{ 125} (2020) 091801},
  [\href{https://arxiv.org/abs/2003.04886}{\texttt{2003.04886}}].

\bibitem{Keshavarzi:2020bfy}
A.~Keshavarzi, W.~J. Marciano, M.~Passera and A.~Sirlin, \textit{{Muon $g-2$
  and $\Delta \alpha$ connection}},
  \href{http://dx.doi.org/10.1103/PhysRevD.102.033002}{\emph{Phys. Rev. D}
  \textbf{ 102} (2020) 033002},
  [\href{https://arxiv.org/abs/2006.12666}{\texttt{2006.12666}}].

\bibitem{Malaescu:2020zuc}
B.~Malaescu and M.~Schott, \textit{{Impact of correlations between $a_{\mu }$
  and $\alpha _\text {QED}$ on the EW fit}},
  \href{http://dx.doi.org/10.1140/epjc/s10052-021-08848-9}{\emph{Eur. Phys. J.
  C} \textbf{ 81} (2021) 46},
  [\href{https://arxiv.org/abs/2008.08107}{\texttt{2008.08107}}].

\bibitem{Czarnecki:2001pv}
A.~Czarnecki and W.~J. Marciano, \textit{{The Muon anomalous magnetic moment: A
  Harbinger for 'new physics'}},
  \href{http://dx.doi.org/10.1103/PhysRevD.64.013014}{\emph{Phys. Rev. D}
  \textbf{ 64} (2001) 013014},
  [\href{https://arxiv.org/abs/hep-ph/0102122}{\texttt{hep-ph/0102122}}].

\bibitem{Baer:2021aax}
H.~Baer, V.~Barger and H.~Serce, \textit{{Anomalous muon magnetic moment,
  supersymmetry, naturalness, LHC search limits and the landscape}},
  \href{https://arxiv.org/abs/2104.07597}{\texttt{2104.07597}}.

\bibitem{Aboubrahim:2021rwz}
A.~Aboubrahim, M.~Klasen and P.~Nath, \textit{{What Fermilab $(g-2)_{\mu}$
  experiment tells us about discovering SUSY at HL-LHC and HE-LHC}},
  \href{https://arxiv.org/abs/2104.03839}{\texttt{2104.03839}}.

\bibitem{Cao:2021tuh}
J.~Cao, J.~Lian, Y.~Pan, D.~Zhang and P.~Zhu, \textit{{Imporved $(g-2)_\mu$
  Measurement and Singlino dark matter in the general NMSSM}},
  \href{https://arxiv.org/abs/2104.03284}{\texttt{2104.03284}}.

\bibitem{Wang:2021bcx}
F.~Wang, L.~Wu, Y.~Xiao, J.~M. Yang and Y.~Zhang, \textit{{GUT-scale
  constrained SUSY in light of E989 muon g-2 measurement}},
  \href{https://arxiv.org/abs/2104.03262}{\texttt{2104.03262}}.

\bibitem{VanBeekveld:2021tgn}
M.~Van~Beekveld, W.~Beenakker, M.~Schutten and J.~De~Wit, \textit{{Dark matter,
  fine-tuning and $(g-2)_{\mu}$ in the pMSSM}},
  \href{https://arxiv.org/abs/2104.03245}{\texttt{2104.03245}}.

\bibitem{Abdughani:2021pdc}
M.~Abdughani, Y.-Z. Fan, L.~Feng, Y.-L. Sming~Tsai, L.~Wu and Q.~Yuan,
  \textit{{A common origin of muon g-2 anomaly, Galaxy Center GeV excess and
  AMS-02 anti-proton excess in the NMSSM}},
  \href{https://arxiv.org/abs/2104.03274}{\texttt{2104.03274}}.

\bibitem{Baum:2021qzx}
S.~Baum, M.~Carena, N.~R. Shah and C.~E.~M. Wagner, \textit{{The Tiny (g-2)
  Muon Wobble from Small-$\mu$ Supersymmetry}},
  \href{https://arxiv.org/abs/2104.03302}{\texttt{2104.03302}}.

\bibitem{Ahmed:2021htr}
W.~Ahmed, I.~Khan, J.~Li, T.~Li, S.~Raza and W.~Zhang, \textit{{The Natural
  Explanation of the Muon Anomalous Magnetic Moment via the Electroweak
  Supersymmetry from the GmSUGRA in the MSSM}},
  \href{https://arxiv.org/abs/2104.03491}{\texttt{2104.03491}}.

\bibitem{Zhang:2021gun}
H.-B. Zhang, C.-X. Liu, J.-L. Yang and T.-F. Feng, \textit{{Muon anomalous
  magnetic dipole moment in the $\mu\nu$SSM}},
  \href{https://arxiv.org/abs/2104.03489}{\texttt{2104.03489}}.

\bibitem{Chakraborti:2021bmv}
M.~Chakraborti, L.~Roszkowski and S.~Trojanowski, \textit{{GUT-constrained
  supersymmetry and dark matter in light of the new $(g-2)_\mu$
  determination}},
  \href{https://arxiv.org/abs/2104.04458}{\texttt{2104.04458}}.

\bibitem{Athron:2021iuf}
P.~Athron, C.~Bal\'azs, D.~H. Jacob, W.~Kotlarski, D.~St\"ockinger and
  H.~St\"ockinger-Kim, \textit{{New physics explanations of $a_\mu$ in light of
  the FNAL muon $g-2$ measurement}},
  \href{https://arxiv.org/abs/2104.03691}{\texttt{2104.03691}}.

\bibitem{Yin:2021mls}
W.~Yin, \textit{{Muon $g-2$ Anomaly in Anomaly Mediation}},
  \href{https://arxiv.org/abs/2104.03259}{\texttt{2104.03259}}.

\bibitem{Buras:2021btx}
A.~J. Buras, A.~Crivellin, F.~Kirk, C.~A. Manzari and M.~Montull,
  \textit{{Global Analysis of Leptophilic Z' Bosons}},
  \href{https://arxiv.org/abs/2104.07680}{\texttt{2104.07680}}.

\bibitem{Chun:2021dwx}
E.~J. Chun and T.~Mondal, \textit{{Leptophilic bosons and muon g-2 at lepton
  colliders}},  \href{https://arxiv.org/abs/2104.03701}{\texttt{2104.03701}}.

\bibitem{Liu:2018xkx}
J.~Liu, C.~E.~M. Wagner and X.-P. Wang, \textit{{A light complex scalar for the
  electron and muon anomalous magnetic moments}},
  \href{http://dx.doi.org/10.1007/JHEP03(2019)008}{\emph{JHEP} \textbf{ 03}
  (2019) 008}, [\href{https://arxiv.org/abs/1810.11028}{\texttt{1810.11028}}].

\bibitem{CarcamoHernandez:2021qhf}
A.~E. C\'arcamo~Hern\'andez, S.~Kovalenko, M.~Maniatis and I.~Schmidt,
  \textit{{Fermion mass hierarchy and g-2 anomalies in an extended 3HDM
  Model}},  \href{https://arxiv.org/abs/2104.07047}{\texttt{2104.07047}}.

\bibitem{Ban:2021tos}
K.~Ban, Y.~Jho, Y.~Kwon, S.~C. Park, S.~Park and P.-Y. Tseng, \textit{{A
  comprehensive study of vector leptoquark on the $B$-meson and Muon g-2
  anomalies}},  \href{https://arxiv.org/abs/2104.06656}{\texttt{2104.06656}}.

\bibitem{Du:2021zkq}
M.~Du, J.~Liang, Z.~Liu and V.~Q. Tran, \textit{{A vector leptoquark
  interpretation of the muon $g-2$ and $B$ anomalies}},
  \href{https://arxiv.org/abs/2104.05685}{\texttt{2104.05685}}.

\bibitem{Borah:2021jzu}
D.~Borah, M.~Dutta, S.~Mahapatra and N.~Sahu, \textit{{Muon $(g-2)$ and XENON1T
  Excess with Boosted Dark Matter in $L_{\mu}-L_{\tau}$ Model}},
  \href{https://arxiv.org/abs/2104.05656}{\texttt{2104.05656}}.

\bibitem{Zu:2021odn}
L.~Zu, X.~Pan, L.~Feng, Q.~Yuan and Y.-Z. Fan, \textit{{Constraining
  $U(1)_{L_{\mu}-L_{\tau}}$ charged dark matter model for muon $g-2$ anomaly
  with AMS-02 electron and positron data}},
  \href{https://arxiv.org/abs/2104.03340}{\texttt{2104.03340}}.

\bibitem{Yang:2021duj}
J.-L. Yang, H.-B. Zhang, C.-X. Liu, X.-X. Dong and T.-F. Feng, \textit{{Muon
  $(g-2)$ in the B-LSSM}},
  \href{https://arxiv.org/abs/2104.03542}{\texttt{2104.03542}}.

\bibitem{Greljo:2021xmg}
A.~Greljo, P.~Stangl and A.~E. Thomsen, \textit{{A Model of Muon Anomalies}},
  \href{https://arxiv.org/abs/2103.13991}{\texttt{2103.13991}}.

\bibitem{Zhu:2021vlz}
B.~Zhu and X.~Liu, \textit{{Probing light dark matter with scalar mediator:
  muon $(g-2)$ deviation, the proton radius puzzle}},
  \href{https://arxiv.org/abs/2104.03238}{\texttt{2104.03238}}.

\bibitem{Escribano:2021css}
P.~Escribano, J.~Terol-Calvo and A.~Vicente,
  \textit{{$\boldsymbol{(g-2)_{e,\mu}}$ in an extended inverse type-III
  seesaw}},  \href{https://arxiv.org/abs/2104.03705}{\texttt{2104.03705}}.

\bibitem{Arcadi:2021cwg}
G.~Arcadi, L.~Calibbi, M.~Fedele and F.~Mescia, \textit{{Muon $g-2$ and
  $B$-anomalies from Dark Matter}},
  \href{https://arxiv.org/abs/2104.03228}{\texttt{2104.03228}}.

\bibitem{Crivellin:2021rbq}
A.~Crivellin and M.~Hoferichter, \textit{{Consequences of chirally enhanced
  explanations of $(g-2)_\mu$ for $h\to \mu\mu$ and $Z\to \mu\mu$}},
  \href{https://arxiv.org/abs/2104.03202}{\texttt{2104.03202}}.

\bibitem{Buen-Abad:2021fwq}
M.~A. Buen-Abad, J.~Fan, M.~Reece and C.~Sun, \textit{{Challenges for an axion
  explanation of the muon $g-2$ measurement}},
  \href{https://arxiv.org/abs/2104.03267}{\texttt{2104.03267}}.

\bibitem{Ge:2021cjz}
S.-F. Ge, X.-D. Ma and P.~Pasquini, \textit{{Probing the Dark Axion Portal with
  Muon Anomalous Magnetic Moment}},
  \href{https://arxiv.org/abs/2104.03276}{\texttt{2104.03276}}.

\bibitem{Ferreira:2021gke}
P.~M. Ferreira, B.~L. Gon\c{c}alves, F.~R. Joaquim and M.~Sher,
  \textit{{$(g-2)_\mu$ in the 2HDM and slightly beyond -- an updated view}},
  \href{https://arxiv.org/abs/2104.03367}{\texttt{2104.03367}}.

\bibitem{Han:2021gfu}
X.-F. Han, T.~Li, H.-X. Wang, L.~Wang and Y.~Zhang, \textit{{Lepton-specific
  inert two-Higgs-doublet model confronted with the new results for muon and
  electron g-2 anomalies and multi-lepton searches at the LHC}},
  \href{https://arxiv.org/abs/2104.03227}{\texttt{2104.03227}}.

\bibitem{Chen:2021jok}
C.-H. Chen, C.-W. Chiang and T.~Nomura, \textit{{Muon $g-2$ in
  two-Higgs-doublet model with type-II seesaw mechanism}},
  \href{https://arxiv.org/abs/2104.03275}{\texttt{2104.03275}}.

\bibitem{Ghosh:2020tfq}
N.~Ghosh and J.~Lahiri, \textit{{Revisiting a generalized two-Higgs-doublet
  model in light of the muon anomaly and lepton flavor violating decays at the
  HL-LHC}}, \href{http://dx.doi.org/10.1103/PhysRevD.103.055009}{\emph{Phys.
  Rev. D} \textbf{ 103} (2021) 055009},
  [\href{https://arxiv.org/abs/2010.03590}{\texttt{2010.03590}}].

\bibitem{Ghosh:2021jeg}
N.~Ghosh and J.~Lahiri, \textit{{Generalized 2HDM with wrong-sign lepton Yukawa
  coupling, in light of $g_{\mu}-2$ and lepton flavor violation at the future
  LHC}},  \href{https://arxiv.org/abs/2103.10632}{\texttt{2103.10632}}.

\bibitem{Li:2020dbg}
S.-P. Li, X.-Q. Li, Y.-Y. Li, Y.-D. Yang and X.~Zhang, \textit{{Power-aligned
  2HDM: a correlative perspective on $(g-2)_{e,\mu}$}},
  \href{http://dx.doi.org/10.1007/JHEP01(2021)034}{\emph{JHEP} \textbf{ 01}
  (2021) 034}, [\href{https://arxiv.org/abs/2010.02799}{\texttt{2010.02799}}].

\bibitem{Botella:2020xzf}
F.~J. Botella, F.~Cornet-Gomez and M.~Nebot, \textit{{Electron and muon $g-2$
  anomalies in general flavour conserving two Higgs doublets models}},
  \href{http://dx.doi.org/10.1103/PhysRevD.102.035023}{\emph{Phys. Rev. D}
  \textbf{ 102} (2020) 035023},
  [\href{https://arxiv.org/abs/2006.01934}{\texttt{2006.01934}}].

\bibitem{Jana:2020pxx}
S.~Jana, V.~P. K. and S.~Saad, \textit{{Resolving electron and muon $g-2$
  within the 2HDM}},
  \href{http://dx.doi.org/10.1103/PhysRevD.101.115037}{\emph{Phys. Rev. D}
  \textbf{ 101} (2020) 115037},
  [\href{https://arxiv.org/abs/2003.03386}{\texttt{2003.03386}}].

\bibitem{Jana:2020joi}
S.~Jana, P.~K. Vishnu, W.~Rodejohann and S.~Saad, \textit{{Dark matter assisted
  lepton anomalous magnetic moments and neutrino masses}},
  \href{http://dx.doi.org/10.1103/PhysRevD.102.075003}{\emph{Phys. Rev. D}
  \textbf{ 102} (2020) 075003},
  [\href{https://arxiv.org/abs/2008.02377}{\texttt{2008.02377}}].

\bibitem{Anselmi:2021chp}
D.~Anselmi, K.~Kannike, C.~Marzo, L.~Marzola, A.~Melis, K.~M\"u\"ursepp et~al.,
  \textit{{A fake doublet solution to the muon anomalous magnetic moment}},
  \href{https://arxiv.org/abs/2104.03249}{\texttt{2104.03249}}.

\bibitem{Keus:2017ioh}
V.~Keus, N.~Koivunen and K.~Tuominen, \textit{{Singlet scalar and 2HDM
  extensions of the Standard Model: CP-violation and constraints from
  $(g-2)_\mu$ and $e$EDM}},
  \href{http://dx.doi.org/10.1007/JHEP09(2018)059}{\emph{JHEP} \textbf{ 09}
  (2018) 059}, [\href{https://arxiv.org/abs/1712.09613}{\texttt{1712.09613}}].

\bibitem{Sabatta:2019nfg}
D.~Sabatta, A.~S. Cornell, A.~Goyal, M.~Kumar, B.~Mellado and X.~Ruan,
  \textit{{Connecting muon anomalous magnetic moment and multi-lepton anomalies
  at LHC}}, \href{http://dx.doi.org/10.1088/1674-1137/44/6/063103}{\emph{Chin.
  Phys. C} \textbf{ 44} (2020) 063103},
  [\href{https://arxiv.org/abs/1909.03969}{\texttt{1909.03969}}].

\bibitem{Schmidt-Hoberg:2013hba}
K.~Schmidt-Hoberg, F.~Staub and M.~W. Winkler, \textit{{Constraints on light
  mediators: confronting dark matter searches with B physics}},
  \href{http://dx.doi.org/10.1016/j.physletb.2013.11.015}{\emph{Phys. Lett. B}
  \textbf{ 727} (2013) 506--510},
  [\href{https://arxiv.org/abs/1310.6752}{\texttt{1310.6752}}].

\bibitem{Cao:2009as}
J.~Cao, P.~Wan, L.~Wu and J.~M. Yang, \textit{{Lepton-Specific Two-Higgs
  Doublet Model: Experimental Constraints and Implication on Higgs
  Phenomenology}},
  \href{http://dx.doi.org/10.1103/PhysRevD.80.071701}{\emph{Phys. Rev. D}
  \textbf{ 80} (2009) 071701},
  [\href{https://arxiv.org/abs/0909.5148}{\texttt{0909.5148}}].

\bibitem{Broggio:2014mna}
A.~Broggio, E.~J. Chun, M.~Passera, K.~M. Patel and S.~K. Vempati,
  \textit{{Limiting two-Higgs-doublet models}},
  \href{http://dx.doi.org/10.1007/JHEP11(2014)058}{\emph{JHEP} \textbf{ 11}
  (2014) 058}, [\href{https://arxiv.org/abs/1409.3199}{\texttt{1409.3199}}].

\bibitem{Wang:2014sda}
L.~Wang and X.-F. Han, \textit{{A light pseudoscalar of 2HDM confronted with
  muon g-2 and experimental constraints}},
  \href{http://dx.doi.org/10.1007/JHEP05(2015)039}{\emph{JHEP} \textbf{ 05}
  (2015) 039}, [\href{https://arxiv.org/abs/1412.4874}{\texttt{1412.4874}}].

\bibitem{Abe:2015oca}
T.~Abe, R.~Sato and K.~Yagyu, \textit{{Lepton-specific two Higgs doublet model
  as a solution of muon g \ensuremath{-} 2 anomaly}},
  \href{http://dx.doi.org/10.1007/JHEP07(2015)064}{\emph{JHEP} \textbf{ 07}
  (2015) 064}, [\href{https://arxiv.org/abs/1504.07059}{\texttt{1504.07059}}].

\bibitem{Chun:2017yob}
E.~J. Chun, S.~Dwivedi, T.~Mondal and B.~Mukhopadhyaya, \textit{{Reconstructing
  a light pseudoscalar in the Type-X Two Higgs Doublet Model}},
  \href{http://dx.doi.org/10.1016/j.physletb.2017.09.037}{\emph{Phys. Lett. B}
  \textbf{ 774} (2017) 20--25},
  [\href{https://arxiv.org/abs/1707.07928}{\texttt{1707.07928}}].

\bibitem{Chun:2016hzs}
E.~J. Chun and J.~Kim, \textit{{Leptonic Precision Test of Leptophilic
  Two-Higgs-Doublet Model}},
  \href{http://dx.doi.org/10.1007/JHEP07(2016)110}{\emph{JHEP} \textbf{ 07}
  (2016) 110}, [\href{https://arxiv.org/abs/1605.06298}{\texttt{1605.06298}}].

\bibitem{Cherchiglia:2017uwv}
A.~Cherchiglia, D.~St\"ockinger and H.~St\"ockinger-Kim, \textit{{Muon g-2 in
  the 2HDM: maximum results and detailed phenomenology}},
  \href{http://dx.doi.org/10.1103/PhysRevD.98.035001}{\emph{Phys. Rev. D}
  \textbf{ 98} (2018) 035001},
  [\href{https://arxiv.org/abs/1711.11567}{\texttt{1711.11567}}].

\bibitem{Wang:2018hnw}
L.~Wang, J.~M. Yang, M.~Zhang and Y.~Zhang, \textit{{Revisiting lepton-specific
  2HDM in light of muon $g−2$ anomaly}},
  \href{http://dx.doi.org/10.1016/j.physletb.2018.11.045}{\emph{Phys. Lett. B}
  \textbf{ 788} (2019) 519--529},
  [\href{https://arxiv.org/abs/1809.05857}{\texttt{1809.05857}}].

\bibitem{CMS:2018cyk}
{\scshape CMS} collaboration, A.~M. Sirunyan et~al., \textit{{Search for a
  standard model-like Higgs boson in the mass range between 70 and 110 GeV in
  the diphoton final state in proton-proton collisions at $\sqrt{s}=$ 8 and 13
  TeV}}, \href{http://dx.doi.org/10.1016/j.physletb.2019.03.064}{\emph{Phys.
  Lett. B} \textbf{ 793} (2019) 320--347},
  [\href{https://arxiv.org/abs/1811.08459}{\texttt{1811.08459}}].

\bibitem{Barr:1990vd}
S.~M. Barr and A.~Zee, \textit{{Electric Dipole Moment of the Electron and of
  the Neutron}}, \href{http://dx.doi.org/10.1103/PhysRevLett.65.21}{\emph{Phys.
  Rev. Lett.} \textbf{ 65} (1990) 21--24}.

\bibitem{Ilisie:2015tra}
V.~Ilisie, \textit{{New Barr-Zee contributions to $\mathbf{(g-2)_\mu}$ in
  two-Higgs-doublet models}},
  \href{http://dx.doi.org/10.1007/JHEP04(2015)077}{\emph{JHEP} \textbf{ 04}
  (2015) 077}, [\href{https://arxiv.org/abs/1502.04199}{\texttt{1502.04199}}].

\bibitem{Parker:2018vye}
R.~H. Parker, C.~Yu, W.~Zhong, B.~Estey and H.~M\"uller, \textit{{Measurement
  of the fine-structure constant as a test of the Standard Model}},
  \href{http://dx.doi.org/10.1126/science.aap7706}{\emph{Science} \textbf{ 360}
  (2018) 191}, [\href{https://arxiv.org/abs/1812.04130}{\texttt{1812.04130}}].

\bibitem{Morel:2020dww}
L.~Morel, Z.~Yao, P.~Clad\'e and S.~Guellati-Kh\'elifa, \textit{{Determination
  of the fine-structure constant with an accuracy of 81 parts per trillion}},
  \href{http://dx.doi.org/10.1038/s41586-020-2964-7}{\emph{Nature} \textbf{
  588} (2020) 61--65}.

\bibitem{HFLAV:2019otj}
{\scshape HFLAV} collaboration, Y.~S. Amhis et~al., \textit{{Averages of
  b-hadron, c-hadron, and $\tau $-lepton properties as of 2018}},
  \href{http://dx.doi.org/10.1140/epjc/s10052-020-8156-7}{\emph{Eur. Phys. J.
  C} \textbf{ 81} (2021) 226},
  [\href{https://arxiv.org/abs/1909.12524}{\texttt{1909.12524}}].

\bibitem{ALEPH:2001gaj}
{\scshape ALEPH} collaboration, A.~Heister et~al., \textit{{Measurement of the
  Michel parameters and the nu/tau helicity in tau lepton decays}},
  \href{http://dx.doi.org/10.1007/s100520100813}{\emph{Eur. Phys. J. C}
  \textbf{ 22} (2001) 217--230}.

\bibitem{Branco:2011iw}
G.~C. Branco, P.~M. Ferreira, L.~Lavoura, M.~N. Rebelo, M.~Sher and J.~P.
  Silva, \textit{{Theory and phenomenology of two-Higgs-doublet models}},
  \href{http://dx.doi.org/10.1016/j.physrep.2012.02.002}{\emph{Phys. Rept.}
  \textbf{ 516} (2012) 1--102},
  [\href{https://arxiv.org/abs/1106.0034}{\texttt{1106.0034}}].

\bibitem{Glashow:1976nt}
S.~L. Glashow and S.~Weinberg, \textit{{Natural Conservation Laws for Neutral
  Currents}}, \href{http://dx.doi.org/10.1103/PhysRevD.15.1958}{\emph{Phys.
  Rev.} \textbf{ D15} (1977) 1958}.

\bibitem{Paschos:1976ay}
E.~A. Paschos, \textit{{Diagonal Neutral Currents}},
  \href{http://dx.doi.org/10.1103/PhysRevD.15.1966}{\emph{Phys. Rev.} \textbf{
  D15} (1977) 1966}.

\bibitem{Aoki:2009ha}
M.~Aoki, S.~Kanemura, K.~Tsumura and K.~Yagyu, \textit{{Models of Yukawa
  interaction in the two Higgs doublet model, and their collider
  phenomenology}},
  \href{http://dx.doi.org/10.1103/PhysRevD.80.015017}{\emph{Phys. Rev.}
  \textbf{ D80} (2009) 015017},
  [\href{https://arxiv.org/abs/0902.4665}{\texttt{0902.4665}}].

\bibitem{Song:2019aav}
J.~Song and Y.~W. Yoon, \textit{{$W\gamma$ decay of the elusive charged Higgs
  boson in the two-Higgs-doublet model with vectorlike fermions}},
  \href{http://dx.doi.org/10.1103/PhysRevD.100.055006}{\emph{Phys. Rev.}
  \textbf{ D100} (2019) 055006},
  [\href{https://arxiv.org/abs/1904.06521}{\texttt{1904.06521}}].

\bibitem{ATLAS:2020qdt}
{\scshape ATLAS} collaboration, \textit{{A combination of measurements of Higgs
  boson production and decay using up to $139$ fb$^{-1}$ of proton--proton
  collision data at $\sqrt{s}=$ 13 TeV collected with the ATLAS experiment}}, .

\bibitem{Carena:2013ooa}
M.~Carena, I.~Low, N.~R. Shah and C.~E.~M. Wagner, \textit{{Impersonating the
  Standard Model Higgs Boson: Alignment without Decoupling}},
  \href{http://dx.doi.org/10.1007/JHEP04(2014)015}{\emph{JHEP} \textbf{ 04}
  (2014) 015}, [\href{https://arxiv.org/abs/1310.2248}{\texttt{1310.2248}}].

\bibitem{Celis:2013rcs}
A.~Celis, V.~Ilisie and A.~Pich, \textit{{LHC constraints on two-Higgs doublet
  models}}, \href{http://dx.doi.org/10.1007/JHEP07(2013)053}{\emph{JHEP}
  \textbf{ 07} (2013) 053},
  [\href{https://arxiv.org/abs/1302.4022}{\texttt{1302.4022}}].

\bibitem{Bernon:2015qea}
J.~Bernon, J.~F. Gunion, H.~E. Haber, Y.~Jiang and S.~Kraml,
  \textit{{Scrutinizing the alignment limit in two-Higgs-doublet models:
  m$_h$=125 GeV}},
  \href{http://dx.doi.org/10.1103/PhysRevD.92.075004}{\emph{Phys. Rev. D}
  \textbf{ 92} (2015) 075004},
  [\href{https://arxiv.org/abs/1507.00933}{\texttt{1507.00933}}].

\bibitem{Chang:2015goa}
S.~Chang, S.~K. Kang, J.-P. Lee and J.~Song, \textit{{Higgs potential and
  hidden light Higgs scenario in two Higgs doublet models}},
  \href{http://dx.doi.org/10.1103/PhysRevD.92.075023}{\emph{Phys. Rev. D}
  \textbf{ 92} (2015) 075023},
  [\href{https://arxiv.org/abs/1507.03618}{\texttt{1507.03618}}].

\bibitem{Das:2015mwa}
D.~Das and I.~Saha, \textit{{Search for a stable alignment limit in
  two-Higgs-doublet models}},
  \href{http://dx.doi.org/10.1103/PhysRevD.91.095024}{\emph{Phys. Rev. D}
  \textbf{ 91} (2015) 095024},
  [\href{https://arxiv.org/abs/1503.02135}{\texttt{1503.02135}}].

\bibitem{Bernon:2015wef}
J.~Bernon, J.~F. Gunion, H.~E. Haber, Y.~Jiang and S.~Kraml,
  \textit{{Scrutinizing the alignment limit in two-Higgs-doublet models. II.
  m$_H$=125 GeV}},
  \href{http://dx.doi.org/10.1103/PhysRevD.93.035027}{\emph{Phys. Rev. D}
  \textbf{ 93} (2016) 035027},
  [\href{https://arxiv.org/abs/1511.03682}{\texttt{1511.03682}}].

\bibitem{Aad:2019mbh}
{\scshape ATLAS} collaboration, G.~Aad et~al., \textit{{Combined measurements
  of Higgs boson production and decay using up to $80$ fb$^{-1}$ of
  proton-proton collision data at $\sqrt{s}=$ 13 TeV collected with the ATLAS
  experiment}},
  \href{http://dx.doi.org/10.1103/PhysRevD.101.012002}{\emph{Phys. Rev. D}
  \textbf{ 101} (2020) 012002},
  [\href{https://arxiv.org/abs/1909.02845}{\texttt{1909.02845}}].

\bibitem{Ivanov:2006yq}
I.~P. Ivanov, \textit{{Minkowski space structure of the Higgs potential in
  2HDM}}, \href{http://dx.doi.org/10.1103/PhysRevD.75.035001}{\emph{Phys. Rev.
  D} \textbf{ 75} (2007) 035001},
  [\href{https://arxiv.org/abs/hep-ph/0609018}{\texttt{hep-ph/0609018}}].

\bibitem{Barroso:2013awa}
A.~Barroso, P.~M. Ferreira, I.~P. Ivanov and R.~Santos, \textit{{Metastability
  bounds on the two Higgs doublet model}},
  \href{http://dx.doi.org/10.1007/JHEP06(2013)045}{\emph{JHEP} \textbf{ 06}
  (2013) 045}, [\href{https://arxiv.org/abs/1303.5098}{\texttt{1303.5098}}].

\bibitem{Arhrib:2000is}
A.~Arhrib, \textit{{Unitarity constraints on scalar parameters of the standard
  and two Higgs doublets model}},  in \emph{{Workshop on Noncommutative
  Geometry, Superstrings and Particle Physics}}, 12, 2000.
\newblock \href{https://arxiv.org/abs/hep-ph/0012353}{\texttt{hep-ph/0012353}}.

\bibitem{Peskin:1991sw}
M.~E. Peskin and T.~Takeuchi, \textit{{Estimation of oblique electroweak
  corrections}}, \href{http://dx.doi.org/10.1103/PhysRevD.46.381}{\emph{Phys.
  Rev. D} \textbf{ 46} (1992) 381--409}.

\bibitem{PDG2020}
{\scshape Particle Data Group} collaboration, P.~A. Zyla et~al.,
  \textit{{Review of Particle Physics}},
  \href{http://dx.doi.org/10.1093/ptep/ptaa104}{\emph{PTEP} \textbf{ 2020}
  (2020) 083C01}.

\bibitem{He:2001tp}
H.-J. He, N.~Polonsky and S.-f. Su, \textit{{Extra families, Higgs spectrum and
  oblique corrections}},
  \href{http://dx.doi.org/10.1103/PhysRevD.64.053004}{\emph{Phys. Rev. D}
  \textbf{ 64} (2001) 053004},
  [\href{https://arxiv.org/abs/hep-ph/0102144}{\texttt{hep-ph/0102144}}].

\bibitem{Grimus:2008nb}
W.~Grimus, L.~Lavoura, O.~M. Ogreid and P.~Osland, \textit{{The Oblique
  parameters in multi-Higgs-doublet models}},
  \href{http://dx.doi.org/10.1016/j.nuclphysb.2008.04.019}{\emph{Nucl. Phys. B}
  \textbf{ 801} (2008) 81--96},
  [\href{https://arxiv.org/abs/0802.4353}{\texttt{0802.4353}}].

\bibitem{Bechtle:2013xfa}
P.~Bechtle, S.~Heinemeyer, O.~St\r{a}l, T.~Stefaniak and G.~Weiglein,
  \textit{{$HiggsSignals$: Confronting arbitrary Higgs sectors with
  measurements at the Tevatron and the LHC}},
  \href{http://dx.doi.org/10.1140/epjc/s10052-013-2711-4}{\emph{Eur. Phys. J.
  C} \textbf{ 74} (2014) 2711},
  [\href{https://arxiv.org/abs/1305.1933}{\texttt{1305.1933}}].

\bibitem{Bechtle:2020uwn}
P.~Bechtle, S.~Heinemeyer, T.~Klingl, T.~Stefaniak, G.~Weiglein and
  J.~Wittbrodt, \textit{{HiggsSignals-2: Probing new physics with precision
  Higgs measurements in the LHC 13 TeV era}},
  \href{http://dx.doi.org/10.1140/epjc/s10052-021-08942-y}{\emph{Eur. Phys. J.
  C} \textbf{ 81} (2021) 145},
  [\href{https://arxiv.org/abs/2012.09197}{\texttt{2012.09197}}].

\bibitem{Bechtle:2020pkv}
P.~Bechtle, D.~Dercks, S.~Heinemeyer, T.~Klingl, T.~Stefaniak, G.~Weiglein
  et~al., \textit{{HiggsBounds-5: Testing Higgs Sectors in the LHC 13 TeV
  Era}}, \href{http://dx.doi.org/10.1140/epjc/s10052-020-08557-9}{\emph{Eur.
  Phys. J. C} \textbf{ 80} (2020) 1211},
  [\href{https://arxiv.org/abs/2006.06007}{\texttt{2006.06007}}].

\bibitem{Schael:2006cr}
{\scshape ALEPH, DELPHI, L3, OPAL, LEP Working Group for Higgs Boson Searches}
  collaboration, S.~Schael et~al., \textit{{Search for neutral MSSM Higgs
  bosons at LEP}},
  \href{http://dx.doi.org/10.1140/epjc/s2006-02569-7}{\emph{Eur. Phys. J. C}
  \textbf{ 47} (2006) 547--587},
  [\href{https://arxiv.org/abs/hep-ex/0602042}{\texttt{hep-ex/0602042}}].

\bibitem{Aaboud:2018esj}
{\scshape ATLAS} collaboration, M.~Aaboud et~al., \textit{{Search for Higgs
  boson decays into a pair of light bosons in the $bb\mu\mu$ final state in
  $pp$ collision at $\sqrt{s} = $13 TeV with the ATLAS detector}},
  \href{http://dx.doi.org/10.1016/j.physletb.2018.10.073}{\emph{Phys. Lett. B}
  \textbf{ 790} (2019) 1--21},
  [\href{https://arxiv.org/abs/1807.00539}{\texttt{1807.00539}}].

\bibitem{Aaboud:2018iil}
{\scshape ATLAS} collaboration, M.~Aaboud et~al., \textit{{Search for the Higgs
  boson produced in association with a vector boson and decaying into two
  spin-zero particles in the $H \rightarrow aa \rightarrow 4b$ channel in $pp$
  collisions at $\sqrt{s} = 13$ TeV with the ATLAS detector}},
  \href{http://dx.doi.org/10.1007/JHEP10(2018)031}{\emph{JHEP} \textbf{ 10}
  (2018) 031}, [\href{https://arxiv.org/abs/1806.07355}{\texttt{1806.07355}}].

\bibitem{Sirunyan:2018mbx}
{\scshape CMS} collaboration, A.~M. Sirunyan et~al., \textit{{Search for an
  exotic decay of the Higgs boson to a pair of light pseudoscalars in the final
  state of two muons and two $\tau$ leptons in proton-proton collisions at $
  \sqrt{s}=13 $ TeV}},
  \href{http://dx.doi.org/10.1007/JHEP11(2018)018}{\emph{JHEP} \textbf{ 11}
  (2018) 018}, [\href{https://arxiv.org/abs/1805.04865}{\texttt{1805.04865}}].

\bibitem{Sirunyan:2018mot}
{\scshape CMS} collaboration, A.~M. Sirunyan et~al., \textit{{Search for an
  exotic decay of the Higgs boson to a pair of light pseudoscalars in the final
  state with two muons and two b quarks in pp collisions at 13 TeV}},
  \href{http://dx.doi.org/10.1016/j.physletb.2019.06.021}{\emph{Phys. Lett. B}
  \textbf{ 795} (2019) 398--423},
  [\href{https://arxiv.org/abs/1812.06359}{\texttt{1812.06359}}].

\bibitem{Sirunyan:2018pzn}
{\scshape CMS} collaboration, A.~M. Sirunyan et~al., \textit{{Search for an
  exotic decay of the Higgs boson to a pair of light pseudoscalars in the final
  state with two b quarks and two $\tau$ leptons in proton-proton collisions at
  $\sqrt{s}=$ 13 TeV}},
  \href{http://dx.doi.org/10.1016/j.physletb.2018.08.057}{\emph{Phys. Lett. B}
  \textbf{ 785} (2018) 462},
  [\href{https://arxiv.org/abs/1805.10191}{\texttt{1805.10191}}].

\bibitem{Sirunyan:2019gou}
{\scshape CMS} collaboration, A.~M. Sirunyan et~al., \textit{{Search for light
  pseudoscalar boson pairs produced from decays of the 125 GeV Higgs boson in
  final states with two muons and two nearby tracks in pp collisions at
  $\sqrt{s}=$ 13 TeV}},
  \href{http://dx.doi.org/10.1016/j.physletb.2019.135087}{\emph{Phys. Lett. B}
  \textbf{ 800} (2020) 135087},
  [\href{https://arxiv.org/abs/1907.07235}{\texttt{1907.07235}}].

\bibitem{Aaboud:2018bun}
{\scshape ATLAS} collaboration, M.~Aaboud et~al., \textit{{Combination of
  searches for heavy resonances decaying into bosonic and leptonic final states
  using 36 fb$^{-1}$ of proton-proton collision data at $\sqrt{s} = 13$ TeV
  with the ATLAS detector}},
  \href{http://dx.doi.org/10.1103/PhysRevD.98.052008}{\emph{Phys. Rev. D}
  \textbf{ 98} (2018) 052008},
  [\href{https://arxiv.org/abs/1808.02380}{\texttt{1808.02380}}].

\bibitem{Sirunyan:2018qlb}
{\scshape CMS} collaboration, A.~M. Sirunyan et~al., \textit{{Search for a new
  scalar resonance decaying to a pair of Z bosons in proton-proton collisions
  at $\sqrt{s}=13 $ TeV}},
  \href{http://dx.doi.org/10.1007/JHEP06(2018)127}{\emph{JHEP} \textbf{ 06}
  (2018) 127}, [\href{https://arxiv.org/abs/1804.01939}{\texttt{1804.01939}}].

\bibitem{Sirunyan:2019pqw}
{\scshape CMS} collaboration, A.~M. Sirunyan et~al., \textit{{Search for a
  heavy Higgs boson decaying to a pair of W bosons in proton-proton collisions
  at $\sqrt{s} =$ 13 TeV}},
  \href{http://dx.doi.org/10.1007/JHEP03(2020)034}{\emph{JHEP} \textbf{ 03}
  (2020) 034}, [\href{https://arxiv.org/abs/1912.01594}{\texttt{1912.01594}}].

\bibitem{Sirunyan:2018ayu}
{\scshape CMS} collaboration, A.~M. Sirunyan et~al., \textit{{Combination of
  searches for Higgs boson pair production in proton-proton collisions at
  $\sqrt{s} = $ 13 TeV}},
  \href{http://dx.doi.org/10.1103/PhysRevLett.122.121803}{\emph{Phys. Rev.
  Lett.} \textbf{ 122} (2019) 121803},
  [\href{https://arxiv.org/abs/1811.09689}{\texttt{1811.09689}}].

\bibitem{Aaboud:2018ewm}
{\scshape ATLAS} collaboration, M.~Aaboud et~al., \textit{{Search for Higgs
  boson pair production in the $\gamma\gamma WW^{*}$ channel using $pp$
  collision data recorded at $\sqrt{s} = 13$ TeV with the ATLAS detector}},
  \href{http://dx.doi.org/10.1140/epjc/s10052-018-6457-x}{\emph{Eur. Phys. J.
  C} \textbf{ 78} (2018) 1007},
  [\href{https://arxiv.org/abs/1807.08567}{\texttt{1807.08567}}].

\bibitem{Aaboud:2018knk}
{\scshape ATLAS} collaboration, M.~Aaboud et~al., \textit{{Search for pair
  production of Higgs bosons in the $b\bar{b}b\bar{b}$ final state using
  proton-proton collisions at $\sqrt{s} = 13$ TeV with the ATLAS detector}},
  \href{http://dx.doi.org/10.1007/JHEP01(2019)030}{\emph{JHEP} \textbf{ 01}
  (2019) 030}, [\href{https://arxiv.org/abs/1804.06174}{\texttt{1804.06174}}].

\bibitem{Aaboud:2018ksn}
{\scshape ATLAS} collaboration, M.~Aaboud et~al., \textit{{Search for Higgs
  boson pair production in the $WW^{(*)}WW^{(*)}$ decay channel using ATLAS
  data recorded at $\sqrt{s}=13$ TeV}},
  \href{http://dx.doi.org/10.1007/JHEP05(2019)124}{\emph{JHEP} \textbf{ 05}
  (2019) 124}, [\href{https://arxiv.org/abs/1811.11028}{\texttt{1811.11028}}].

\bibitem{Aaboud:2018sfw}
{\scshape ATLAS} collaboration, M.~Aaboud et~al., \textit{{Search for resonant
  and non-resonant Higgs boson pair production in the ${b\bar{b}\tau^+\tau^-}$
  decay channel in $pp$ collisions at $\sqrt{s}=13$ TeV with the ATLAS
  detector}},
  \href{http://dx.doi.org/10.1103/PhysRevLett.121.191801}{\emph{Phys. Rev.
  Lett.} \textbf{ 121} (2018) 191801},
  [\href{https://arxiv.org/abs/1808.00336}{\texttt{1808.00336}}].

\bibitem{Aad:2019uzh}
{\scshape ATLAS} collaboration, G.~Aad et~al., \textit{{Combination of searches
  for Higgs boson pairs in $pp$ collisions at $\sqrt{s} = $13 TeV with the
  ATLAS detector}},
  \href{http://dx.doi.org/10.1016/j.physletb.2019.135103}{\emph{Phys. Lett. B}
  \textbf{ 800} (2020) 135103},
  [\href{https://arxiv.org/abs/1906.02025}{\texttt{1906.02025}}].

\bibitem{Aad:2020kub}
{\scshape ATLAS} collaboration, G.~Aad et~al., \textit{{Search for the $HH
  \rightarrow b \bar{b} b \bar{b}$ process via vector-boson fusion production
  using proton-proton collisions at $\sqrt{s} = 13$ TeV with the ATLAS
  detector}}, \href{http://dx.doi.org/10.1007/JHEP07(2020)108}{\emph{JHEP}
  \textbf{ 07} (2020) 108},
  [\href{https://arxiv.org/abs/2001.05178}{\texttt{2001.05178}}].

\bibitem{Aaboud:2017yyg}
{\scshape ATLAS} collaboration, M.~Aaboud et~al., \textit{{Search for new
  phenomena in high-mass diphoton final states using 37 fb$^{-1}$ of
  proton--proton collisions collected at $\sqrt{s}=13$ TeV with the ATLAS
  detector}},
  \href{http://dx.doi.org/10.1016/j.physletb.2017.10.039}{\emph{Phys. Lett. B}
  \textbf{ 775} (2017) 105--125},
  [\href{https://arxiv.org/abs/1707.04147}{\texttt{1707.04147}}].

\bibitem{Sirunyan:2018aui}
{\scshape CMS} collaboration, A.~M. Sirunyan et~al., \textit{{Search for a
  standard model-like Higgs boson in the mass range between 70 and 110 GeV in
  the diphoton final state in proton-proton collisions at $\sqrt{s}=$ 8 and 13
  TeV}}, \href{http://dx.doi.org/10.1016/j.physletb.2019.03.064}{\emph{Phys.
  Lett. B} \textbf{ 793} (2019) 320--347},
  [\href{https://arxiv.org/abs/1811.08459}{\texttt{1811.08459}}].

\bibitem{Sirunyan:2018zut}
{\scshape CMS} collaboration, A.~M. Sirunyan et~al., \textit{{Search for
  additional neutral MSSM Higgs bosons in the $\tau\tau$ final state in
  proton-proton collisions at $\sqrt{s}=$ 13 TeV}},
  \href{http://dx.doi.org/10.1007/JHEP09(2018)007}{\emph{JHEP} \textbf{ 09}
  (2018) 007}, [\href{https://arxiv.org/abs/1803.06553}{\texttt{1803.06553}}].

\bibitem{Aad:2020zxo}
{\scshape ATLAS} collaboration, G.~Aad et~al., \textit{{Search for heavy Higgs
  bosons decaying into two tau leptons with the ATLAS detector using $pp$
  collisions at $\sqrt{s}=13$ TeV}},
  \href{http://dx.doi.org/10.1103/PhysRevLett.125.051801}{\emph{Phys. Rev.
  Lett.} \textbf{ 125} (2020) 051801},
  [\href{https://arxiv.org/abs/2002.12223}{\texttt{2002.12223}}].

\bibitem{Aad:2019fac}
{\scshape ATLAS} collaboration, G.~Aad et~al., \textit{{Search for high-mass
  dilepton resonances using 139 fb$^{-1}$ of $pp$ collision data collected at
  $\sqrt{s}=$13 TeV with the ATLAS detector}},
  \href{http://dx.doi.org/10.1016/j.physletb.2019.07.016}{\emph{Phys. Lett. B}
  \textbf{ 796} (2019) 68--87},
  [\href{https://arxiv.org/abs/1903.06248}{\texttt{1903.06248}}].

\bibitem{Aaboud:2019sgt}
{\scshape ATLAS} collaboration, M.~Aaboud et~al., \textit{{Search for scalar
  resonances decaying into $\mu^{+}\mu^{-}$ in events with and without
  $b$-tagged jets produced in proton-proton collisions at $\sqrt{s}=13$ TeV
  with the ATLAS detector}},
  \href{http://dx.doi.org/10.1007/JHEP07(2019)117}{\emph{JHEP} \textbf{ 07}
  (2019) 117}, [\href{https://arxiv.org/abs/1901.08144}{\texttt{1901.08144}}].

\bibitem{Sirunyan:2019tkw}
{\scshape CMS} collaboration, A.~M. Sirunyan et~al., \textit{{Search for MSSM
  Higgs bosons decaying to \ensuremath{\mu} + \ensuremath{\mu} \ensuremath{-}
  in proton-proton collisions at s=13TeV}},
  \href{http://dx.doi.org/10.1016/j.physletb.2019.134992}{\emph{Phys. Lett. B}
  \textbf{ 798} (2019) 134992},
  [\href{https://arxiv.org/abs/1907.03152}{\texttt{1907.03152}}].

\bibitem{Sirunyan:2018ikr}
{\scshape CMS} collaboration, A.~M. Sirunyan et~al., \textit{{Search for
  low-mass resonances decaying into bottom quark-antiquark pairs in
  proton-proton collisions at $\sqrt{s} =$ 13 TeV}},
  \href{http://dx.doi.org/10.1103/PhysRevD.99.012005}{\emph{Phys. Rev. D}
  \textbf{ 99} (2019) 012005},
  [\href{https://arxiv.org/abs/1810.11822}{\texttt{1810.11822}}].

\bibitem{Sirunyan:2018taj}
{\scshape CMS} collaboration, A.~M. Sirunyan et~al., \textit{{Search for beyond
  the standard model Higgs bosons decaying into a $\mathrm{b\overline{b}}$ pair
  in pp collisions at $\sqrt{s} =$ 13 TeV}},
  \href{http://dx.doi.org/10.1007/JHEP08(2018)113}{\emph{JHEP} \textbf{ 08}
  (2018) 113}, [\href{https://arxiv.org/abs/1805.12191}{\texttt{1805.12191}}].

\bibitem{Aad:2019zwb}
{\scshape ATLAS} collaboration, G.~Aad et~al., \textit{{Search for heavy
  neutral Higgs bosons produced in association with $b$-quarks and decaying
  into $b$-quarks at $\sqrt{s}=13$ TeV with the ATLAS detector}},
  \href{http://dx.doi.org/10.1103/PhysRevD.102.032004}{\emph{Phys. Rev. D}
  \textbf{ 102} (2020) 032004},
  [\href{https://arxiv.org/abs/1907.02749}{\texttt{1907.02749}}].

\bibitem{Sirunyan:2019wph}
{\scshape CMS} collaboration, A.~M. Sirunyan et~al., \textit{{Search for heavy
  Higgs bosons decaying to a top quark pair in proton-proton collisions at
  $\sqrt{s} =$ 13 TeV}},
  \href{http://dx.doi.org/10.1007/JHEP04(2020)171}{\emph{JHEP} \textbf{ 04}
  (2020) 171}, [\href{https://arxiv.org/abs/1908.01115}{\texttt{1908.01115}}].

\bibitem{Aad:2015wra}
{\scshape ATLAS} collaboration, G.~Aad et~al., \textit{{Search for a CP-odd
  Higgs boson decaying to Zh in pp collisions at $\sqrt{s} = 8$ TeV with the
  ATLAS detector}},
  \href{http://dx.doi.org/10.1016/j.physletb.2015.03.054}{\emph{Phys. Lett. B}
  \textbf{ 744} (2015) 163--183},
  [\href{https://arxiv.org/abs/1502.04478}{\texttt{1502.04478}}].

\bibitem{Sirunyan:2019xls}
{\scshape CMS} collaboration, A.~M. Sirunyan et~al., \textit{{Search for a
  heavy pseudoscalar boson decaying to a Z and a Higgs boson at $\sqrt{s} =$ 13
  TeV}}, \href{http://dx.doi.org/10.1140/epjc/s10052-019-7058-z}{\emph{Eur.
  Phys. J. C} \textbf{ 79} (2019) 564},
  [\href{https://arxiv.org/abs/1903.00941}{\texttt{1903.00941}}].

\bibitem{Aaboud:2018eoy}
{\scshape ATLAS} collaboration, M.~Aaboud et~al., \textit{{Search for a heavy
  Higgs boson decaying into a $Z$ boson and another heavy Higgs boson in the
  $\ell\ell bb$ final state in $pp$ collisions at $\sqrt{s}=13$ TeV with the
  ATLAS detector}},
  \href{http://dx.doi.org/10.1016/j.physletb.2018.07.006}{\emph{Phys. Lett. B}
  \textbf{ 783} (2018) 392--414},
  [\href{https://arxiv.org/abs/1804.01126}{\texttt{1804.01126}}].

\bibitem{Sirunyan:2019wrn}
{\scshape CMS} collaboration, A.~M. Sirunyan et~al., \textit{{Search for new
  neutral Higgs bosons through the H$\to$ ZA $\to \ell^{+}\ell^{-}
  \mathrm{b\bar{b}}$ process in pp collisions at $\sqrt{s} =$ 13 TeV}},
  \href{http://dx.doi.org/10.1007/JHEP03(2020)055}{\emph{JHEP} \textbf{ 03}
  (2020) 055}, [\href{https://arxiv.org/abs/1911.03781}{\texttt{1911.03781}}].

\bibitem{Aaboud:2018cwk}
{\scshape ATLAS} collaboration, M.~Aaboud et~al., \textit{{Search for charged
  Higgs bosons decaying into top and bottom quarks at $\sqrt{s}$ = 13 TeV with
  the ATLAS detector}},
  \href{http://dx.doi.org/10.1007/JHEP11(2018)085}{\emph{JHEP} \textbf{ 11}
  (2018) 085}, [\href{https://arxiv.org/abs/1808.03599}{\texttt{1808.03599}}].

\bibitem{Sirunyan:2020hwv}
{\scshape CMS} collaboration, A.~M. Sirunyan et~al., \textit{{Search for
  charged Higgs bosons decaying into a top and a bottom quark in the all-jet
  final state of pp collisions at $ \sqrt{s} $ = 13 TeV}},
  \href{http://dx.doi.org/10.1007/JHEP07(2020)126}{\emph{JHEP} \textbf{ 07}
  (2020) 126}, [\href{https://arxiv.org/abs/2001.07763}{\texttt{2001.07763}}].

\bibitem{Aaboud:2018gjj}
{\scshape ATLAS} collaboration, M.~Aaboud et~al., \textit{{Search for charged
  Higgs bosons decaying via $H^{\pm} \to \tau^{\pm}\nu_{\tau}$ in the
  $\tau$+jets and $\tau$+lepton final states with 36 fb$^{-1}$ of $pp$
  collision data recorded at $\sqrt{s} = 13$ TeV with the ATLAS experiment}},
  \href{http://dx.doi.org/10.1007/JHEP09(2018)139}{\emph{JHEP} \textbf{ 09}
  (2018) 139}, [\href{https://arxiv.org/abs/1807.07915}{\texttt{1807.07915}}].

\bibitem{Sirunyan:2019hkq}
{\scshape CMS} collaboration, A.~M. Sirunyan et~al., \textit{{Search for
  charged Higgs bosons in the H$^{\pm}$ $\to$ $\tau^{\pm}\nu_\tau$ decay
  channel in proton-proton collisions at $\sqrt{s} =$ 13 TeV}},
  \href{http://dx.doi.org/10.1007/JHEP07(2019)142}{\emph{JHEP} \textbf{ 07}
  (2019) 142}, [\href{https://arxiv.org/abs/1903.04560}{\texttt{1903.04560}}].

\bibitem{Khachatryan:2017mnf}
{\scshape CMS} collaboration, V.~Khachatryan et~al., \textit{{Search for light
  bosons in decays of the 125 GeV Higgs boson in proton-proton collisions at $
  \sqrt{s}=8 $ TeV}},
  \href{http://dx.doi.org/10.1007/JHEP10(2017)076}{\emph{JHEP} \textbf{ 10}
  (2017) 076}, [\href{https://arxiv.org/abs/1701.02032}{\texttt{1701.02032}}].

\bibitem{Abbiendi:2013hk}
{\scshape ALEPH, DELPHI, L3, OPAL, LEP} collaboration, G.~Abbiendi et~al.,
  \textit{{Search for Charged Higgs bosons: Combined Results Using LEP Data}},
  \href{http://dx.doi.org/10.1140/epjc/s10052-013-2463-1}{\emph{Eur. Phys. J.
  C} \textbf{ 73} (2013) 2463},
  [\href{https://arxiv.org/abs/1301.6065}{\texttt{1301.6065}}].

\bibitem{Aoyama:2017uqe}
T.~Aoyama, T.~Kinoshita and M.~Nio, \textit{{Revised and Improved Value of the
  QED Tenth-Order Electron Anomalous Magnetic Moment}},
  \href{http://dx.doi.org/10.1103/PhysRevD.97.036001}{\emph{Phys. Rev. D}
  \textbf{ 97} (2018) 036001},
  [\href{https://arxiv.org/abs/1712.06060}{\texttt{1712.06060}}].

\bibitem{Laporta:2017okg}
S.~Laporta, \textit{{High-precision calculation of the 4-loop contribution to
  the electron g-2 in QED}},
  \href{http://dx.doi.org/10.1016/j.physletb.2017.06.056}{\emph{Phys. Lett. B}
  \textbf{ 772} (2017) 232--238},
  [\href{https://arxiv.org/abs/1704.06996}{\texttt{1704.06996}}].


\bibitem{Grazzini:2015hta}
M.~Grazzini, S.~Kallweit and D.~Rathlev, \textit{{ZZ production at the LHC:
  fiducial cross sections and distributions in NNLO QCD}},
  \href{http://dx.doi.org/10.1016/j.physletb.2015.09.055}{\emph{Phys. Lett. B}
  \textbf{ 750} (2015) 407--410},
  [\href{https://arxiv.org/abs/1507.06257}{\texttt{1507.06257}}].

\bibitem{Aad:2015zqe}
{\scshape ATLAS} collaboration, G.~Aad et~al., \textit{{Measurement of the $ZZ$
  Production Cross Section in $pp$ Collisions at $\sqrt{s}$ = 13 TeV with the
  ATLAS Detector}},
  \href{http://dx.doi.org/10.1103/PhysRevLett.116.101801}{\emph{Phys. Rev.
  Lett.} \textbf{ 116} (2016) 101801},
  [\href{https://arxiv.org/abs/1512.05314}{\texttt{1512.05314}}].

\bibitem{Aaboud:2019nkz}
{\scshape ATLAS} collaboration, M.~Aaboud et~al., \textit{{Measurement of
  fiducial and differential $W^+W^-$ production cross-sections at $\sqrt{s}=13$
  TeV with the ATLAS detector}},
  \href{http://dx.doi.org/10.1140/epjc/s10052-019-7371-6}{\emph{Eur. Phys. J.
  C} \textbf{ 79} (2019) 884},
  [\href{https://arxiv.org/abs/1905.04242}{\texttt{1905.04242}}].

\bibitem{Michel:1949qe}
L.~Michel, \textit{{Interaction between four half spin particles and the decay
  of the $\mu$ meson}},
  \href{http://dx.doi.org/10.1088/0370-1298/63/5/311}{\emph{Proc. Phys. Soc. A}
  \textbf{ 63} (1950) 514--531}.

\bibitem{Logan:2009uf}
H.~E. Logan and D.~MacLennan, \textit{{Charged Higgs phenomenology in the
  lepton-specific two Higgs doublet model}},
  \href{http://dx.doi.org/10.1103/PhysRevD.79.115022}{\emph{Phys. Rev. D}
  \textbf{ 79} (2009) 115022},
  [\href{https://arxiv.org/abs/0903.2246}{\texttt{0903.2246}}].

\bibitem{Stahl:1999ui}
A.~Stahl, \textit{{Michel parameters: Averages and interpretation}},
  \href{http://dx.doi.org/10.1016/S0920-5632(99)00454-5}{\emph{Nucl. Phys. B
  Proc. Suppl.} \textbf{ 76} (1999) 173--181}.

\bibitem{Belle:2017wxw}
{\scshape Belle} collaboration, N.~Shimizu et~al., \textit{{Measurement of the
  tau Michel parameters $\bar{\eta}$ and $\xi\kappa$ in the radiative leptonic
  decay $\tau^- \rightarrow \ell^- \nu_{\tau} \bar{\nu}_{\ell}\gamma$}},
  \href{http://dx.doi.org/10.1093/ptep/pty003}{\emph{PTEP} \textbf{ 2018}
  (2018) 023C01},
  [\href{https://arxiv.org/abs/1709.08833}{\texttt{1709.08833}}].

\bibitem{Gentile:1995ue}
S.~Gentile and M.~Pohl, \textit{{Physics of $\tau$ leptons}},
  \href{http://dx.doi.org/10.1016/0370-1573(96)00002-6}{\emph{Phys. Rept.}
  \textbf{ 274} (1996) 287--376}.

\bibitem{ALEPH:2005ab}
{\scshape ALEPH, DELPHI, L3, OPAL, SLD, LEP Electroweak Working Group, SLD
  Electroweak Group, SLD Heavy Flavour Group} collaboration, S.~Schael et~al.,
  \textit{{Precision electroweak measurements on the $Z$ resonance}},
  \href{http://dx.doi.org/10.1016/j.physrep.2005.12.006}{\emph{Phys. Rept.}
  \textbf{ 427} (2006) 257--454},
  [\href{https://arxiv.org/abs/hep-ex/0509008}{\texttt{hep-ex/0509008}}].

\end{thebibliography}


\end{document}